\documentclass{emulateapj}
\shorttitle{AGES III: NGC 7332 \& NGC 1156}
\shortauthors{R. Minchin et al.}
\begin{document}

\title{The Arecibo Galaxy Environment Survey III: observations toward the 
galaxy pair NGC 7332/7339 and the isolated galaxy NGC 1156.}

\author{R. F. Minchin,\altaffilmark{1}
E. Momjian,\altaffilmark{2}
R. Auld,\altaffilmark{3}
J. I. Davies,\altaffilmark{3}
D. Valls-Gabaud,\altaffilmark{4}
I. D. Karachentsev,\altaffilmark{5}
P. A. Henning,\altaffilmark{6}
K. L. O'Neil,\altaffilmark{7}
S. Schneider,\altaffilmark{8}
M. W. L. Smith,\altaffilmark{3}
M. D. Stage,\altaffilmark{8}
R. Taylor,\altaffilmark{3}
and W. van Driel\altaffilmark{4}
\altaffiltext{1}{Arecibo Observatory, HC03 Box 53995, Arecibo, PR 00612.}
\altaffiltext{2}{NRAO, Dominici Science Operations Center, PO Box O, 1003 Lopezville Rd, Socorro, NM 87801.}
\altaffiltext{3}{School of Physics and Astronomy, Cardiff University, Cardiff, CF24 3YB, United Kingdom.}
\altaffiltext{4}{GEPI, CNRS UMR 8111, Observatoire de Paris, 5 Place Jules Janssen, 92195 Meudon Cedex, France}
\altaffiltext{5}{Special Astrophysical Observatory, Russian Academy of Sciences, Nizhnii Arkhyz, 369167, Zelencukskaya, Karachai-Cherkessia, Russia.}
\altaffiltext{6}{Department of Physics and Astronomy, University of New Mexico, 800 Yale Blvd NE, Albuquerque, NM 87131.}
\altaffiltext{7}{NRAO Green Bank, PO Box 2, Rt 28/92, Green Bank, WV 24944}
\altaffiltext{8}{Department of Astronomy, University of Massachusetts, 710 North Pleasent St, Amherst, MA 01003.}
}\noaffiliation

\begin{abstract}
Two  5 square degree regions around the NGC 7332/9 galaxy pair and the isolated
galaxy NGC 1156 have been mapped in the 21-cm line of neutral hydrogen 
(H\,{\sc i}) with the Arecibo L-band Feed Array out to a redshift of $\sim 
0.065$ ($\sim 20 000$ km\,s$^{-1}$) as part of the Arecibo Galaxy Environment 
Survey.  One of the aims of this survey is to investigate the environment 
of galaxies by identifying dwarf companions and interaction remnants; both of 
these areas provide the potential for such discoveries.  The neutral hydrogen 
observations were complemented by optical and radio follow-up observations
with a number of telescopes.  A total of 87 galaxies were found, of which 39
(45 per cent) were previously cataloged and 15 (17 per cent) have prior 
redshifts.  Two dwarf galaxies have been discovered in the NGC 7332 group 
and a single dwarf galaxy in the vicinity NGC 1156 .  A parallel optical search
of the area revealed one further possible dwarf galaxy near NGC 7332.
\end{abstract}

\keywords{galaxies: individual (NGC 7332) -- galaxies: individual (NGC 7339) 
-- galaxies: individual (NGC 1156) -- radio lines: galaxies -- surveys}

\section{Introduction}

The NGC 7332 and NGC 1156 regions are the first two completely surveyed areas
from the Arecibo Galaxy Environment Survey (AGES).  The survey aims to address
a number of scientific goals, including the H\,{\sc i} mass function in 
different environments, the contribution of neutral gas to the baryonic mass 
density, the identification of gaseous tidal features, and the identification
of isolated neutral gas clouds.  In order to do this, AGES is investigating
galaxies in a number of different environments, from the Local Void through
to the Virgo Cluster.  These environments include three isolated galaxies
(NGC 1156, described here, UGC 2082, and NGC 5233) and two galaxy pairs
(NGC 7332/9, described here, and NGC 2577/UGC 4375).  A full description of 
the aims of the survey, which is expected to be concluded in around four
to five years, can be found in Auld et al. (2006) and on the AGES 
website\footnote{http://www.naic.edu/\~{}ages}.

\subsection{The NGC 7332 region}

NGC 7332 and NGC 7339 form a close pair (5.2 arc minutes separation) with
velocities of 1172 and 1313 km\,s$^{-1}$ respectively.  They lie at a distance 
of 23 Mpc (Tonry et al. 2001, from surface-brightness fluctuations), 
giving a projected separation of 35 kpc. We adopt the Tonry et al. distance 
throughout this paper and convert masses and fluxes from the literature to 
this distance.

NGC 7332 itself is an S0 galaxy;
whether or not it contains neutral hydrogen has been debated in the literature
for many years. Knapp, Kerr \& Williams (1978), using Arecibo with the
circular-polarised L-band feed (FWHM $3.3 \pm 0.1$ arcmin), claimed a 
detection with an H\,{\sc i} mass 
$1.1\times 10^8 M_\odot$.  Biermann, Clarke \& Fricke (1979), using the same
receiver, also see this signal but attribute it to side-lobe contamination from
NGC 7339.  Haynes (1981) and Burstein, Krumm \& Salpeter (1987) both observed
the galaxy using Arecibo with the linear-polarised L-band feed (FWHM $3.9\pm 
0.1$ arcmin), which had significantly lower side-lobes than the 
circular-polarised feed but also less sensitivity (6 K Jy$^{-1}$ versus 8 K 
Jy$^{-1}$), and Balkowski \& Chamaraux (1983) observed it with the Nan\c{c}ay 
Radio Telescope (FWHM $4 \times 22$ arcmin).  None of these observations had
the sensitivity to detect the signal seen by Knapp et al., although Haynes
(1983) discusses the Knapp et al. detection and attributes it to side-lobes.

Ionized hydrogen has been detected in NGC 7332 (Plana \& Boulesteix 1996) 
which extends over the velocity range of the Knapp et al. neutral hydrogen 
detection.  This ionized gas is organised into two dynamically separate discs, 
one of which (the more massive) is counter-rotating with respect to the stars.

The ionized gas in NGC 7332 was also examined by Falc\'on-Barroso et al.
(2004).  They found that ``NGC 7332 is young everywhere''
and that there was ``a significant amount of unsettled gas'', suggesting that
``NGC 7332 is still evolving''.  Both  Falc\'on-Barroso et al. and
Plana \& Boulesteix find that a recent accretion event is the likely source
of the counter-rotating gas disc.   Falc\'on-Barroso et al. also suggest
that the same event could have led to the formation of the bar in this 
galaxy, and that later accretion may be required to explain some other
features of the ionized gas distribution.

%It therefore appears from the ionized gas that there has been at least one 
%relatively recent accretion event, and possibly repeated recent accretion 
%events.  Given this, it may well be reasonably expected that some neutral
%gas would be found in
%the vicinity of NGC 7332, and quite possible that, like the claimed
%Knapp et al. detection, it would not be at the same velocity as the stellar
%disc.

Morganti et al. (2006) observed NGC 7332 with the Westerbork Synthesis
Radio Telescope (WSRT).  They did not detect any gas at the position of 
NGC 7332 but did detect the gas in NGC 7339 and also found a cloud with an 
H\,{\sc i} mass of $6\times 10^6 M_\odot$ around 3 arc minutes from NGC 7332, 
in the direction of NGC 7339, which they attribute to an interaction between 
the two galaxies.  

%The detection and dynamics of ionized gas in NGC 7332 and the recent 
%high-resolution  H\,{\sc i} observations of 
%Morganti et al. (2006) with the WSRT hint at a recent interaction between
%the two galaxies.  The ionized gas is organised in two dynamically separate
%discs, one of which is counter-rotating with respect to the stars, implying
%recent accretion.  The high-resolution H\,{\sc i} observations found no
%neutral hydrogen in NGC 7332 but detected a small cloud ($6\times 10^6 
%M_\odot$) around 3 arc minutes (20 kpc in projection) from NGC 7332, in the 
%direction of NGC 7339.

NGC 7339 is a nearly edge-on Sbc galaxy.  It has been observed a number of 
times in the H\,{\sc i} line.  Springob et al. (2005) find a neutral hydrogen
mass of $1.3 \times 10^9 M_\odot$ after correction for the size of the galaxy,
which is consistent with the mass of $1.2\pm 0.2 \times 10^9 M_\odot$ found by 
Staveley-Smith \& Davies (1987) with the Lovell Telescope at Jodrell Bank.

Besides the two giant galaxies,
two dwarf spheroidal galaxies, KKR 72 and KKR 73, have 
been noted by Karachentseva, Karachentsev \& Richter (1999) as being possibly 
associated with the group following a careful search of POSS II films.  
Attempts to observe H\,{\sc i} in these galaxies by Huchtmeier, Karachentsev \&
Karachentseva (2000) reached a limit of around $10^8 M_\odot$ but did not 
detect anything. 

The region behind NGC 7332 is poorly studied.  There are only nine redshifts
for galaxies in NED\footnote{The NASA/IPAC Extragalactic Database (NED) is 
operated by the Jet Propulsion Laboratory, California Institute of Technology, 
under contract with the National Aeronautics and Space Administration.}.  
Directly behind NGC 7332
lie a succession of void regions, identified by Fairall (1998) as parts of 
the Delphinus, Cygnus and Pegasus voids.  The AGES region lies within, but 
close to the edge of, these voids out to $\sim 7000$ km\,s$^{-1}$.

%This paper reports on the first observations of this galaxy pair with the
%upgraded Arecibo Gregorian Telescope.  The sensitivity reached is better than 
%any previous observations and the map covers a $2.5^\circ\times 2^\circ$ 
%region, centred approximately on NGC 7332.  The velocity range extends well
%beyond the group, with similar sensitivity out to around 20 000 km\,s$^{-1}$,
%where the roll off of the band-pass causes an increase in noise. 

The survey area of $2.5^\circ \times 2^\circ$ equates to a physical area of 
0.8 Mpc$^2$ at the distance of NGC 7332, allowing us to see a wide region 
around the galaxy pair.  AGES observations of this region are particularly 
motivated by the possibility of finding interaction remnants from NGC 7332 
and NGC 7339 along with searching their local environment for dwarf galaxies.

\subsection{The NGC 1156 region}

NGC 1156 is a nearby, star-forming, irregular galaxy at a redshift of 375 
km\,s$^{-1}$.
It is similar in both optical
appearance and H\,{\sc i} mass to the Large Magellanic Cloud.  The galaxy
is listed in the isolated galaxy catalogue of Karachentseva, Lebedev \&
Shcherbanovskij (1973) as one of the most isolated galaxies in the nearby 
Universe, with no companion within 10$^\circ$.  Karachentsev, Musella \& 
Grimaldi (1996) estimated its distance as 7.8 Mpc (on the basis of the 
brightest stars); we have adopted this estimate in this paper.  The nearest
significant companion (taken to be one with at least 10 per cent of the
$B$-band luminosity) is UGC 2259 which lies at an angular separation of
12.5 degrees and a distance from Earth of 10 Mpc (Tully \& Fisher 1988), 
giving it
a $B$-band luminosity of 11 per cent of NGC 1156 and a distance from
NGC 1156 of 2.9 Mpc.  The closest galaxy of a similar or larger size (one
with at least 50 per cent of the $B$-band luminosity) is NGC 1023 at an
angular separation of 14.4 degrees and a distance from Earth of 10.4 Mpc
(Ajhar et al. 1997), giving it a $B$-band luminosity around 400 times larger
than NGC 1156 and a distance from NGC 1156 of 3.4 Mpc.

A number of studies have included the galaxy, due to its isolation and its 
status as a relatively close, irregular galaxy with fairly active star 
formation.  Relevant data from these studies is described below.

Swaters et al. (2002) observed NGC 1156 with the WSRT as part of the WHISP
(Westerbork H\,{\sc i} Survery of Spiral and Irregular Galaxies) project.  
These data show a clumpy H\,{\sc i} morphology at small scales, with holes
and knots particularly visible in the outskirts, and possible
signs of a warp in the central regions.  Swaters et al. found an H\,{\sc i}
flux of 71.3 Jy km\,s$^{-1}$, giving $M_{HI} = 1.02 \times 10^9 M_\odot$.
This is consistent with the 71.27 Jy km\,s$^{-1}$ Green Bank 43-m 
measurement of Haynes et al. (1998) which, with a beam-size of 21 arcminutes,
should not resolve the galaxy.  Haynes et al. estimate a finite source size
correction of 2 per cent for their NGC 1156 measurement, giving a final flux
of 72.72 Jy km\,s$^{-1}$.

Barazza, Binggelli \& Prugniel (2001) included NGC 1156 in their sample of 
nearby field galaxies.  They measured a extinction-corrected $B$ 
magnitude of $11.78 \pm 0.10$ and extinction-corrected colors of $B - V = 
0.46$ and $B - R = 0.87$.  This 
gives an absolute $B$ magnitude of $-17.68$ and an H\,{\sc i} mass to light 
ratio (using the Swaters et al. H\,{\sc i} mass) of $M_{HI}/L_B = 0.56
 M_\odot/L_\odot$, which is fairly typical for a Magellanic irregular.

NGC 1156 was included in the sample of galaxies studied in H$\alpha$ by James 
et al. (2004).  They found an $R$-band magnitude of $11.91 \pm 0.04$ 
(uncorrected for extinction), and a 
star formation rate (using our adopted distance of 7.8 Mpc) of $0.71\pm 
0.07 M_\odot$ yr$^{-1}$.

The kinematics of NGC 1156 were studied in detail by Hunter et al. (2002) as
part of a study of star-forming irregular galaxies.  They found that the
kinematical axes of the ionized gas, neutral gas and stellar discs were 
probably aligned at a position angle of $84^\circ$.  This is markedly
different from the morphological position angle of $39^\circ$, due to the
light of the galaxy being dominated by its bar.  In the H\,{\sc i}, they
report a structure resembling a tiny tidal tail to the north-east of the
galaxy, terminating in an H\,{\sc i} complex with a large velocity width.

The background region behind NGC 1156 passes through the outskirts of the
Taurus void (Fairall 1998).  It is only marginally better studied than the NGC
7332 region, with 12 redshifts in NED (not including NGC 1156 itself), many of 
them from the Springob et al. (2005) Arecibo General Catalog.  Two groups
have been identified at just over 10 000 km\,s$^{-1}$: WBL 091 (White et al.
1999) and PPS2 187 (Trasarti-Battistoni 1998).

The AGES observations of the NGC 1156 region cover an area of $2.5^\circ 
\times 2^\circ$, equivalent to 0.09 Mpc$^2$ at the distance of NGC 1156.  The
observations are particularly motivated by the possibility of finding low
surface-brightness companions and dwarf galaxies that have evaded optical 
detection but that may have interacted (or even be interacting) with NGC 1156.

\section{Observations and data reduction}

The H\,{\sc i} survey observations of the NGC 1156 region were carried out
using the 305-m telescope at Arecibo Observatory\footnote{Arecibo Observatory 
is part of the National Astronomy and Ionosphere Center, which is operated by 
Cornell University under a cooperative agreement with the NSF.} 
between December 2005
and February 2006 and those of the NGC 7332 region between July and
 September, 2006.  Both sets of observations used the
Arecibo L-band Feed Array multibeam 
system\footnote{see http://www.naic.edu/alfa for more details}
and the fixed-azimuth drift
observing mode to reach an integration time of around 300 s point$^{-1}$.
The data reduction used the standard AGES pipeline (Cortese et al.
2008), which is based on that used for HIPASS (Barnes et al. 2001) and uses
the Livedata and Gridzilla multibeam processing 
packages\footnote{http://www.atnf.csiro.au/computing/software/livedata.html}.  
Baseline
estimation and calibration were carried out using Livedata, which estimates
the baseline to be removed by median combining an entire drift scan, and the 
reduced spectra were gridded into a datacube using Gridzilla.

The reduced data cubes were Hanning-smoothed with a width of
three channels, giving a velocity resolution of 10 km\,s$^{-1}$ and a noise of 
0.75 mJy channel$^{-1}$ beam$^{-1}$.  The final beam resolution (from 2-D
Gaussian fitting on the 670 mJy continuum source 4C +24.59 which is close to
NGC 7332; Condon et al. 1998) was 3.4
arc minutes (the same as the input beam).

%One background source from the NGC 7332 region (AGES J224005+244154) was 
%observed with the Very Large Array (VLA) of the 
%NRAO\footnote{The National Radio Astronomy Observatory is a facility of the 
%National Science Foundation operated under cooperative agreement by Associated 
%Universities, Inc.}.  These observations and their results are described in 
%Section \ref{vlasect}.

\subsection{Extended Sources}

The standard AGES pipeline is optimized for point sources and, as such,
does not give the correct flux for extended sources such as NGC 7339 (in the
NGC 7332 field) and NGC 1156.  There are two reasons for this: firstly the
baseline estimation carried out in Livedata can remove flux if a strong
source fills a significant portion of a scan and secondly the gridding
is optimised to correctly reconstruct the flux of a point-source at the
pixel center.

In order to overcome this problem, we have formed a extended-source cubes
of the regions around NGC 1156 and NGC 7332 using the MinMed bandpass estimator
in Livedata (Putman et al. 2002) and without applying the beam normalization in
Gridzilla.  This gives a reduction in the flux of point sources but gives
correct fluxes for gas that overfills the beam.  Following Hanning smoothing, 
the noise in the extended-source cubes was 0.4 mJy beam$^{-1}$ for the NGC 7332
and 0.7 mJy beam$^{-1}$ for NGC 1156 (implying that the bandpass estimate from
MinMed was not as effective for NGC 1156 as for NGC 7332).  The gridded 
beam remained at 3.4 arc minutes, giving a 3$\sigma$ sensitivity to extended
emission that fills or overfills the beam of $5 \times 
10^{17} (\Delta V/10)$ atoms cm$^{-2}$ in NGC 7332 and $8 \times 
10^{17} (\Delta V/10)$ atoms cm$^{-2}$ in NGC 1156.

\begin{figure}
\plottwo{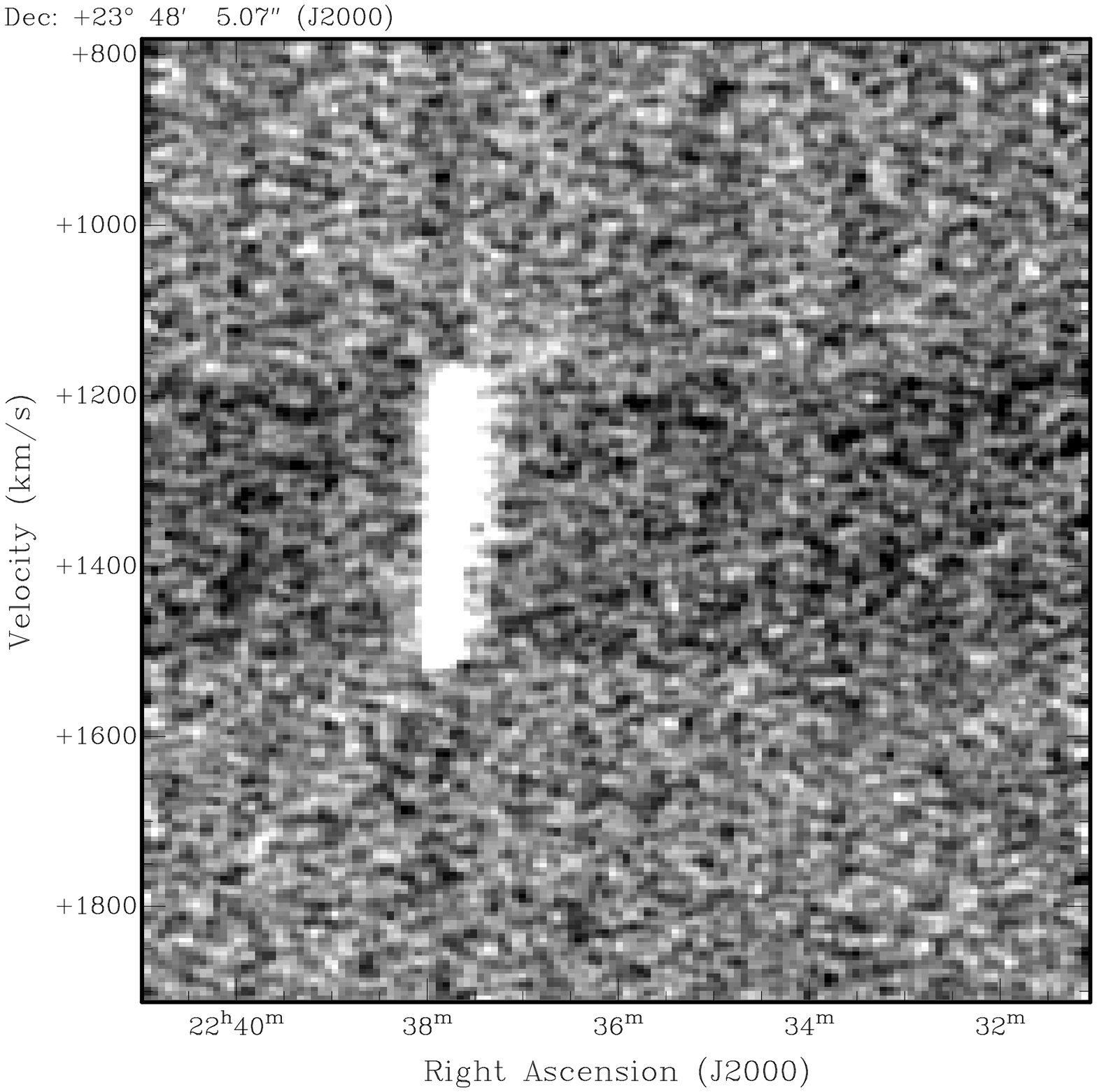}{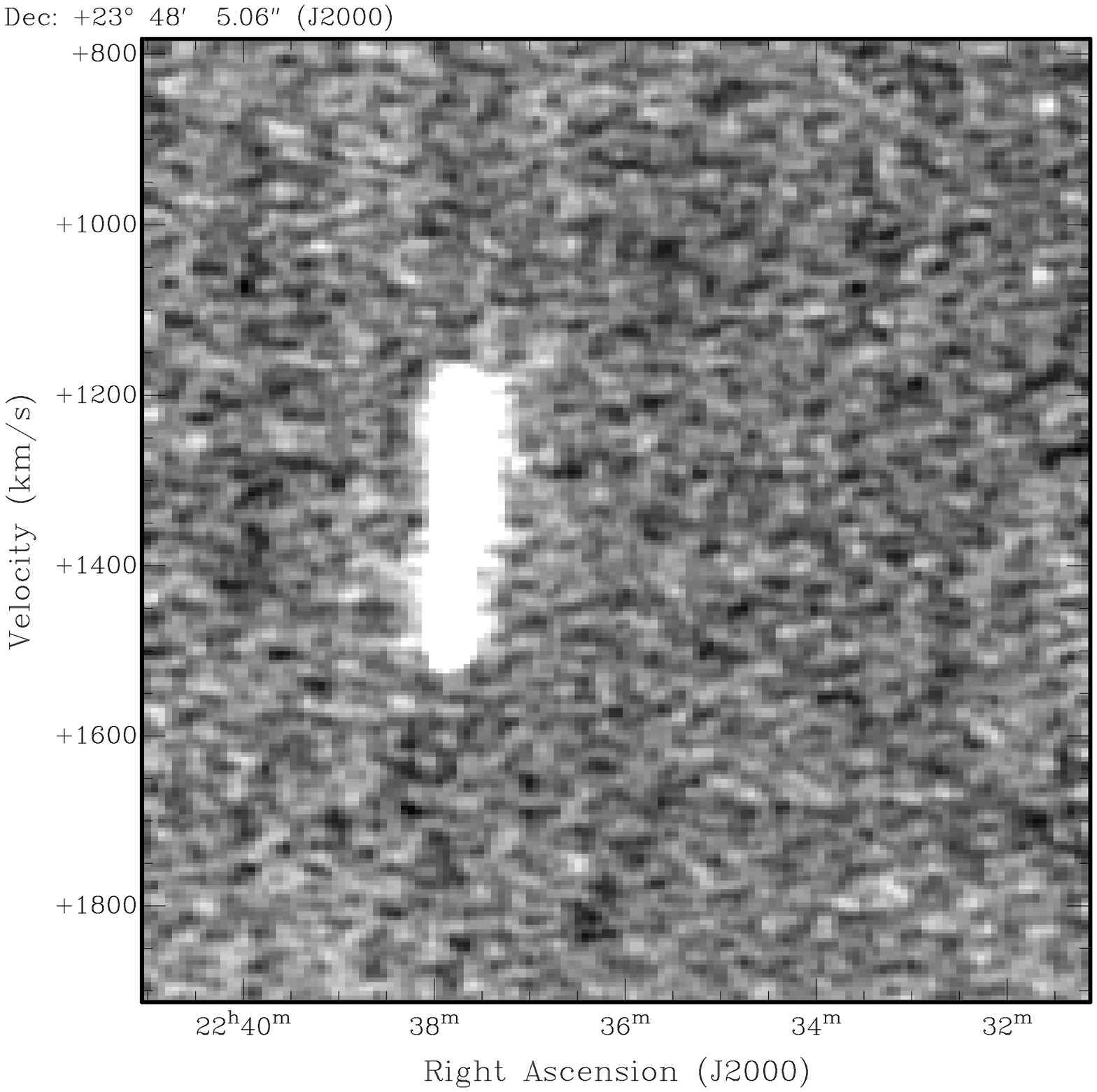}
\caption{Comparison between the standard (left) and extended-source optimised
(right) cubes, showing NGC 7339.  Both have been Hanning smoothed and show 
the same R.A.-velocity slice through the NGC 7332 cube, clipped at equivalent 
3$\sigma$ levels.  It can be seen that the standard cube has a `shadow' in
the scan direction, which has removed flux from NGC 7339.  This is not present
in the extended-source optimised cube.}
\label{cube-comparison}
\end{figure}

The difference between the standard and extended cubes can be seen in Fig. 
\ref{cube-comparison}.  The standard cube has a `shadow' in the scan direction
(R.A.) caused by NGC 7332 being included in the determination of the baseline.
This causes the extension of NGC 7332 in the R.A. direction to be visibly
truncated on the standard cube as well as removing flux from the whole galaxy.

The fluxes of the extended sources (NGC 7339 and NGC 1156) were found by fitting
a 2D Gaussian (using the {\sc Miriad} routine {\sc Imfit} to a moment 0 map
of each source made from the relevant extended-source optimised cube.    The 
measured flux was then beam-corrected using a size for the beam of 3.4 arc 
minutes.  In addition to the flux, this method fits the position , deconvolved 
size, and position angle of the source.  As described in the sections on NGC 
7339 and NGC 1156, this gives an estimate of the flux of these galaxies that is
consistent with previous measurements.  The spectral parameters of the
extended sources were measured on summed and beam-corrected spectra formed in an
11 by 11 arc minute box, also on the extended-source optimised cubes

\subsection{Radio Frequency Interference}

The survey data is affected by radio frequency interference (RFI) at 1350 MHz
from the Federal Aviation Authority radar at Pico del Este.  This is
seen at $\sim 15 000$ km\,s$^{-1}$.  An intermod of the Punta Salinas radar
at 1241.7 MHz appears in the data at $\sim 7 000$ km\,s$^{-1}$.  As these radar 
signals are present in all or nearly all of the observations, they are not 
removed by median combining the data. 

The other major source of RFI is the GPS L3 beacon at 1380 MHz, which is seen
at $\sim 8 800$ km\,s$^{-1}$ in the data.  However, this only transmits
intermittently and is therefore mostly removed when the data are combined.
Extra care is taken when identifying previously uncataloged sources at this
redshift to ensure that they are true H\,{\sc i} sources and not GPS 
transmissions.

\section{The AGES catalog}
\label{cataloguesection}

The NGC 7332 cube was searched independently by eye by two members of the
team (RM \& EM).
Detections were classified as `solid' or `dubious'.  Any detections not
classified as `solid' by both searchers were targetted for follow-up
observations.  These were carried out with the L-Band Wide and ALFA receivers 
at Arecibo Observatory during the summer of 2008. 

The NGC 1156 cube was searched by two members of the team (RA \& Luca
Cortese) and using 
an automated searching algorithm
(see Cortese et al. 2008 for details).  Any detections with a signal to noise
of less than 7 (calculated retrospectively from the measured parameters, see
below for details) were followed up with the L-band Wide receiver at 
Arecibo Observatory during the winter of 2006 and the fall of 2009.

All of the sources in the catalog presented here (Table \ref{catalogue})
are either therefore considered solid detections on the original
AGES cubes or have been confirmed by follow-up observations.  The data is
presented in the same format as Cortese et al. (2008): Column 1 gives
the AGES H\,{\sc i} source ID.  Columns 2 and 3 give the right ascension of
the H\,{\sc i} source and its associated error (in seconds of right
ascension). Columns 3 and 4 give the declination of the H\,{\sc i} source and 
its associated error (in seconds of arc).  Column 6 gives the heliocentric
velocity ($cz$) and associated error, measured as the mid-point of the 
H\,{\sc i}
profile at 50 per cent of the peak flux.  Columns 7 and 8 give the observed 
width of the H\,{\sc i} profile (not corrected for instrumental broadening,
turbulent motions, etc.) at 50 and 20 per cent respectively of the peak flux,
along with their associated errors.  Column 9 gives the peak flux density and 
its associated error.  Column 10 gives the integrated flux and its associated
error.  Column 11 gives the object flags, defined as follows: flag 0 is
a source securely detected in the survey data; flag 1 is a source confirmed
by follow-up observations; flag 2 is a source contaminated by RFI (none
of the sources in the regions reported in this paper have this flag); flag 3 
marks extended sources, measured as described in Section 2.1.

The H\,{\sc i} parameters in Table \ref{catalogue} were derived as in Cortese 
et al. 2008 using the {\sc Miriad} routine {\sc MBSpect}.  Uncertainties were 
derived using the equations from Koribalski et al. (2004), e.g. $\sigma(F_{HI})
 = 4 \times (\sigma(S_{peak})/S_{peak}) \times (S_{peak} F_{HI} 
\delta v)^{1/2}$, where $\delta v$ is the channel width and 
$\sigma(S_{peak})$ is taken to be the quadrature sum
of the rms measured off-source and $0.05 S_{peak}$, e.g. $\sigma(S_{peak})^2 =
rms^2 + (0.05 S_{peak})^2$.  The uncertainty is the sytematic velocity is given
by $\sigma(V = 3 \times \sigma(S_{peak})/S_{peak} \times (\frac{W_{50} - W_{20}
}{2} \delta v)^(1/2)$; errors in the widths are given by $\sigma(W_{50} = 
2 \sigma(V)$ and $\sigma(W_{20} = 3 \sigma(V)$.  The positions and their
associated errors are from the fitting of a gaussian to the source in 
{\sc MBSpect}, the errors are the formal errors on the fit and, while they
are indicative of the quality of the fit, do not correspond directly to 
the offset between the H{\sc i} position and the optical position and appear
to underestimate the true uncertainty in the position by a factor of around 
three (see Section 5.3).

\begin{deluxetable*}{lllllllllll}
%\begin{minipage}{200mm}
\tablecaption{H\,{\sc i} Properties of the AGES galaxies in the NGC 7332 and 
NGC 1156 regions.\label{catalogue}}
%\begin{tabular}{lllllllllll}
%\hline
\tabletypesize{\footnotesize}
\tablewidth{0pt}
\tablehead{
\colhead{H\,{\sc i} ID}&
\colhead{ R.A.}&
\colhead{$\sigma_{R.A}$}&
\colhead{Dec.}&
\colhead{$\sigma_{Dec.}$}&
\colhead{V}&
\colhead{W50}&
\colhead{W20}&
\colhead{$S_{Peak}$}&
\colhead{$F_{HI}$}& 
\colhead{Flag}\\
\colhead{}&
\colhead{(J2000)}&
\colhead{(s)}&
\colhead{(J2000)}&
\colhead{(arc sec)}&
\colhead{(km\,s$^{-1}$)}&
\colhead{(km\,s$^{-1}$)}&
\colhead{(km\,s$^{-1}$)}&
\colhead{(mJy)}&
\colhead{(Jy km\,s$^{-1}$)}&
\colhead{}}
%\hline
\startdata
AGES J025512+243812  &02:55:11.9 & 1.1  &24:38:12 & 19  & 6988 $\pm$  6  & 98 $\pm$ 12  &180 $\pm$ 18  & 8.0 $\pm$ 0.8  &0.679 $\pm$ 0.094  &0 \\
AGES J025626+254614  &02:56:26.0 & 1.4  &25:46:14 & 14  &10364 $\pm$  7  & 80 $\pm$ 15  &107 $\pm$ 22  & 2.4 $\pm$ 0.5  &0.080 $\pm$ 0.037  &1 \\
AGES J025737+244321  &02:57:36.5 & 0.7  &24:43:21 & 12  &10452 $\pm$ 11  &265 $\pm$ 21  &434 $\pm$ 32  & 4.5 $\pm$ 0.5  &0.824 $\pm$ 0.094  &0 \\
AGES J025742+261755  &02:57:42.4 & 0.8  &26:17:55 & 12  &10383 $\pm$  4  &231 $\pm$  8  &239 $\pm$ 12  & 7.9 $\pm$ 1.6  &1.093 $\pm$ 0.245  &1 \\
AGES J025753+255737  &02:57:53.2 & 0.7  &25:57:37 & 11  &10421 $\pm$  3  &234 $\pm$  5  &254 $\pm$  8  & 6.9 $\pm$ 0.6  &1.088 $\pm$ 0.097  &0 \\
AGES J025801+252556  &02:58:00.7 & 0.7  &25:25:56 & 12  & 7002 $\pm$  5  &125 $\pm$ 11  &298 $\pm$ 16  &15.4 $\pm$ 0.9  &1.936 $\pm$ 0.130  &0 \\
AGES J025818+252711  &02:58:17.6 & 0.7  &25:27:11 & 11  &10514 $\pm$  2  &204 $\pm$  5  &233 $\pm$  7  &12.5 $\pm$ 0.8  &1.956 $\pm$ 0.127  &0 \\
AGES J025826+241836  &02:58:25.6 & 2.3  &24:18:36 & 13  &10254 $\pm$  7  &208 $\pm$ 14  &274 $\pm$ 21  & 5.2 $\pm$ 0.7  &0.815 $\pm$ 0.104  &0 \\
AGES J025835+241844  &02:58:34.8 & 0.7  &24:18:44 & 10  &10147 $\pm$  5  &317 $\pm$ 10  &394 $\pm$ 15  & 7.0 $\pm$ 0.6  &1.678 $\pm$ 0.119  &0 \\
AGES J025836+251656  &02:58:35.5 & 0.7  &25:16:56 & 10  &10459 $\pm$  2  &171 $\pm$  3  &191 $\pm$  5  &30.2 $\pm$ 1.6  &3.866 $\pm$ 0.228  &0 \\
AGES J025842+252348  &02:58:42.2 & 1.3  &25:23:48 & 29  &10369 $\pm$  7  &236 $\pm$ 15  &415 $\pm$ 22  & 9.2 $\pm$ 0.8  &1.854 $\pm$ 0.136  &0 \\
AGES J025843+254521  &02:58:42.6 & 0.7  &25:45:21 & 10  & 7225 $\pm$  2  &218 $\pm$  4  &239 $\pm$  6  &12.8 $\pm$ 0.8  &1.955 $\pm$ 0.127  &0 \\
AGES J025902+253518  &02:59:01.8 & 1.2  &25:35:18 & 13  & 7176 $\pm$ 10  &272 $\pm$ 19  &354 $\pm$ 29  & 3.3 $\pm$ 0.5  &0.262 $\pm$ 0.059  &1 \\
AGES J025917+244756  &02:59:17.4 & 0.9  &24:47:56 & 12  & 4658 $\pm$  3  & 56 $\pm$  6  & 77 $\pm$  8  & 6.7 $\pm$ 0.6  &0.342 $\pm$ 0.054  &0 \\
AGES J025930+255419  &02:59:30.3 & 0.7  &25:54:19 & 14  &10407 $\pm$  7  &221 $\pm$ 13  &306 $\pm$ 20  & 5.2 $\pm$ 0.6  &0.928 $\pm$ 0.095  &0 \\
AGES J025936+253446  &02:59:36.0 & 0.8  &25:34:46 & 11  &10253 $\pm$  7  & 68 $\pm$ 15  &125 $\pm$ 22  & 3.6 $\pm$ 0.5  &0.216 $\pm$ 0.052  &1 \\
AGES J025942+251430  &02:59:42.3 & 0.7  &25:14:30 & 10  &  379 $\pm$  2  &75  $\pm$  4  &113 $\pm$  6  &925.0 $\pm$ 46.3 &75.6 $\pm$ 6.4 &3\\
AGES J025953+254350  &02:59:53.0 & 0.8  &25:43:50 & 13  & 7192 $\pm$ 11  & 19 $\pm$ 23  &117 $\pm$ 34  & 3.1 $\pm$ 0.5  &0.114 $\pm$ 0.040  &0 \\
AGES J025954+241323  &02:59:54.1 & 0.7  &24:13:23 & 10  &10224 $\pm$  2  &303 $\pm$  5  &329 $\pm$  7  &15.8 $\pm$ 1.1  &3.302 $\pm$ 0.193  &0 \\
AGES J030008+241600  &03:00:08.3 & 0.7  &24:16:00 & 10  &15081 $\pm$  5  &430 $\pm$ 10  &441 $\pm$ 15  & 6.2 $\pm$ 1.4  &1.483 $\pm$ 0.280  &0 \\
AGES J030014+250315  &03:00:14.0 & 0.7  &25:03:15 & 10  &11308 $\pm$  2  & 76 $\pm$  4  & 92 $\pm$  7  & 7.7 $\pm$ 0.6  &0.492 $\pm$ 0.064  &0 \\
AGES J030025+255335  &03:00:24.9 & 0.7  &25:53:35 & 15  &10612 $\pm$  5  &104 $\pm$  9  &111 $\pm$ 14  & 2.3 $\pm$ 0.6  &0.169 $\pm$ 0.066  &1 \\
AGES J030027+241301  &03:00:27.2 & 0.8  &24:13:01 & 10  & 9877 $\pm$  5  & 57 $\pm$ 10  &170 $\pm$ 14  &15.4 $\pm$ 1.0  &0.933 $\pm$ 0.102  &0 \\
AGES J030036+241156  &03:00:36.4 & 1.1  &24:11:56 & 11  &15183 $\pm$  9  &112 $\pm$ 17  &158 $\pm$ 26  & 4.9 $\pm$ 0.9  &0.427 $\pm$ 0.110  &1 \\
AGES J030039+254656  &03:00:39.3 & 1.6  &25:46:56 & 15  &  308 $\pm$  3  & 19 $\pm$  7  & 44 $\pm$ 10  & 5.8 $\pm$ 0.6  &0.114 $\pm$ 0.032  &0 \\
AGES J030112+242411  &03:01:11.5 & 0.7  &24:24:11 & 10  & 9781 $\pm$  4  &392 $\pm$  7  &429 $\pm$ 11  & 6.8 $\pm$ 0.6  &1.594 $\pm$ 0.117  &0 \\
AGES J030136+245602  &03:01:36.1 & 0.7  &24:56:02 & 11  &10417 $\pm$  8  &188 $\pm$ 16  &243 $\pm$ 24  & 3.3 $\pm$ 0.5  &0.366 $\pm$ 0.070  &0 \\
AGES J030139+254442  &03:01:39.2 & 0.7  &25:44:42 & 13  & 6722 $\pm$  4  &149 $\pm$  7  &187 $\pm$ 11  & 7.0 $\pm$ 0.6  &0.952 $\pm$ 0.090  &0 \\
AGES J030146+254314  &03:01:45.8 & 0.8  &25:43:14 & 14  &11216 $\pm$  3  &312 $\pm$  6  &332 $\pm$  8  & 6.5 $\pm$ 0.6  &1.378 $\pm$ 0.110  &0 \\
AGES J030200+250030  &03:01:59.7 & 0.7  &25:00:30 & 11  & 7025 $\pm$  2  &264 $\pm$  5  &286 $\pm$  7  & 9.2 $\pm$ 0.7  &1.808 $\pm$ 0.120  &0 \\
AGES J030204+254745  &03:02:04.0 & 0.7  &25:47:45 & 10  & 6791 $\pm$  3  &253 $\pm$  6  &280 $\pm$  9  & 7.4 $\pm$ 0.6  &1.421 $\pm$ 0.109  &0 \\
AGES J030234+244938  &03:02:34.1 & 0.7  &24:49:38 & 10  & 3260 $\pm$  1  & 48 $\pm$  3  & 63 $\pm$  4  &22.4 $\pm$ 1.2  &0.983 $\pm$ 0.103  &0 \\
AGES J030254+260028  &03:02:54.2 & 0.7  &26:00:28 & 10  &10625 $\pm$  2  &117 $\pm$  4  &135 $\pm$  5  &13.9 $\pm$ 0.9  &1.555 $\pm$ 0.114  &0 \\
AGES J030309+260407  &03:03:08.6 & 0.7  &26:04:07 & 11  &14014 $\pm$  9  &209 $\pm$ 18  &281 $\pm$ 27  & 3.4 $\pm$ 0.5  &0.433 $\pm$ 0.075  &0 \\
AGES J030325+241510  &03:03:25.3 & 0.7  &24:15:10 & 11  &14750 $\pm$  8  &158 $\pm$ 17  &292 $\pm$ 25  & 7.4 $\pm$ 0.8  &0.933 $\pm$ 0.112  &0 \\
AGES J030355+241922  &03:03:54.7 & 0.7  &24:19:22 & 12  &14517 $\pm$ 11  &102 $\pm$ 23  &286 $\pm$ 34  & 5.3 $\pm$ 0.7  &0.496 $\pm$ 0.080  &0 \\
AGES J030450+260045  &03:04:50.3 & 0.8  &26:00:45 & 11  & 9424 $\pm$ 10  &134 $\pm$ 20  &183 $\pm$ 31  & 2.9 $\pm$ 0.6  &0.239 $\pm$ 0.071  &1 \\
AGES J030453+251532  &03:04:53.1 & 0.7  &25:15:32 & 10  &14206 $\pm$  5  &239 $\pm$ 10  &267 $\pm$ 16  & 5.1 $\pm$ 0.7  &0.735 $\pm$ 0.113  &0 \\
\enddata
\end{deluxetable*}

\addtocounter{table}{-1}
\begin{deluxetable*}{lllllllllll}
%\begin{minipage}{200mm}
\tablecaption{(continued)}
%\begin{tabular}{lllllllllll}
%\hline
\tabletypesize{\footnotesize}
\tablewidth{0pt}
\tablehead{
\colhead{H\,{\sc i} ID}&
\colhead{ R.A.}&
\colhead{$\sigma_{R.A}$}&
\colhead{Dec.}&
\colhead{$\sigma_{Dec.}$}&
\colhead{V}&
\colhead{W50}&
\colhead{W20}&
\colhead{$S_{Peak}$}&
\colhead{$F_{HI}$}& 
\colhead{Flag}\\
\colhead{}&
\colhead{(J2000)}&
\colhead{(s)}&
\colhead{(J2000)}&
\colhead{(arc sec)}&
\colhead{(km\,s$^{-1}$)}&
\colhead{(km\,s$^{-1}$)}&
\colhead{(km\,s$^{-1}$)}&
\colhead{(mJy)}&
\colhead{(Jy km\,s$^{-1}$)}&
\colhead{}}
%\hline
\startdata
AGES J223111+234146  &22:31:11.0  &0.7  &23:41:46  &14  &16645 $\pm$  3&  266 $\pm$  7&  276 $\pm$ 10&  3.9 $\pm$ 0.6 & 0.491 $\pm$ 0.090 & 1 \\
AGES J223122+230436  &22:31:22.1  &0.7  &23:04:36  &10  &17017 $\pm$  3&  200 $\pm$  7&  214 $\pm$ 10&  5.7 $\pm$ 0.8 & 0.692 $\pm$ 0.105 & 0 \\
AGES J223143+244513  &22:31:43.0  &0.7  &24:45:13  &13  &16130 $\pm$  7&  167 $\pm$ 13&  233 $\pm$ 20&  4.5 $\pm$ 0.5 & 0.565 $\pm$ 0.078 & 1 \\
AGES J223213+232657  &22:32:12.6  &0.7  &23:26:57  &10  &16892 $\pm$  3&  160 $\pm$  7&  193 $\pm$ 10&  8.6 $\pm$ 0.7 & 1.226 $\pm$ 0.111 & 0 \\
AGES J223218+235131  &22:32:17.8  &0.7  &23:51:31  &10  &11950 $\pm$  6&  367 $\pm$ 12&  403 $\pm$ 18&  3.6 $\pm$ 0.5 & 0.731 $\pm$ 0.096 & 0 \\
AGES J223223+232613  &22:32:23.4  &0.7  &23:26:13  &10  &11427 $\pm$  8&  51  $\pm$ 16&  288 $\pm$ 24&  9.8 $\pm$ 0.8 & 0.727 $\pm$ 0.084 & 0 \\
AGES J223231+231601  &22:32:30.5  &0.8  &23:16:01  &11  &11787 $\pm$ 10&  361 $\pm$ 20&  475 $\pm$ 31&  3.7 $\pm$ 0.5 & 0.902 $\pm$ 0.105 & 1 \\
AGES J223236+235555  &22:32:36.3  &0.7  &23:55:55  &10  &7457  $\pm$  2&  329 $\pm$  4&  349 $\pm$  6& 11.3 $\pm$ 0.8 & 2.628 $\pm$ 0.146 & 0 \\
AGES J223237+231209  &22:32:37.2  &0.7  &23:12:09  &11  &16614 $\pm$  6&  209 $\pm$ 11&  236 $\pm$ 17&  3.2 $\pm$ 0.5 & 0.444 $\pm$ 0.079 & 1 \\
AGES J223245+243816  &22:32:45.2  &0.7  &24:38:16  &11  &15968 $\pm$  8&  117 $\pm$ 16&  199 $\pm$ 23&  5.1 $\pm$ 0.7 & 0.515 $\pm$ 0.083 & 0 \\
AGES J223318+244545  &22:33:17.8  &0.8  &24:45:45  &11  &12232 $\pm$  4&  450 $\pm$  8&  462 $\pm$ 12&  4.3 $\pm$ 0.7 & 0.946 $\pm$ 0.137 & 1 \\
AGES J223320+230400  &22:33:20.2  &0.7  &23:04:00  &10  &7623  $\pm$  4&  92  $\pm$  8&  125 $\pm$ 12&  5.4 $\pm$ 0.6 & 0.476 $\pm$ 0.068 & 0 \\
AGES J223329+231109  &22:33:29.1  &0.7  &23:11:09  &10  &13600 $\pm$  6&  187 $\pm$ 11&  248 $\pm$ 17&  6.3 $\pm$ 0.7 & 0.930 $\pm$ 0.104 & 0 \\
AGES J223342+242712  &22:33:42.1  &0.8  &24:27:12  &11  &9975  $\pm$  5&  130 $\pm$ 10&  184 $\pm$ 15&  5.8 $\pm$ 0.6 & 0.687 $\pm$ 0.079 & 0 \\
AGES J223355+243114  &22:33:54.9  &0.9  &24:31:14  &22  &10012 $\pm$  4&  115 $\pm$  9&  132 $\pm$ 13&  3.4 $\pm$ 0.5 & 0.265 $\pm$ 0.059 & 1 \\
AGES J223415+233057  &22:34:15.0  &0.8  &23:30:57  &12  &16476 $\pm$  8&  136 $\pm$ 15&  186 $\pm$ 23&  3.3 $\pm$ 0.5 & 0.370 $\pm$ 0.070 & 1 \\
AGES J223449+240744  &22:34:48.8  &0.7  &24:07:44  &12  &16289 $\pm$  2&  40  $\pm$  5&  58  $\pm$  7&  8.9 $\pm$ 0.7 & 0.315 $\pm$ 0.056 & 0 \\
AGES J223502+235258  &22:35:01.7  &0.7  &23:52:58  &11  &9713  $\pm$  8&  151 $\pm$ 16&  192 $\pm$ 23&  2.9 $\pm$ 0.5 & 0.255 $\pm$ 0.062 & 1 \\
AGES J223502+242131  &22:35:02.4  &0.7  &24:21:31  &12  &12241 $\pm$  9&  170 $\pm$ 18&  235 $\pm$ 28&  3.1 $\pm$ 0.5 & 0.332 $\pm$ 0.069 & 1 \\
AGES J223506+233707  &22:35:05.6  &0.7  &23:37:07  &10  &5655  $\pm$  3&  139 $\pm$  6&  159 $\pm$  8&  6.4 $\pm$ 0.6 & 0.754 $\pm$ 0.082 & 0 \\
AGES J223517+244317  &22:35:16.8  &0.7  &24:43:17  &11  &9954  $\pm$  8&  193 $\pm$ 15&  233 $\pm$ 23&  3.4 $\pm$ 0.6 & 0.348 $\pm$ 0.079 & 1 \\
AGES J223605+242407  &22:36:04.7  &0.7  &24:24:07  &11  &16062 $\pm$  5&  68  $\pm$ 11&  104 $\pm$ 16&  4.0 $\pm$ 0.5 & 0.262 $\pm$ 0.055 & 1 \\
AGES J223613+243504  &22:36:13.0  &0.7  &24:35:04  &10  &12790 $\pm$  3&  265 $\pm$  5&  284 $\pm$  8&  7.9 $\pm$ 0.7 & 1.179 $\pm$ 0.111 & 0 \\
AGES J223627+234258  &22:36:27.2  &0.7  &23:42:58  &10  &1410  $\pm$  2&  59  $\pm$  4&  74  $\pm$  6&  9.5 $\pm$ 0.7 & 0.522 $\pm$ 0.065 & 0 \\
AGES J223628+245307  &22:36:27.8  &2.2  &24:53:07  &54  &12893 $\pm$ 10&  195 $\pm$ 20&  242 $\pm$ 29&  7.2 $\pm$ 1.5 & 0.771 $\pm$ 0.202 & 1 \\
AGES J223631+240823  &22:36:30.6  &0.7  &24:08:23  &10  &12041 $\pm$  6&  68  $\pm$ 11&  124 $\pm$ 17&  6.1 $\pm$ 0.7 & 0.366 $\pm$ 0.066 & 0 \\
AGES J223701+225532  &22:37:00.8  &0.7  &22:55:32  &10  &11519 $\pm$  3&  195 $\pm$  7&  229 $\pm$ 10&  6.8 $\pm$ 0.6 & 1.146 $\pm$ 0.099 & 0 \\
AGES J223715+232957  &22:37:15.0  &0.7  &23:29:57  &12  &18562 $\pm$  9&  142 $\pm$ 17&  188 $\pm$ 26&  3.3 $\pm$ 0.6 & 0.324 $\pm$ 0.078 & 1 \\
AGES J223739+244947  &22:37:39.1  &0.7  &24:49:47  &12  &15071 $\pm$  4&  276 $\pm$  9&  291 $\pm$ 13&  5.9 $\pm$ 0.9 & 0.944 $\pm$ 0.152 & 0 \\
AGES J223741+242520  &22:37:41.0  &0.7  &24:25:20  &10  &8782  $\pm$  2&  165 $\pm$  4&  181 $\pm$  6&  8.2 $\pm$ 0.6 & 1.033 $\pm$ 0.092 & 0 \\
AGES J223745+225309  &22:37:45.0  &0.7  &22:53:09  &11  &11500 $\pm$  5&  63  $\pm$  9&  94  $\pm$ 14&  5.2 $\pm$ 0.7 & 0.240 $\pm$ 0.056 & 1 \\
AGES J223746+234712  &22:37:46.5  &0.7  &23:47:12  &10  &1341  $\pm$  2&  325 $\pm$  4&  348 $\pm$  5& 36.3 $\pm$ 2.0 & 9.1 $\pm$ 0.9 & 3 \\
AGES J223823+245207  &22:38:22.7  &0.7  &24:52:07  &11  &15323 $\pm$  6&  205 $\pm$ 12&  230 $\pm$ 17&  8.4 $\pm$ 1.5 & 0.556 $\pm$ 0.150 & 1 \\
AGES J223829+235135  &22:38:29.4  &0.7  &23:51:35  &10  &1414  $\pm$  3&  36  $\pm$  6&  73  $\pm$  8&  8.3 $\pm$ 0.6 & 0.351 $\pm$ 0.047 & 0 \\
AGES J223834+231114  &22:38:34.4  &0.7  &23:11:14  &11  &8770  $\pm$  6&  154 $\pm$ 12&  192 $\pm$ 18&  5.2 $\pm$ 0.7 & 0.691 $\pm$ 0.108 & 0 \\
AGES J223839+234247  &22:38:39.0  &0.7  &23:42:47  &11  &8797  $\pm$  7&  90  $\pm$ 14&  163 $\pm$ 21&  5.4 $\pm$ 0.7 & 0.422 $\pm$ 0.073 & 0 \\
AGES J223842+233156  &22:38:42.3  &0.7  &23:31:56  &10  &11824 $\pm$  4&  239 $\pm$  8&  265 $\pm$ 12&  4.7 $\pm$ 0.6 & 0.689 $\pm$ 0.085 & 0 \\
AGES J223846+234923  &22:38:45.6  &0.8  &23:49:23  &21  &8788  $\pm$  6&  153 $\pm$ 11&  195 $\pm$ 17&  4.3 $\pm$ 0.5 & 0.436 $\pm$ 0.070 & 0 \\
AGES J223900+244752  &22:39:00.1  &0.8  &24:47:52  &11  &14965 $\pm$  4&  302 $\pm$  8&  328 $\pm$ 12&  6.6 $\pm$ 0.8 & 0.950 $\pm$ 0.117 & 0 \\
AGES J223905+240651  &22:39:05.1  &0.7  &24:06:51  &10  &8842  $\pm$  6&  169 $\pm$ 11&  248 $\pm$ 17&  6.2 $\pm$ 0.6 & 1.024 $\pm$ 0.096 & 0 \\
AGES J223946+242157  &22:39:46.3  &0.7  &24:21:57  &11  &7493  $\pm$  5&  44  $\pm$ 10&  120 $\pm$ 16&  6.8 $\pm$ 0.6 & 0.350 $\pm$ 0.055 & 0 \\
AGES J224005+244154  &22:40:04.6  &0.7  &24:41:54  &10  &12805 $\pm$  4&  379 $\pm$  8&  469 $\pm$ 13& 16.2 $\pm$ 1.1 & 3.981 $\pm$ 0.212 & 0 \\
AGES J224016+244658  &22:40:16.3  &0.7  &24:46:58  &11  &15979 $\pm$  7&  71  $\pm$ 15&  120 $\pm$ 22&  4.8 $\pm$ 0.7 & 0.257 $\pm$ 0.069 & 0 \\
AGES J224024+243019  &22:40:24.0  &0.7  &24:30:19  &11  &8856  $\pm$  5&  43  $\pm$ 11&  90  $\pm$ 16&  5.7 $\pm$ 0.7 & 0.222 $\pm$ 0.052 & 0 \\
AGES J224025+243925  &22:40:24.9  &0.8  &24:39:25  &12  &16026 $\pm$  3&  237 $\pm$  6&  245 $\pm$  9&  4.1 $\pm$ 0.6 & 0.549 $\pm$ 0.093 & 1 \\
AGES J224039+243229  &22:40:39.0  &0.7  &24:32:29  &11  &12855 $\pm$  4&  171 $\pm$  9&  189 $\pm$ 13&  3.5 $\pm$ 0.5 & 0.450 $\pm$ 0.076 & 0 \\
AGES J224052+234635  &22:40:52.3  &0.8  &23:46:35  &11  &11745 $\pm$  2&  85  $\pm$  5&  95  $\pm$  7&  5.6 $\pm$ 0.7 & 0.336 $\pm$ 0.065 & 0 \\
AGES J224110+243500  &22:41:09.9  &0.7  &24:35:00  &10  &9047  $\pm$  2&  192 $\pm$  5&  212 $\pm$  7& 12.3 $\pm$ 0.9 & 1.957 $\pm$ 0.149 & 0 \\
AGES J224125+232228  &22:41:25.5  &0.8  &23:22:28  &11  &7181  $\pm$  5&  62  $\pm$ 10&  181 $\pm$ 15& 29.0 $\pm$ 1.9 & 1.927 $\pm$ 0.201 & 1 \\
%\hline
%\end{tabular}
%\end{minipage}
%\end{table*}
\enddata
\end{deluxetable*}

\begin{figure*}
\centerline{
\resizebox{0.6\columnwidth}{!}{\includegraphics{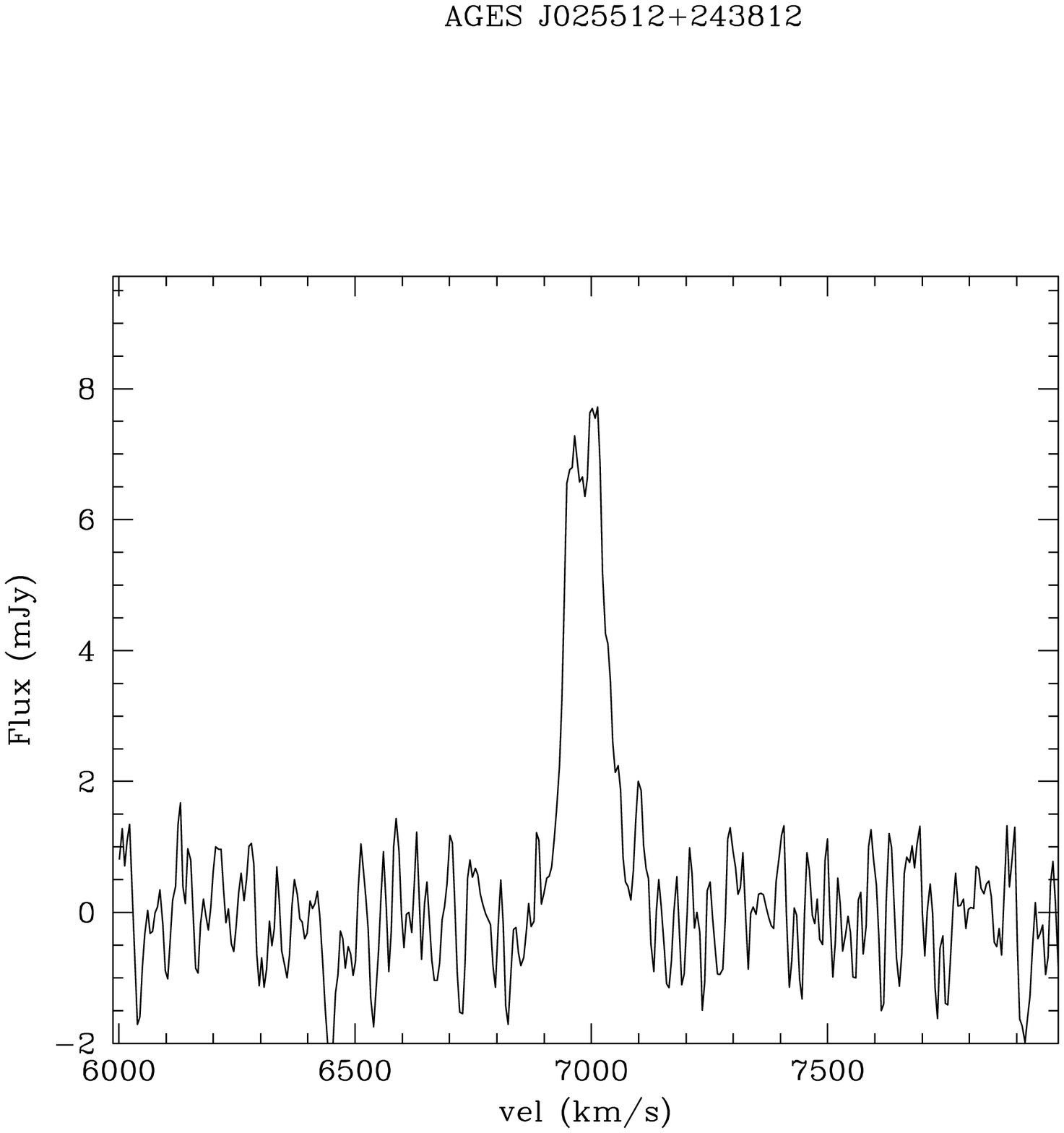}}\hfill
\resizebox{0.6\columnwidth}{!}{\includegraphics{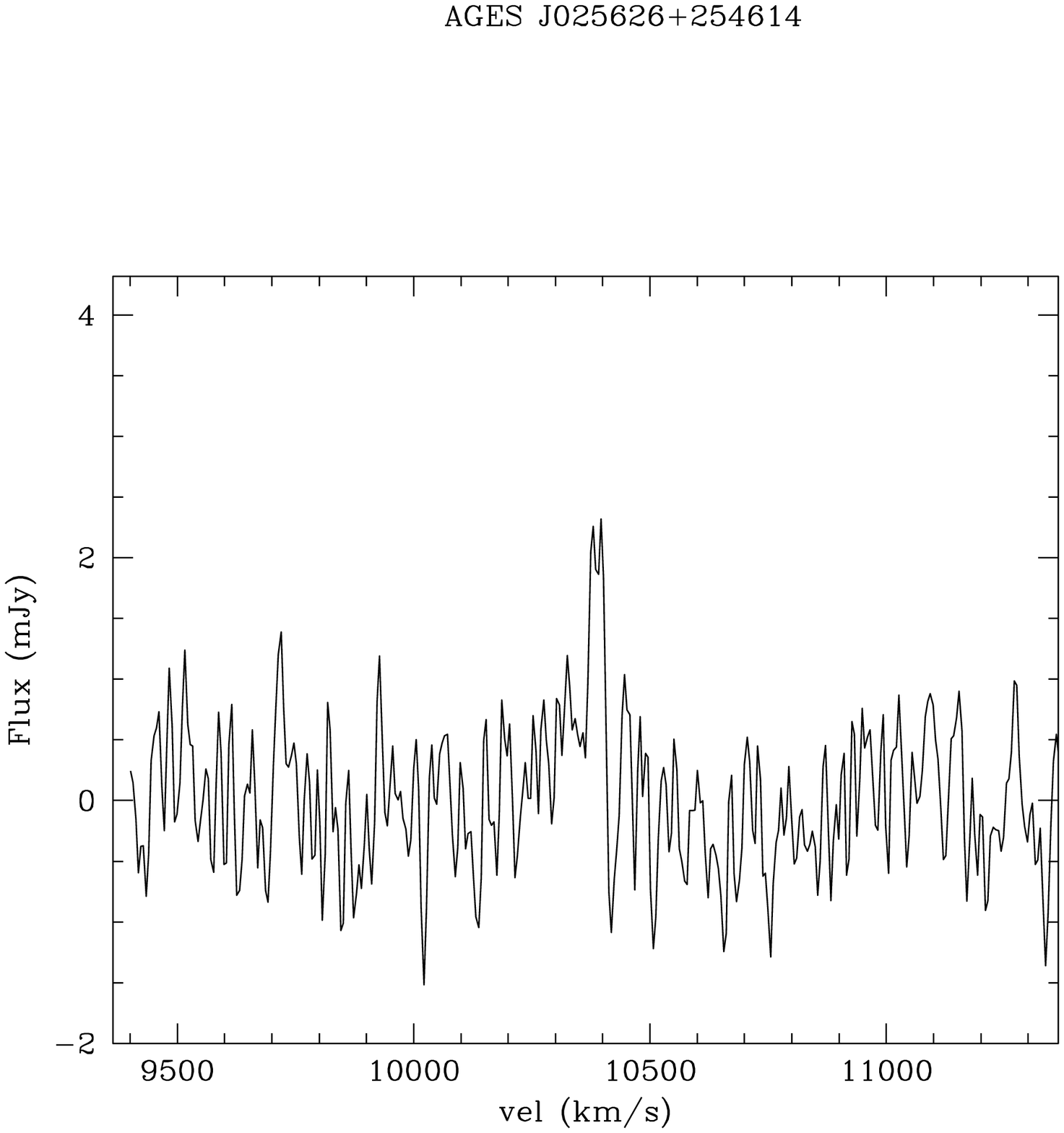}}\hfill
\resizebox{0.6\columnwidth}{!}{\includegraphics{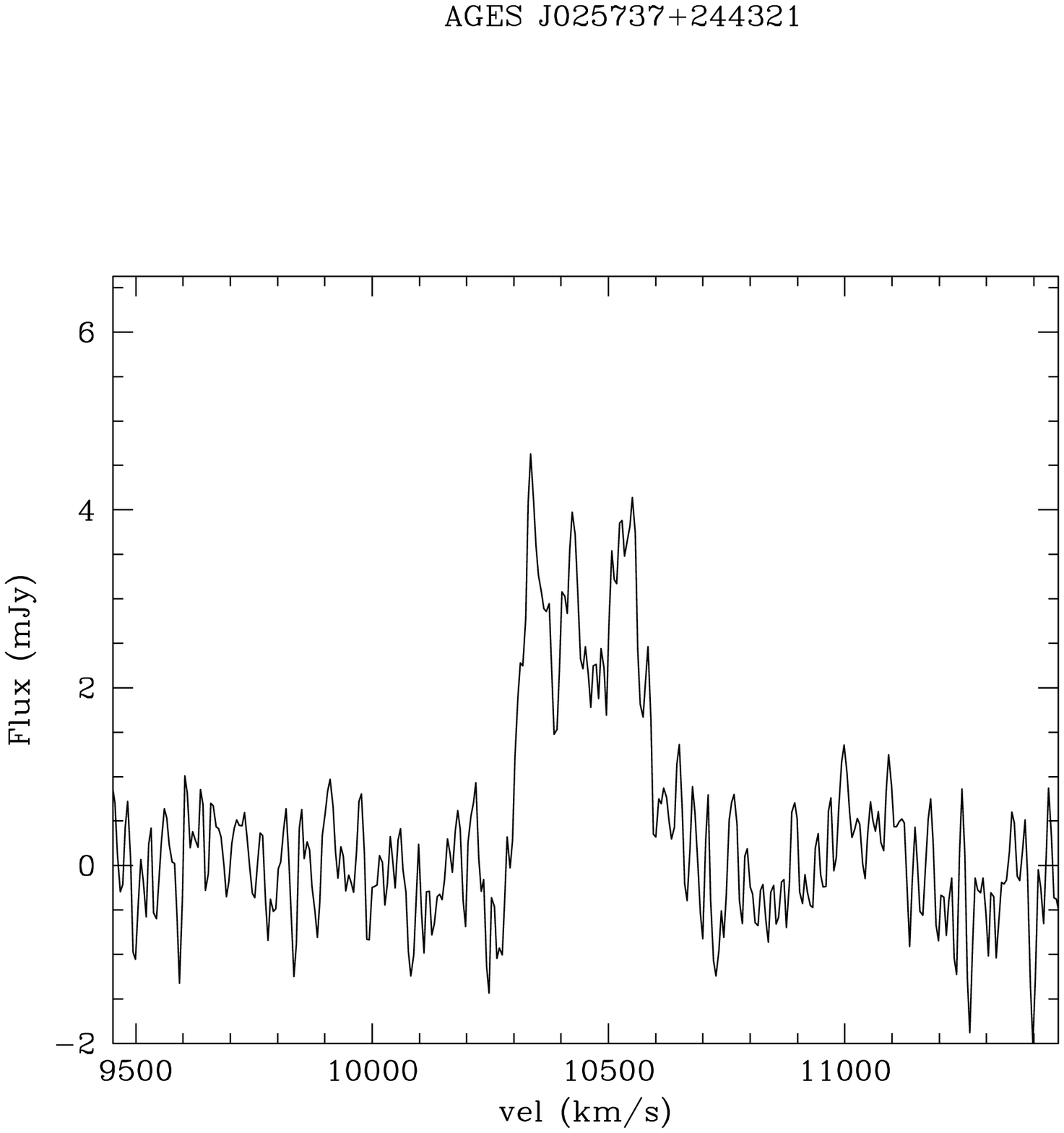}}
}
\centerline{
\resizebox{0.6\columnwidth}{!}{\includegraphics{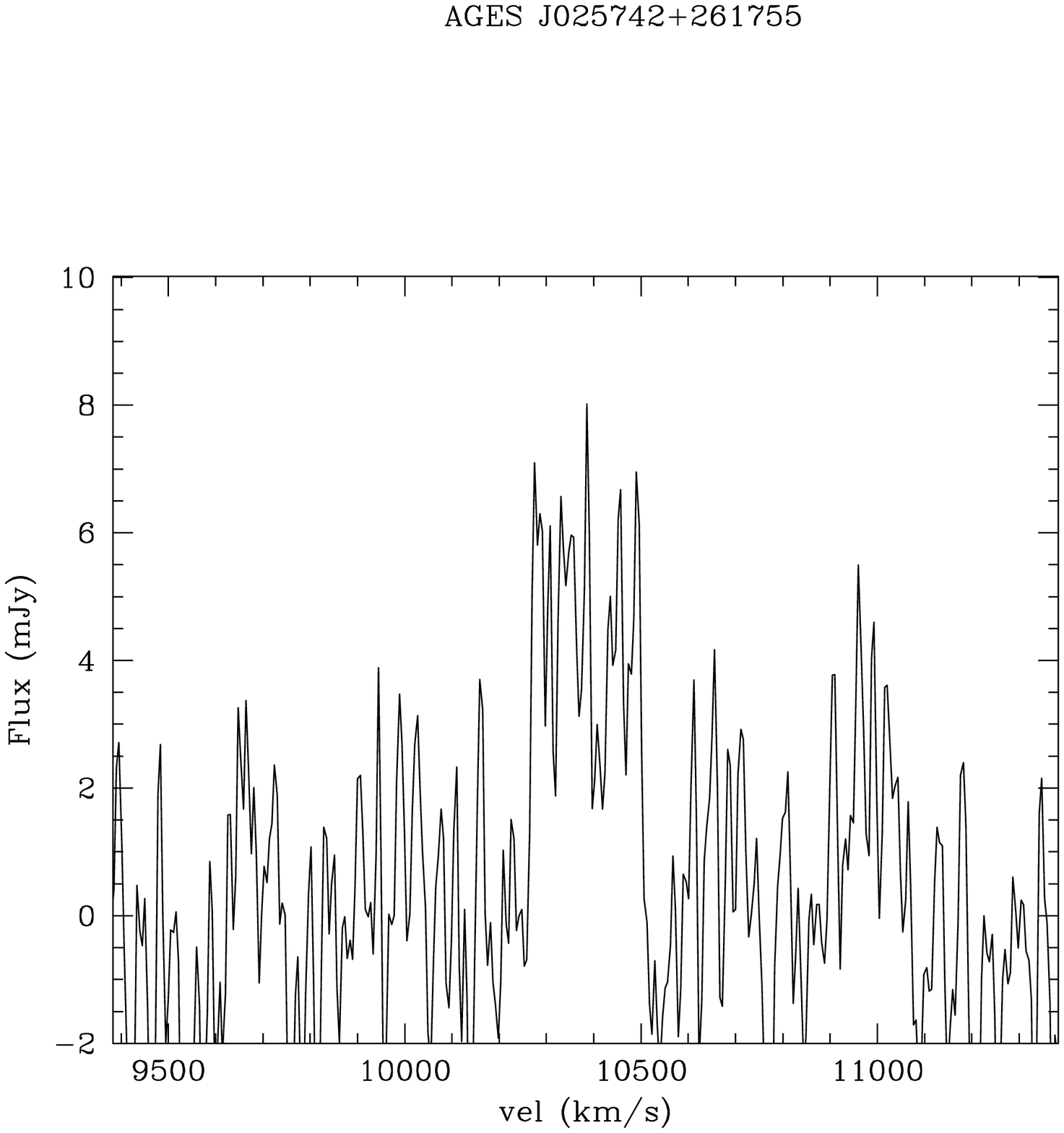}}\hfill
\resizebox{0.6\columnwidth}{!}{\includegraphics{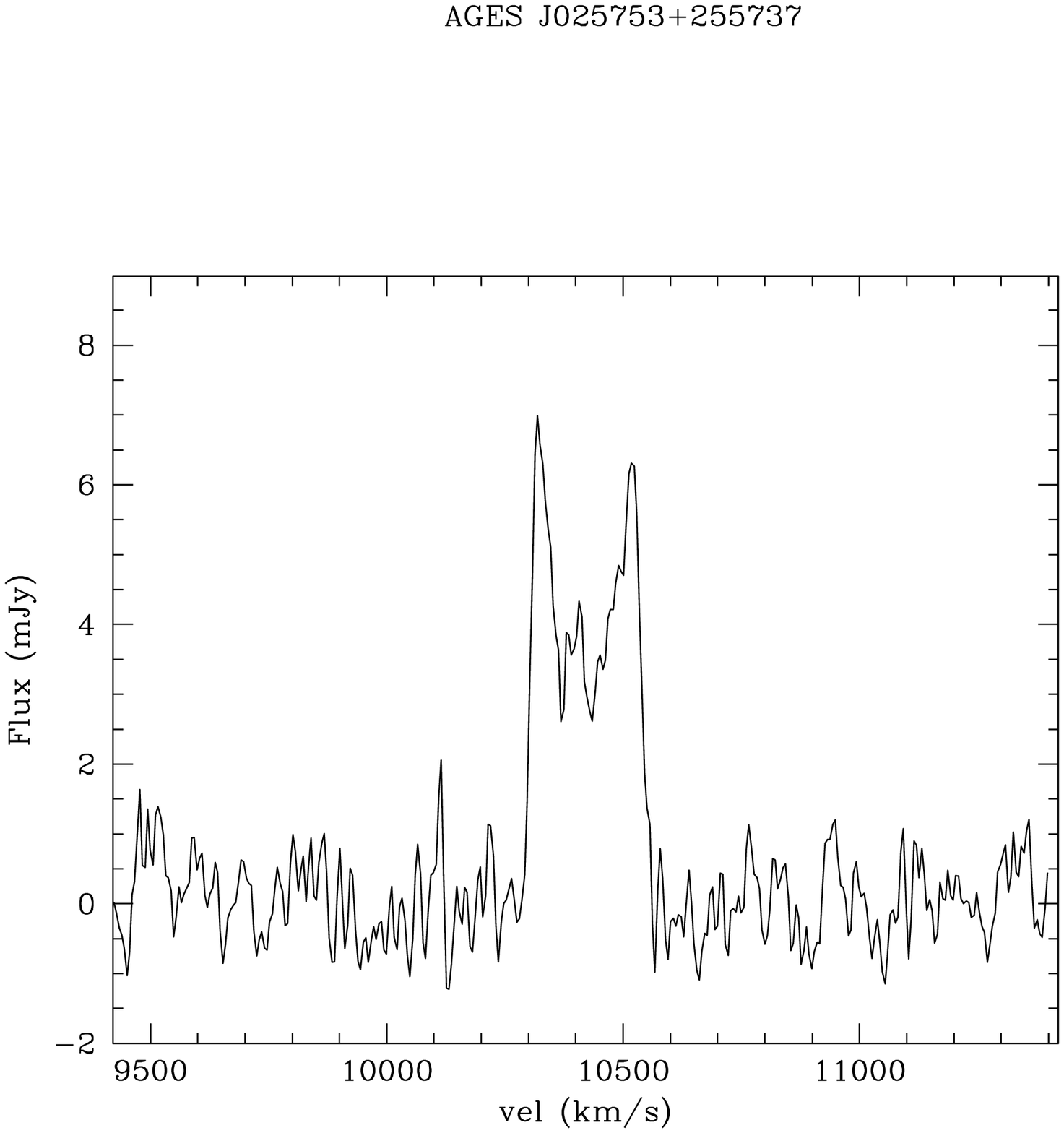}}\hfill
\resizebox{0.6\columnwidth}{!}{\includegraphics{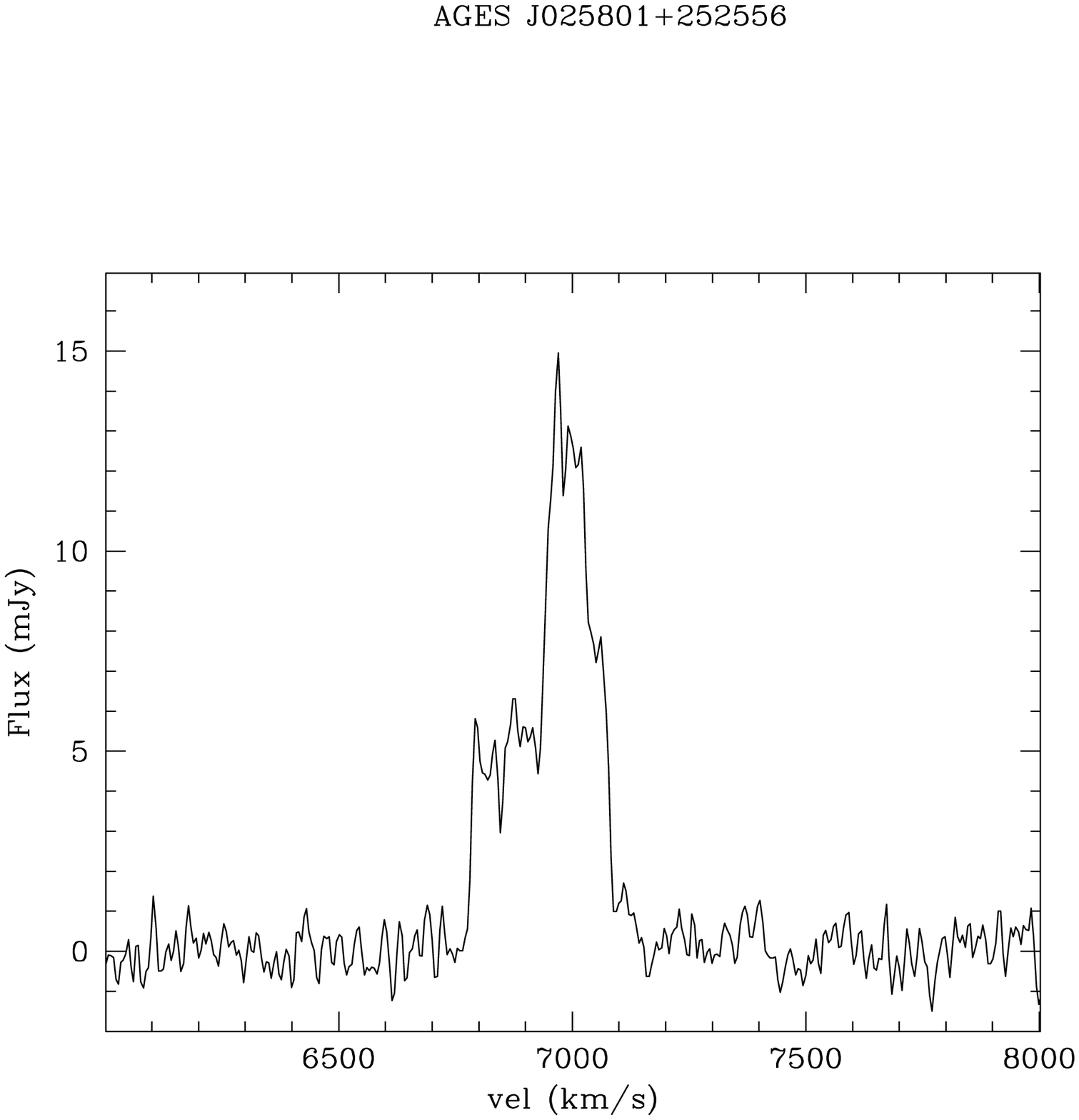}}
}
\centerline{
\resizebox{0.6\columnwidth}{!}{\includegraphics{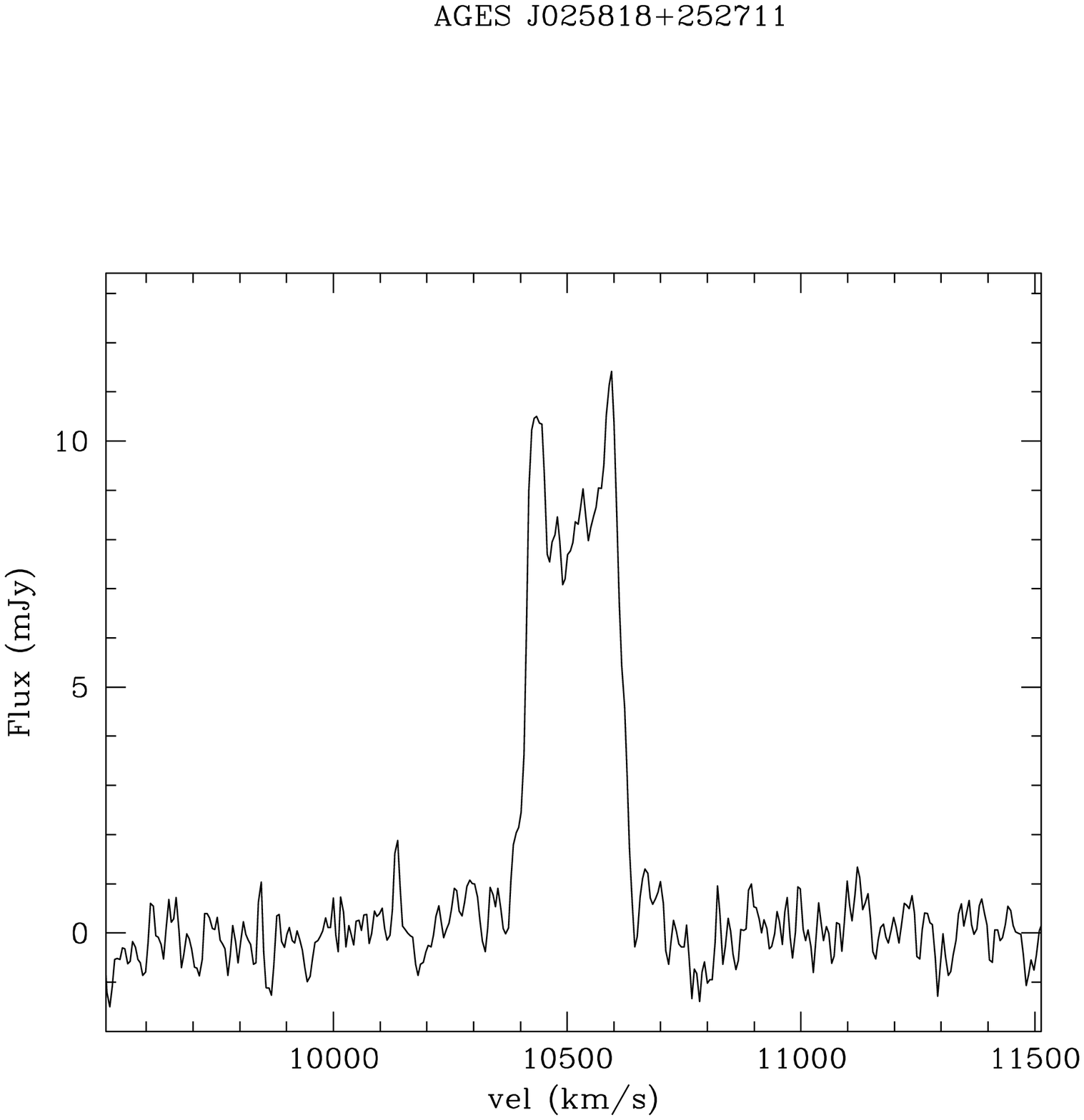}}\hfill
\resizebox{0.6\columnwidth}{!}{\includegraphics{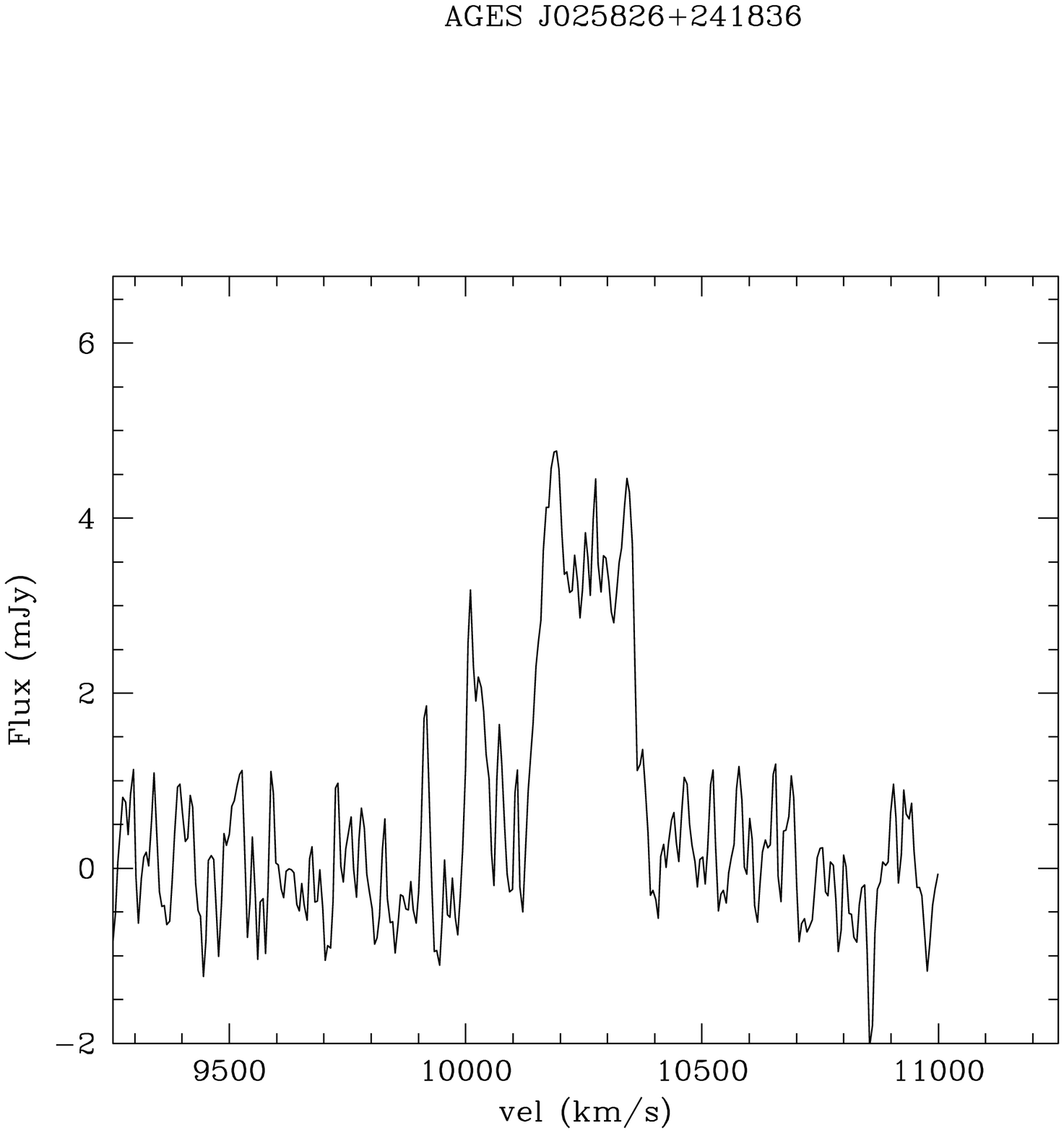}}\hfill
\resizebox{0.6\columnwidth}{!}{\includegraphics{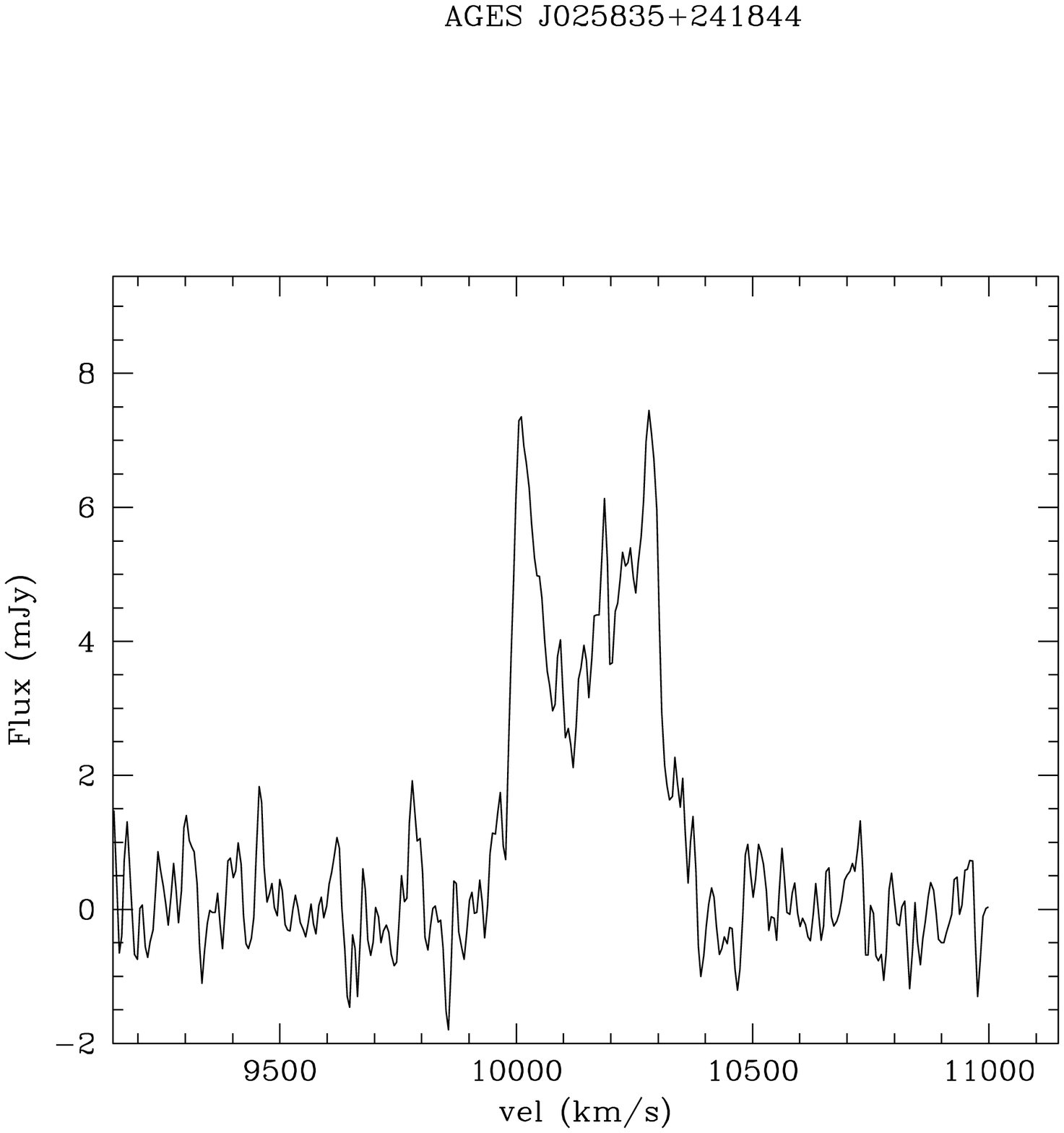}}
}
\centerline{
\resizebox{0.6\columnwidth}{!}{\includegraphics{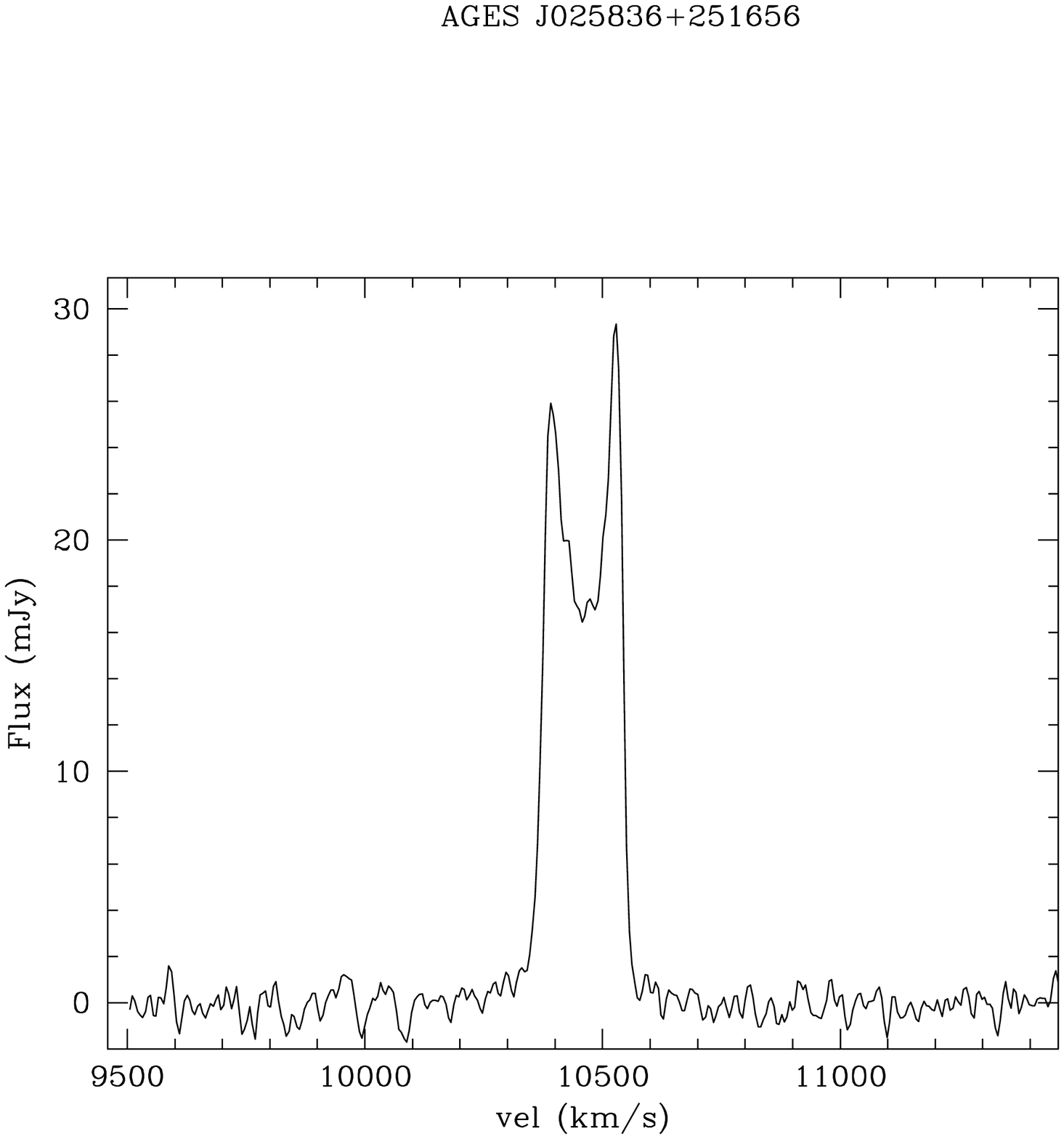}}\hfill
\resizebox{0.6\columnwidth}{!}{\includegraphics{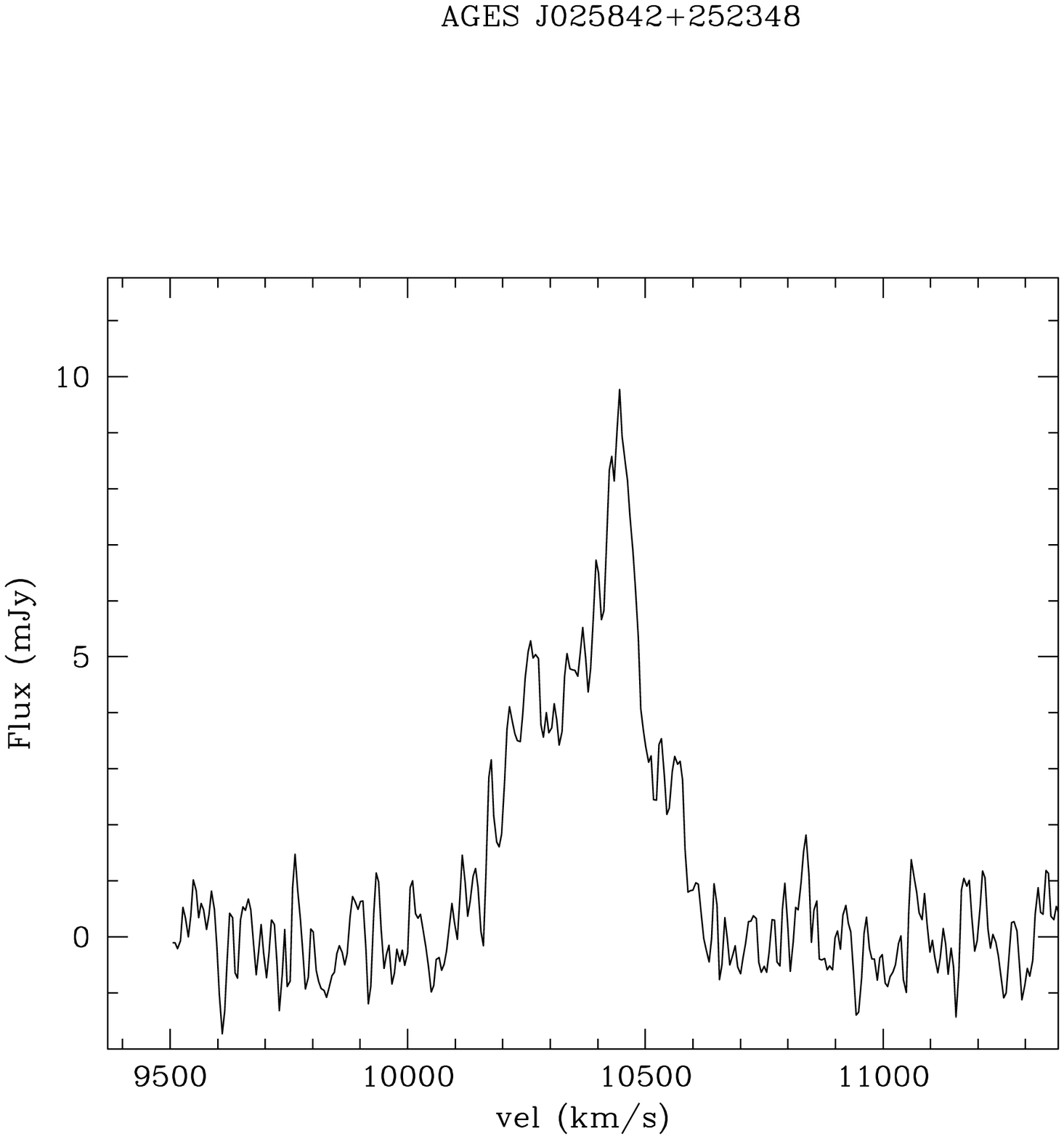}}\hfill
\resizebox{0.6\columnwidth}{!}{\includegraphics{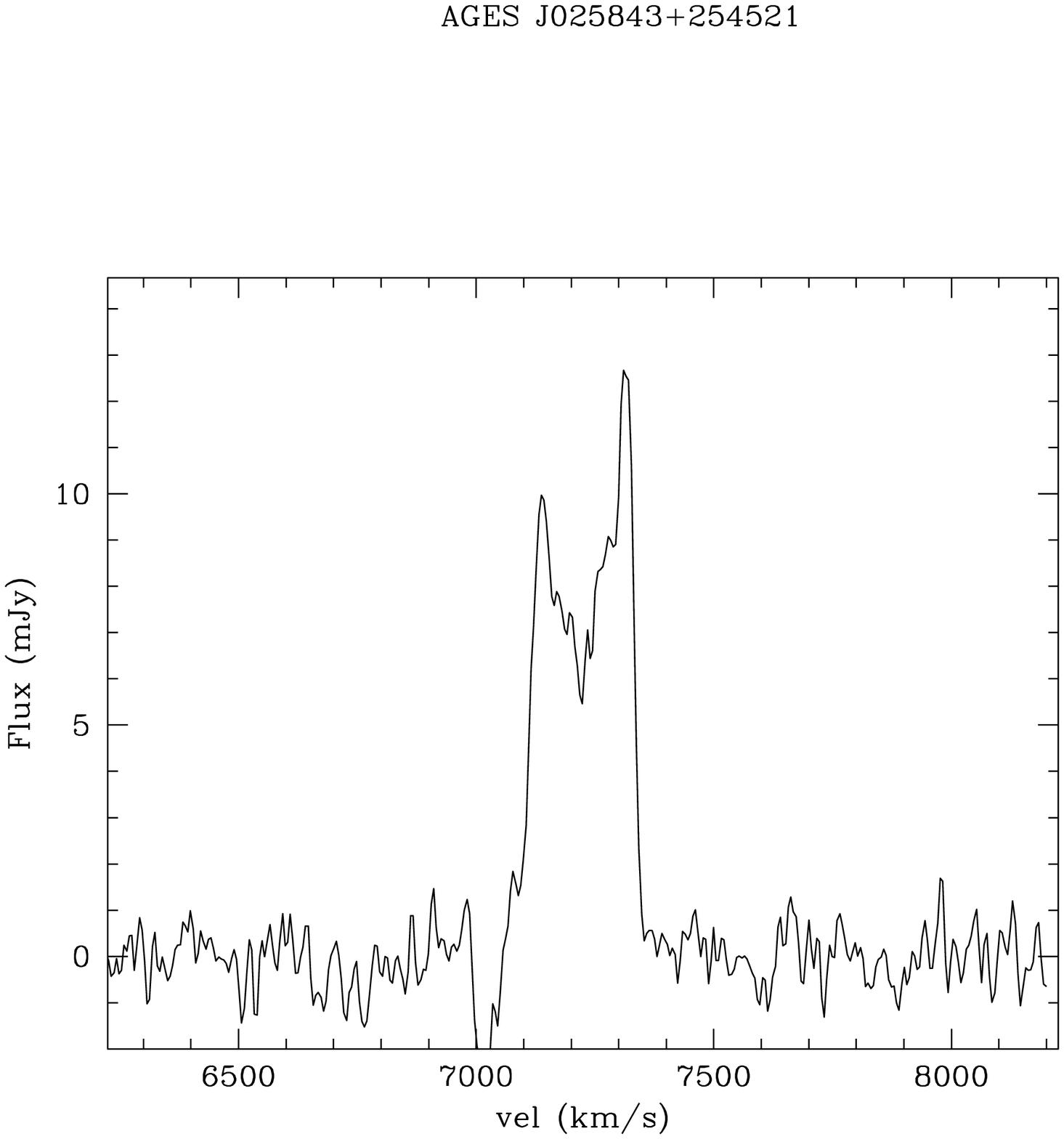}}
}
\caption{Spectra of sources from the NGC 1156 and NGC 7332 regions.  Only 
the first page is shown due to ArXiv size constraints - please see the journal
version for the full figure.}
\label{allspectra}
\end{figure*}

\begin{figure}
\plotone{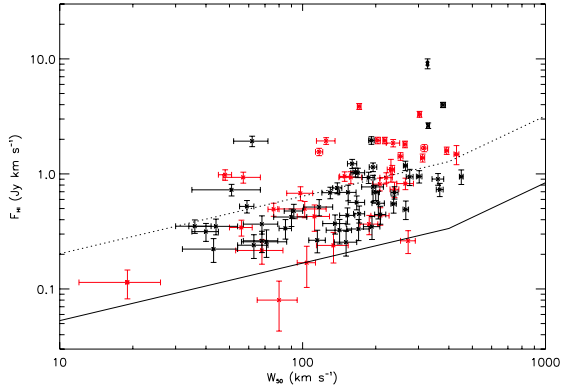}
%\resizebox{\columnwidth}{!}{\includegraphics{dv50-flux2.eps}}
\caption{Velocity width at 50 per cent of the peak flux versus the integrated
flux.  The solid line shows the SNR = 5 (approximate) detection limit, and the 
dotted line shows the ALFALFA `solid detection' limit (Kent et al. 2009).  Black
symbols represent sources from the NGC 7332 region and red symbols sources
from the NGC 1156 region.}
\label{dv-flux}
\end{figure}

\begin{figure}
\plotone{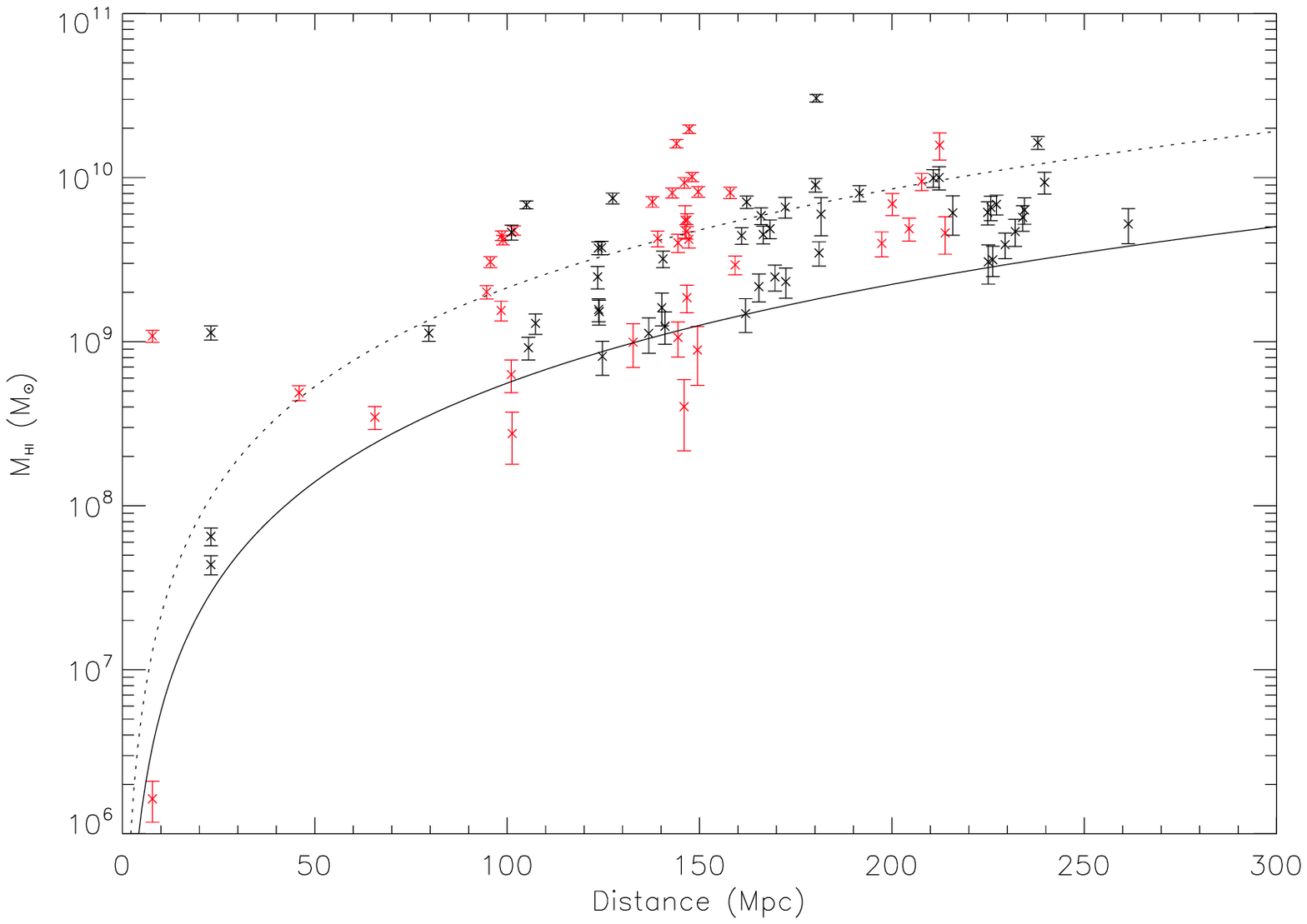}
%\resizebox{\columnwidth}{!}{\includegraphics{mass-dist6.eps}}
\caption{H\,{\sc i} mass versus distance.  The solid line shows the SNR = 5
(approximate) detection limit and the dotted line shows the ALFALFA `solid 
detection' limit (all for $W_{50} = 200$ km\,s$^{-1}$).  Red symbols indicate
sources from the NGC 1156 region and black symbols sources from the NGC 7332
region.}
\label{mass-dist}
\end{figure}

%The average noise levels on the spectra of detections in the
%AGES catalogue have been compared with those in the published ALFALFA catalogue
%in order to compare the sensitivity of the two surveys.  We found that the
%ALFALFA noise level is $\sim \sqrt{25/2}$ times higher than the AGES noise 
%level.  As AGES is covering each point in the sky 25 times and ALFALFA covers
%each point twice, this is the expected difference in noise between the two
%surveys.

Following Giovanelli et al. (2005), Cortese et al. (2008), Kent et al. 
(2009) and Henning et al. (2010), we characterise the effective signal to 
noise ratio of the galaxies retrospectively using the measured parameters 
on sources originally detected by-eye.  The SNR is estimated as:

\begin{equation}
SNR = \frac{F_{HI}}{\sigma}\sqrt{\frac{\omega}{20}}\frac{1}{W_{50}}
\end{equation}

Where $\omega = W_{50}$ (the velocity width at 50 per cent of the peak height)
for $W_{50} < 400$ km\,s$^{-1}$ and $\omega = 400$ km\,s$^{-1}$ for higher
velocity widths.  This corresponds to optimal selection 
($\propto 1/\sqrt{\Delta V}$) at lower velocity widths and peak-flux selection
at higher velocity widths.  We find that most of our data fall above an 
approximate detection 
limit of $SNR = 5$, as can be seen in Fig. \ref{dv-flux}.
Fig. \ref{mass-dist} shows how the sensitivity (in terms of detectable 
H\,{\sc i} mass) varies with distance.  Both also give the ALFALFA limit for
comparison purposes (note that ALFALFA and AGES do not use entirely consistent 
definitions of $W_{50}$, so this limit is necessarily only an approximation).

\section{Follow-up observations}

\subsection{L-band Follow-up Observations}

A number of sources that were not secure detections in the AGES cubes were
confirmed by follow-up observations.  NGC 1156 sources were followed up in 
October 2006 with the L-Band Wide (LBW) receiver using the Interim Correlator 
(IC) backend and in September 2009 with the same receiver using the WAPP
backend.  NGC 7332 sources were followed up in July to September 2008 using
either the L-Band Wide receiver or the central pixel of ALFA, both using the
WAPP backend.  The observations are summarised in Table \ref{lbfu}.  The AGES
ID matches the catalog AGES ID for confirmed sources, for unconfirmed sources
it is merely indicative -- these `sources' are most probably noise fluctuations
or baseline ripples introduced by continuum sources.
`No. obs.' indicates the number of 300s on-source on-off observations that were
made of the source.  The RA and Dec given here are the targetted RA and Dec 
and the velocity is the targetted central velocity.  These may not always 
match precisely with the cataloged RA, Dec and central velocity (Table
\ref{catalogue}).

\begin{deluxetable*}{llllllll}
\tablecaption{L-band Follow-up Observations\label{lbfu}}
\tabletypesize{\footnotesize}
\tablehead{
\colhead{AGES ID}&
\colhead{Receiver}&
\colhead{Backend}&
\colhead{No. Obs.}&
\colhead{R.A.}&
\colhead{Decl.)}&
\colhead{Velocity}&
\colhead{Status}\\
\colhead{}&
\colhead{}&
\colhead{}&
\colhead{}&
\colhead{(J2000)}&
\colhead{(J2000)}&
\colhead{(km\,s$^{-1}$)}&
\colhead{}
}
\startdata
J025542+253222&LBW&IC&4&02:55:42.0&25:32:22&14973&Unconfirmed\\
J025626+254614&LBW&IC&1&02:56:26.0&25:46:14&10385&Confirmed\\
J025711+254925&LBW&IC&2&02:57:11.0&25:49:25&10430&Unconfirmed\\
J025720+245431&LBW&IC&1&02:57:19.5&24:54:31&2310&Unconfirmed\\
J025742+261755&LBW&IC&1&02:57:42.4&26:17:55&10384&Confirmed\\
J025800+252143&LBW&WAPP&2&02:57:59.7&25:21:43&6924&Unconfirmed\\
J025801+255804&LBW&IC&2&02:58:00.8&25:58:04&6989&Unconfirmed\\
J025817+241737&LBW&IC&1&02:58:17.0&24:17:37&10243&Unconfirmed\\
J025824+251514&LBW&IC&1&02:58:24.2&25:15:14&560&Unconfirmed\\
J025828+253058&LBW&IC&2&02:58:27.9&25:30:58&6938&Unconfirmed\\
%J025833+251952&LBW&IC&1&02:58:33.0&25:19:52&10500&Part of J025836+251656\\
J025902+253518&LBW&IC&1&02:59:04.0&25:35:10&7120&Confirmed\\
J025924+243433&LBW&IC&2&02:59:24.0&24:34:33&14050&Unconfirmed\\
J025936+253446&LBW&WAPP&3&02:59:36.0&25:34:46&10253&Confirmed\\
J030025+255335&LBW&WAPP&1&03:00:25.0&25:53:35&10611&Confirmed\\
%J030028+255407&LBW&IC&1&03:00:28.0&25:54:07&10600&Part of J030025+255407\\
J030036+241156&LBW&IC&2&03:00:30.0&24:11:32&15180&Confirmed\\
J030104+243724&LBW&IC&2&03:01:03.5&24:37:24&11109&Unconfirmed\\
J030107+253206&LBW&IC&1&03:01:07.4&25:32:06&436&Unconfirmed\\
J030114+260059&LBW&IC&1&03:01:13.9&26:00:59&14050&Unconfirmed\\
J030116+245206&LBW&IC&1&03:01:16.3&24:52:06&11086&Unconfirmed\\
J030214+250805&LBW&IC&1&03:02:13.8&25:08:05&458&Unconfirmed\\
J030313+245749&LBW&IC&2&03:03:13.0&24:57:49&9790&Unconfirmed\\
J030428+244322&LBW&IC&1&03:04:28.0&24:43:22&15968&Unconfirmed\\
J030450+260045&LBW&IC&1&03:04:50.0&26:00:45&9420&Confirmed\\
J030506+244308&LBW&IC&1&03:05:06.0&24:43:08&14450&Unconfirmed\\
J030528+242441&LBW&IC&2&03:05:28.0&24:24:41&15000&Unconfirmed\\
J223111+234146&LBW&WAPP&2&22:31:11.0&23:41:46&16645&Confirmed\\
J223143+244513&LBW&WAPP&1&22:31:42.9&24:45:11&16130&Confirmed\\
J223226+231111&ALFA+LBW&WAPP&2+1&22:32:26.1&23:11:11&18256&Unconfirmed\\
J223231+231601&LBW&WAPP&1&22:32:31.3&23:16:11&11797&Confirmed\\
J223237+231209&ALFA&WAPP&3&22:32:37.0&23:12:05&16619&Confirmed\\
J223240+224548&ALFA&WAPP&1&22:32:40.6&22:45:48&11481&Unconfirmed\\
J223314+241428&LBW&WAPP&1&22:33:14.8&24:14:28&7463&Unconfirmed\\
J223318+244545&LBW&WAPP&1&22:33:17.5&24:45:38&12233&Confirmed\\
J223355+243114&ALFA&WAPP&1&22:33:54.8&24:31:17&10011&Confirmed\\
J223404+240213&ALFA&WAPP&1&22:34:04.1&24:02:13&16121&Unconfirmed\\
J223415+233057&LBW&WAPP&2&22:34:15.8&23:30:59&16487&Confirmed\\
J223502+235258&ALFA+LBW&WAPP&1+1&22:35:02.5&23:52:33&9637&Confirmed\\
J223502+242142&ALFA&WAPP&1&22:35:02.7&24:21:42&12242&Confirmed\\
J223517+244317&ALFA&WAPP&1&22:35:16.7&24:43:15&9956&Confirmed\\
J223544+235401&ALFA&WAPP&1&22:35:44.0&23:54:01&16471&Unconfirmed\\
J223605+242407&LBW&WAPP&1&22:36:04.6&24:24:06&16067&Confirmed\\
J223628+245307&LBW&WAPP&1&22:36:27.7&24:53:07&12893&Confirmed\\
J223715+232957&LBW&WAPP&2&22:37:14.6&23:30:04&18560&Confirmed\\
J223745+225309&LBW&WAPP&1&22:37:45.2&22:53:04&11499&Confirmed\\
J223823+245207&LBW&WAPP&1&22:28:23.9&24:52:08&15259&Confirmed\\
J223957+233942&ALFA&WAPP&1&22:39:57.0&23:39:42&16574&Unconfirmed\\
J224025+243925&ALFA+LBW&WAPP&1+2&22:40:25.0&24:39:27&16026&Confirmed\\
J224125+232228&LBW&WAPP&1&22:41:25.4&23:22:26&7181&Confirmed\\
\enddata
\end{deluxetable*}

All observations were on-off observations with 300s on-source integration time.
Data reduction was performed using the standard routines in the AO-IDL system.
The data were smoothed to the same resolution as AGES survey observations
(10 km\,s$^{-1}$) and reach noise levels of around 0.7 mJy.  Multiple 
observations were carried out on sources which showed a possible signal at the
location that was too weak to call a confirmation, this was often due to
the presence of continuuum radiation causing baseline ripple and
increased noise levels.

The confirmation rate for follow-up on the NGC 1156 field was 28 per cent (7 out
of 25), while the confirmation rate on the NGC 7332 field was substantially
higher, at 74 per cent (17 out of 23), similar to the confirmation rate in 
the NGC 1156 field (Cortese et al. 2008).  This is due to the difference in the
selection critera for follow-up observations used in the two fields (see \S 
\ref{cataloguesection}) -- these were changed after most of the follow-up of
NGC 1156 had been completed and continue to be modified in order to make
the most efficient use of follow-up time.  

\subsection{Optical survey of the NGC 7332 region}
\label{optsurv}

In addition to the neutral hydrogen observations, we have carried out a 
careful search of the blue and red POSS-II plates within 300 kpc (45 arc 
minutes) of NGC 7332.  This repeated the search of Karachentseva et al.
(1999) but included galaxies with a smaller diameter to account for the 
group being at 23 Mpc while Karachentseva et al. were primarily searching 
for objects within 10 Mpc.  This search turned up a further three candidate 
dwarf galaxies: J223450+240757 lies 41 arc minutes from NGC 7332 and appears 
to be a background galaxy although it could be a dIrr, J223558+234825 is 
20 arc minutes away and has the appearence of a dSph, while J223631+240814 is 
13 arc minutes away and also appears to be a dSph.

On cross-referencing this with the main AGES source list, it became apparent 
that 
J223450+240757 and J223631+240814 were background galaxies.  We therefore
add one candidate dSph, J223558+234825, to the list of possible group 
members.

\subsection{Optical spectroscopy}
\label{optspec}

Optical redshifts were obtained for a number of galaxies in the NGC 1156
and NGC 7332 regions using a dispersion grism on the 1.5-m telescope at 
Loiano Observatory.  Data
reduction was carried out in {\sc iraf}.  The
frames were bias-subtracted and flat-fielded and spectral calibration was
carried out using a template spectrum from an Ar-He lamp.  No flux calibration
was applied.  Post-calibration, the data were sky subtracted to minimise
the impact of night sky lines.  A final 1-D spectrum was produced by 
integrating over pixels that appeared to contain emission from the target
galaxy.  The results of these observations are given in Table
\ref{optspect}, along with the AGES H\,{\sc i} velocities for these galaxies.

\begin{deluxetable*}{lll}
\tablecaption{Optical redshifts\label{optspect}}
\tablehead{
\colhead{AGES ID}&
\colhead{Optical Velocity (km\,s$^{-1}$)}&
\colhead{H\,{\sc i} Velocity (km\,s$^{-1}$)}
}
\startdata
J025737+244321  &    10437 $\pm$ 82  & 10452 $\pm$ 11\\
J025753+255737  &    10422 $\pm$ 82  & 10421 $\pm$ 3\\
J025835+241844  &    10170 $\pm$ 133 & 10147 $\pm$ 5\\
J030008+241600  &    15090 $\pm$ 64  & 15081 $\pm$ 5\\
J030112+242411  &     9892 $\pm$ 142 &  9781 $\pm$ 4\\
J030136+245602  &    10372 $\pm$ 74  & 10417 $\pm$ 8\\
J030146+254314  &    11238 $\pm$ 64  & 11216 $\pm$ 3\\
J030325+241510  &    14786 $\pm$ 76  & 14750 $\pm$ 8\\
J223213+232657  &    16916 $\pm$ 119 & 16892 $\pm$ 3\\
J223449+240744  &    16298 $\pm$ 81  & 16289 $\pm$ 2\\
J223613+243504  &    12742 $\pm$ 69  & 12790 $\pm$ 3\\
J223627+234258  &     1440 $\pm$ 88  &  1410 $\pm$ 2\\
J224005+244154  &    12799 $\pm$ 100 & 12805 $\pm$ 4\\
J224110+243500  &     9030 $\pm$ 94  &  9047 $\pm$ 2\\
\enddata
\end{deluxetable*}

\subsection{VLA observations of AGES J224005+244154}
\label{vlasect}

AGES J224005+244154 is a galaxy in the background of the NGC 7332 field with
a very large H\,{\sc i} mass of $3.1\times 10^{10} M_\odot$.  Its has a
1.4 GHz luminosity of $6.9 \times 10^{22}$ W Hz$^{-1}$ (fron the NVSS; Condon
et al. 1998) and a far IR (FIR) luminosity of $8.1\pm 2.2 \times 10^{10} 
L_\odot$ (from IRAS; Moshir et al. 1990).  Neither the NVSS source nor the
IR source had previously been associated either with each other or with
an optical counterpart.   

The ratio of the FIR and 1.4 GHz fluxes gives a 
$q$ value of $2.07$ (Helou 1985).  A $q$ value of less than 1.64 would 
indicate that a radio-loud AGN dominated the 1.4 GHz continuum, while higher 
values indicate that star formation is dominant.  The $q$ value for AGES 
J224005+244154 thus implies that the source of the 1.4 GHz
continuum is star formation rather than an AGN.  
Using the equations of Yun, Reddy \& Condon
(2001), the FIR flux indicates a current SFR of $21 \pm 9 M_\odot$ yr$^{-1}$
while the 1.4 GHz flux gives a current SFR of $41 \pm 12 M_\odot$ yr$^{-1}$.  
These are both consistent with an SFR of $\sim 30 M_\odot$ yr$^{-1}$; 
at this rate, the galaxy has a gas-consumption timescale of just 1 Gyr,
putting it in the starburst category.

We observed
this galaxy with the VLA of NRAO\footnote{The National Radio Astronomy 
Observatory is a facility of the National Science Foundation operated under 
cooperative agreement by Associated Universities, Inc.}
in order to determine firstly whether it was a
single source and, if so, whether it showed signs of a recent interaction
that might have triggered a star burst.  The galaxy was observed 
in C configuration in April 2008 for a 
total of 6 hours. The observations were centered at a frequency of 1362.22 MHz
(corresponding to 12805 km\,s$^{-1}$, the AGES velocity of the galaxy) with a 
bandwidth of 6.25 MHz over 64
spectral channels, resulting in a velocity resolution of 21.5 km\,s$^{-1}$.
The data were reduced using the Astronomical Image Processing System (AIPS).

\begin{figure}
\plotone{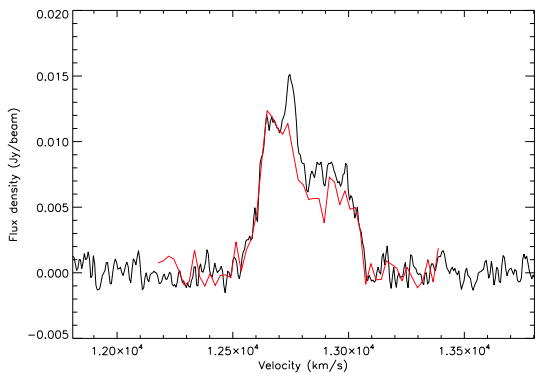}
\caption{Arecibo spectrum of AGES J224005+244154 (black) with the VLA spectrum 
overlaid (red).}
\label{j2240+2441spec}
\end{figure}

The VLA data reached a noise level of $\sim 0.3$ mJy.  Cleaning was carried out
to a depth of 1 mJy, and a moment 0 map was made using a 1 mJy clipping.  This
map was used to mask the part of the image containing signal from the galaxy.
The flux recovered from this region with the VLA was $3.49\pm 0.09$ Jy 
km\,s$^{-1}$, compared to $3.98\pm 0.21$ Jy km\,s$^{-1}$ with Arecibo, a
difference of less than 2.5 $\sigma$. The spectrum from this region of the VLA
data appears consistent with the Arecibo spectrum (Fig. \ref{j2240+2441spec}.  
It thus appears that little extended or lower-level flux has been missed with 
the VLA.

\begin{figure}
\plottwo{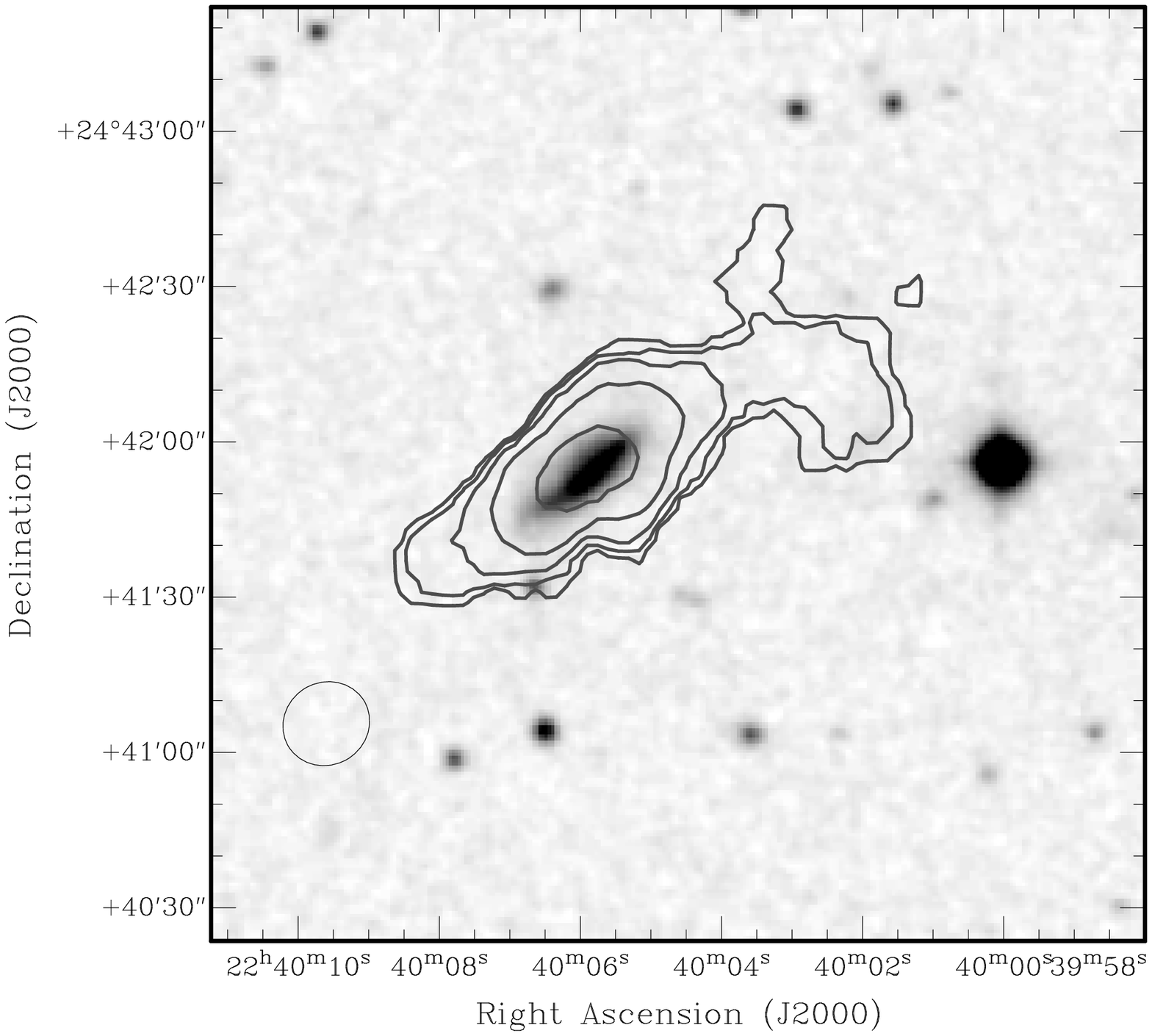}{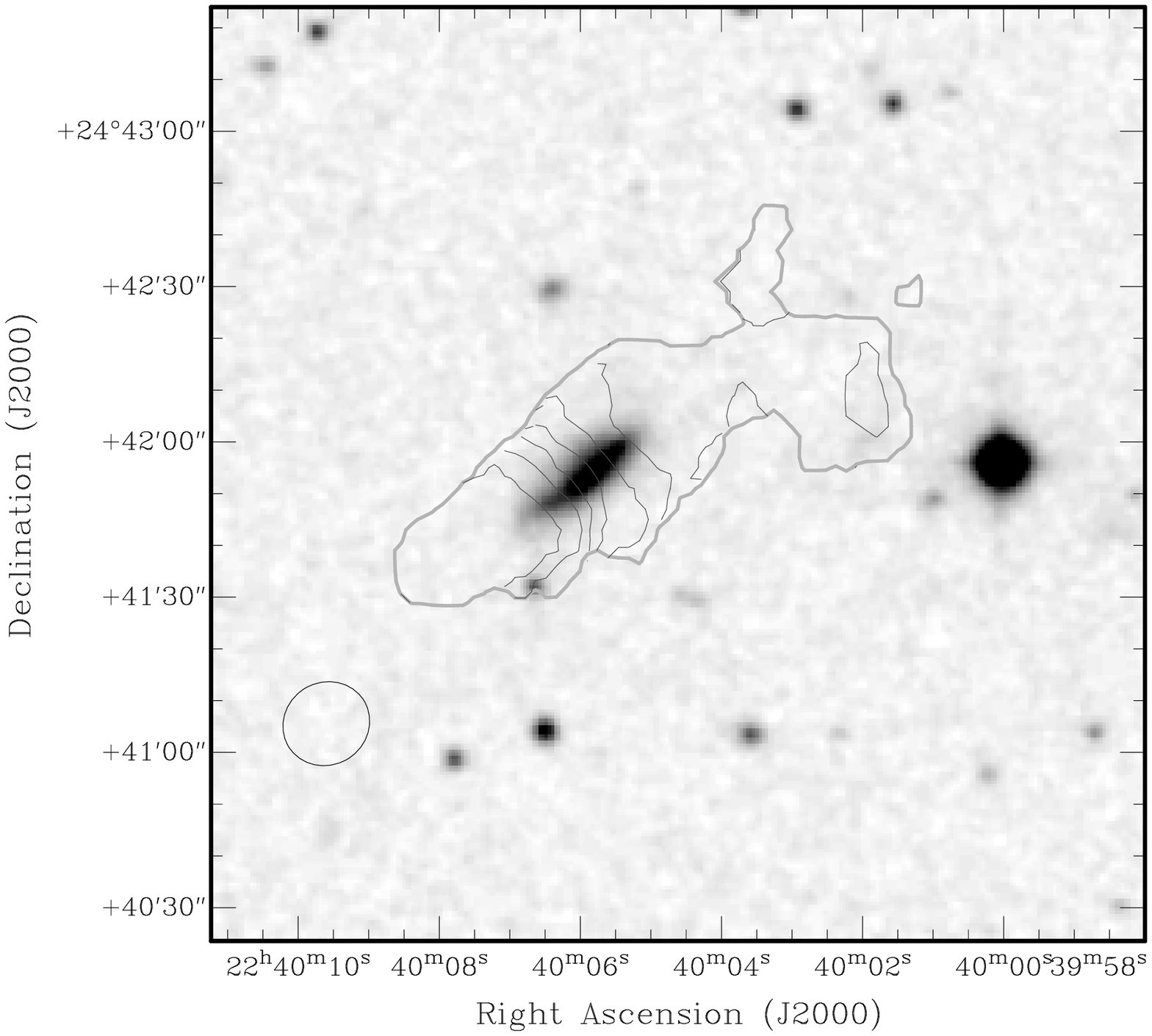}
\caption{Moment 0 (left) and 1 (right) maps of AGES J224005+244154 overlaid
on a DSS $B$-band image.  The 
contour levels for the moment 0 map are 40, 80, 160, 320 and 640 Jy beam$^{-1}$
km\,s$^{-1}$.  The contour levels for the moment 1 map are 12600 to 12850
km\,s$^{-1}$ in steps of 50 km\,s$^{-1}$, with the higher velocities to the
east (left) of the image.}
\label{j2240moms}
\end{figure}

As can be seen from Figure 
\ref{j2240moms}, the galaxy is disturbed in both the H\,{\sc i} and the
optical.  The optical disturbance can be seen as a faint plume at the eastern
end of the galaxy; the H\,{\sc i} disturbance is much more visible and takes
the form of a large plume at the western end of the galaxy.  The velocity field
is fairly smooth over the main disc of the galaxy, but is again disturbed at 
the western end.

AGES J224005+244154 is clearly not evolving in isolation.  It seems most 
likely that it has recently undergone a merger with a smaller galaxy that
has triggered the current starburst phase and has resulted in the disturbances
seen in the optical and H\,{\sc i} images.

\subsection{H$\alpha$ observations of AGES J030039+254656}
\label{newdwarfobs}

AGES J030039+254656 is a new dwarf galaxy discovered by AGES in the 
neighborhood of NGC 1156. CCD images in the $H\alpha$-line and continuum 
were obtained
during an observing run in November 2007 under seeing of $1.8\arcsec$.
The observations were performed by Serafim Kaisin with
the BTA 6-m telescope of the Special Astrophysical Observatory
equipped with the SCORPIO focal reducer.
A CCD chip of 2048$\times$2048 pixels provides a total field of
view of about 6.1$\arcmin$ with a scale of 0.18$\arcsec$/pixel. The images
in $H\alpha$+[NII] and continuum were obtained via observing the galaxy
though a narrow-band interference filter $H\alpha (\Delta\lambda=$75\AA)
with an effective wavelength $\lambda$=6555\AA\ and two medium-band
filters for the continuous spectrum SED607 with
$\Delta\lambda$=167\AA, $\lambda$=6063\AA\, and SED707 with
$\Delta\lambda$=207\AA, $\lambda$=7063\AA,\, respectively. Exposure
times for the galaxy were $2\times300$s in the continuum and
$2\times 600$s in $H\alpha$. 

\begin{figure}
\plottwo{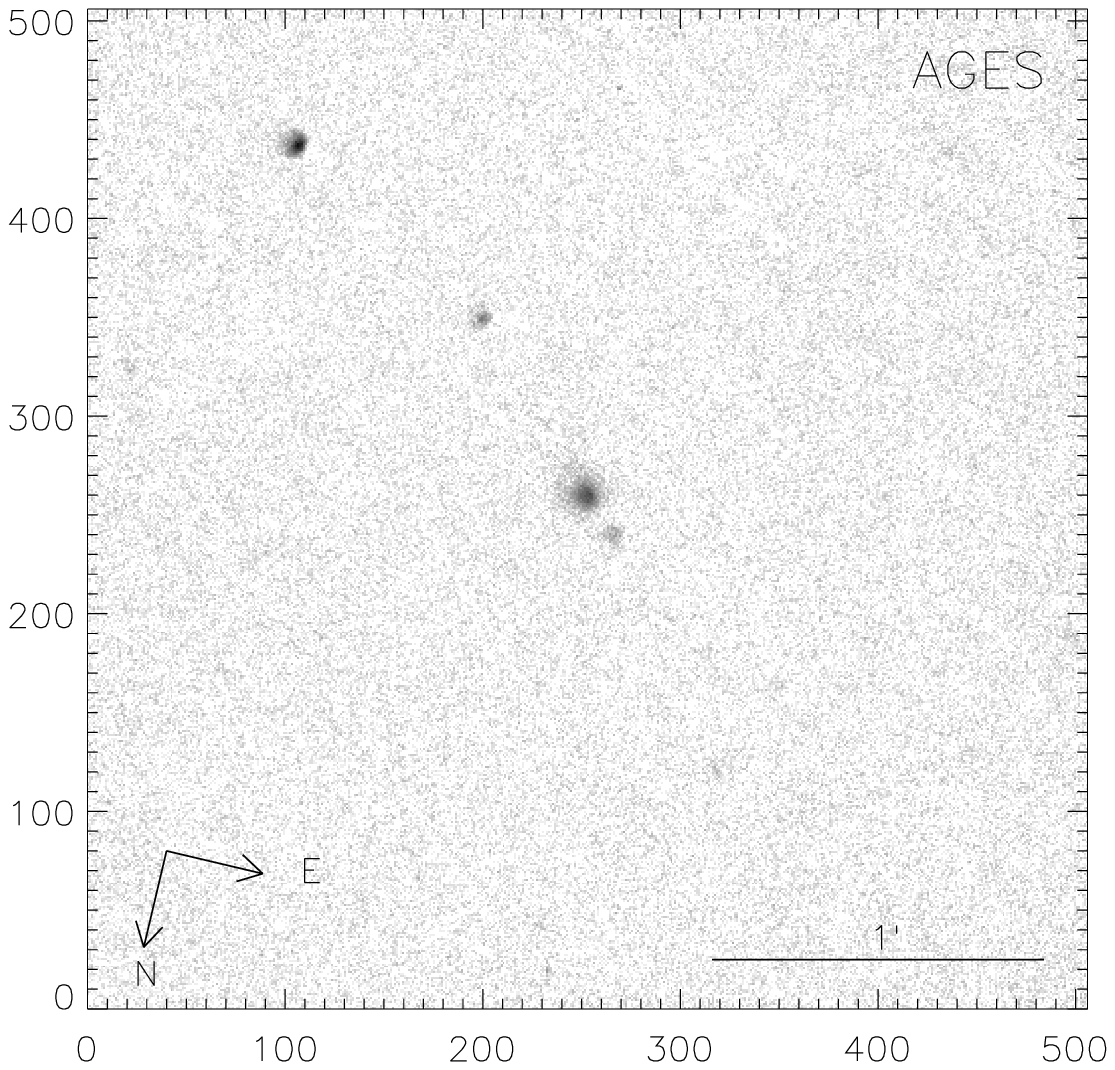}{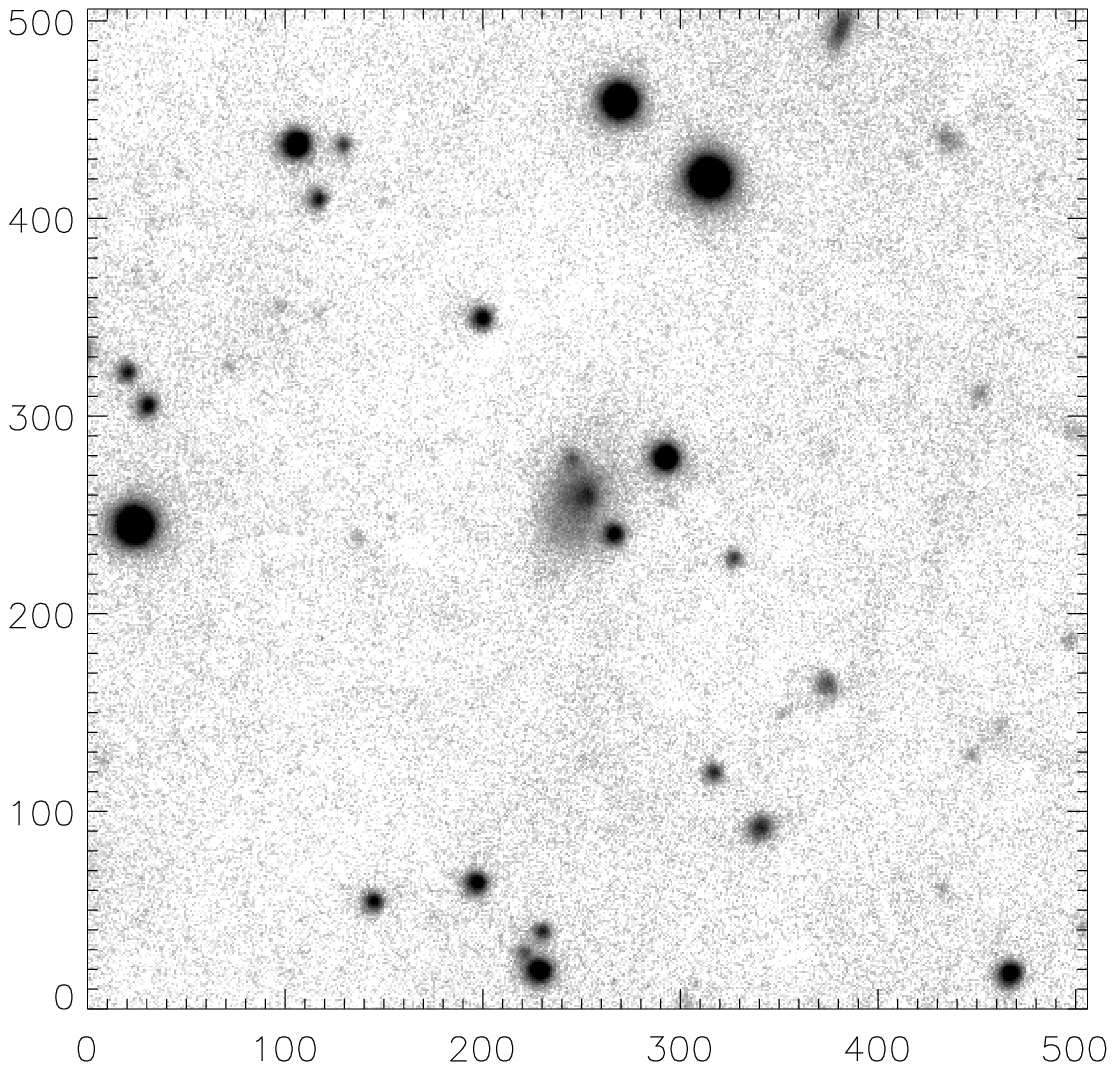}
\caption{Continuum-subtracted AGES H$\alpha$ (left) and $R$-band continuum 
(right) images of AGES J030039+254656.  The orientation and scale are given
in the left panel and are the same for both images.}
\label{halphaobs}
\end{figure}

Our data reduction followed the standard practice and was performed within
the MIDAS package. For all the data bias was subtracted and the
images were flat-fielded by twilight flats. Cosmic particles were
removed and the sky background was subtracted.
Then the images in the continuum were
normalized to $H\alpha$ images using 5--15 field stars and subtracted.
$H\alpha$ fluxes were obtained for the continuum-subtracted images,
using spectrophotometric standard stars
observed in the same nights as the object.  The continuum-subtracted and
$R$-band continuum images are shown in Figure \ref{halphaobs}.
The investigation of measurement errors contributed from the
continuum subtraction, flat-fielding and scatter in the zeropoints
has shown that they have typical values within 10\%. We did not
correct $H\alpha$ fluxes for the contribution of the [NII] lines,
because it is likely to be small for the low-luminosity galaxy.
We find an H$\alpha$ flux of $7.1 \pm 0.7 \times 
10^{-22}$ W cm$^{-2}$, which (after correction for galactic absorption)
gives a star formation rate of $8.7 \pm 0.9 \times 10^{-4}$ M$_\odot$ yr$^{-1}$.

\section{Results and Discussion}

\subsection{The NGC 7332 group}

Three galaxies were detected in neutral hydrogen within the NGC 7332 group.
In addition to NGC 7339, two new sources, AGES J223829+235135 
and AGES J223627+234258, were found in H\,{\sc i}.  Both of these have optical
counterparts in the Digitized Sky Survey but have not been previously 
catalogued.

\begin{figure}
\plotone{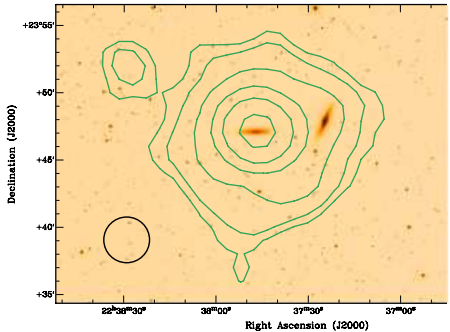}
\caption{Integrated H\,{\sc i} (moment 0) map of the NGC 7332 group overlaid 
on the digitized sky survey image of the region.  Contours are at
3, 5, 10, 25, 50 and 100 $\sigma$, where 1$\sigma$ = 0.04 Jy km\,s$^{-1}$ 
beam$^{-1} = 1.5 \times 10^{18}$ cm$^{-2}$.  The AGES beam is
shown as a black ellipse at the bottom-left.}
\label{mom0}
\end{figure}

Figure \ref{mom0}
shows the integrated H\,{\sc i} (moment 0) map of the group over the velocity 
range 1160 -- 1520 km\,s$^{-1}$, overlaid on the digitized sky survey image of
the region.  The two new sources can be seen, and it is also clear that the 
H\,{\sc i} detected at the position of NGC 7332 is associated with NGC 7339. 

No obvious tidal streams or H\,{\sc i} clouds were found in the group.  For
a velocity width of 50 km\,s$^{-1}$, the 5$\sigma$ detection limit would be
a column density of $4\times 10^{18}$ cm$^{-2}$ for features that fill 
the beam (FWHM = 23 kpc at the distance of NGC 7332).  The limit for features 
that do not fill the beam will rise as the reciprocal of their filing factor;
a feature with a column density of 10$^{20}$ cm$^{-2}$ would be seen
down to a filling factor of 4 per cent, for instance.

\subsubsection{NGC 7332 and NGC 7339}

The flux of NGC 7339 was found by fitting a 2D Gaussian as described in
Section 2.1.  This gave a beam-corrected flux of $9.1 \pm 
0.9$ Jy km\,s$^{-1}$ and deconvolved major and minor axes (FWHM of the 
Gaussian) for the H\,{\sc i}
of $215 \pm 14 \times 167 \pm 11$ arc seconds, at a position angle of $81
\pm 9$ degrees (north through east).  The fitted position for the H\,{\sc i} 
is 22:37:46.2, 23:47:12, 0.2 arc minutes from the position in NED (taken
from the 2MASS catalog, Jarrett et al. 2000).  The position
angle is consistent with alignment of the neutral hydrogen with the optical 
disc of the galaxy.

Our results are consistent with the flux measured using the Lovell
Telescope at Jodrell Bank by Staveley-Smith \& Davies (1987) of $9.8\pm 1.5$ 
Jy km\,s$^{-1}$.  From our determination of the size of the H\,{\sc i}, the
galaxy should appear as a point source to the Lovell Telescope (beam size $\sim
12$ arc minutes), so the Staveley-Smith \& Davies flux should not
be affected by the size of the source.

\begin{figure}
\plotone{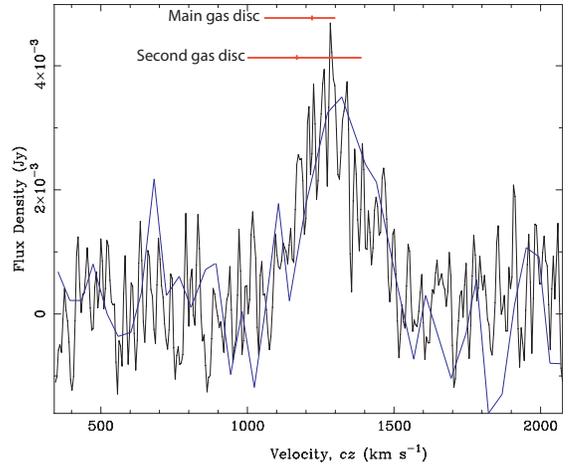}
\caption{H\,{\sc i} spectrum at the position of NGC 7332.  This matches very
well with the detection of Knapp et al. (1978), which is overlaid in blue.
The two red bars at the top indicate the central velocity (vertical tick) and
the velocity extent (horizontal length) of the main and second 
(counter-rotating) ionised gas discs seen in NGC 7339 by Plana \& Boulesteix 
(1996).  It can be seen that the neutral hydrogen is at a generally higher 
velocity than the ionised gas, althogh there is a significant overlap.}
\label{n7332spec}
\end{figure}

\begin{figure}
\plotone{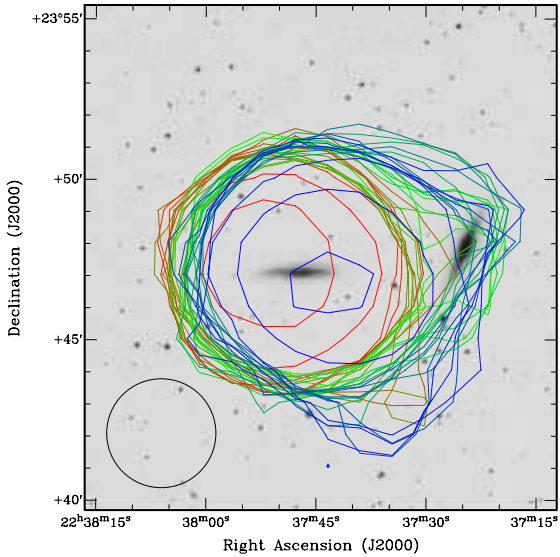}
\caption{Renzogram of the H\,{\sc i} in NGC 7339 overlaid on the Digitized
Sky Survey $B$-band image of the region showing NGC 7332 (right) and NGC 7339
(centre).  This shows contours at the 5$\sigma$ (2 mJy beam$^{-1}$)
level, colour-coded to indicate the velocity channel from which there were
taken (blue for lower recessional velocity, red for higher).
The rotation of NGC 7339 can be clearly seen, with the west side having a 
lower recessional velocity and 
the east higher.  The majority of the distortion over the position of NGC 7332
occurs at intermediate velocities ($\sim 1300$ km\,s$^{-1}$), indicated by
green contours.  The black ellipse at the lower left shows the size of the
Arecibo beam.}
\label{renzogram}
\end{figure}

We find flux at the position of NGC 7332 (Fig. \ref{n7332spec}) which matches 
very well with the detection of Knapp et al. (1978).  As the ALFA 
receiver has much lower 
sidelobes than the old circular feed used by Knapp et al., such a match would
not be expected if their detection were merely NGC 7339 in the sidelobes, as
has been suggested (Biermann, Clarke \& Fricke 1979; Haynes 1981).  Our 
mapping of the region (Figs. \ref{mom0} and 
\ref{renzogram}) shows that the gas disc of NGC 7339 is distorted, presumably
by an interaction with NGC 7332.  The gas at this position appears to be part 
of this distorted disc rather than a separate peak in the H\,{\sc i}
distribution associated with NGC 7332.  

Morganti et al. (2006) detected a small gas cloud between NGC 7332 and
NGC 7339 with the WSRT (FWHM $46^{\prime\prime} \times 31^{\prime\prime}$).  
This is too small and distant to account for the 
neutral hydrogen
we see at the position of NGC 7332.  It seems likely that we are seeing 
lower column-density gas not seen by the WSRT that is at or near the position
of NGC 7332 -- our observations are $\sim 4$ times more sensitive to low
column-density gas that fills the beam (albeit with a larger beam).  It is 
possible that the Morganti et al.
cloud is part of a low column-density
stream that is feeding the counter-rotating disc in NGC 7332.

\subsubsection{Dwarf galaxies in the NGC 7332 group}

\begin{figure}
\plottwo{fig11a.ps}{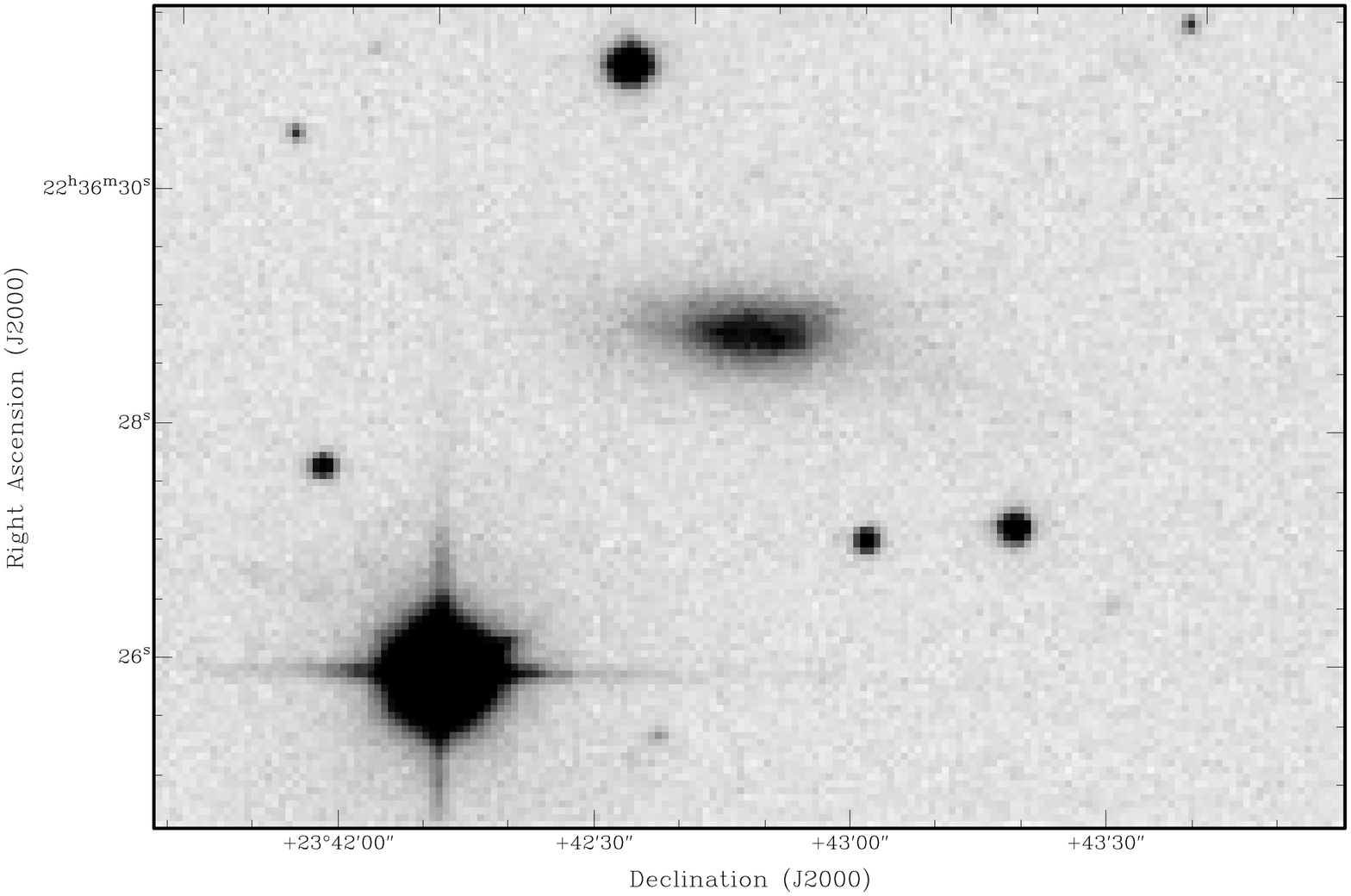}\\
\plottwo{fig11c.ps}{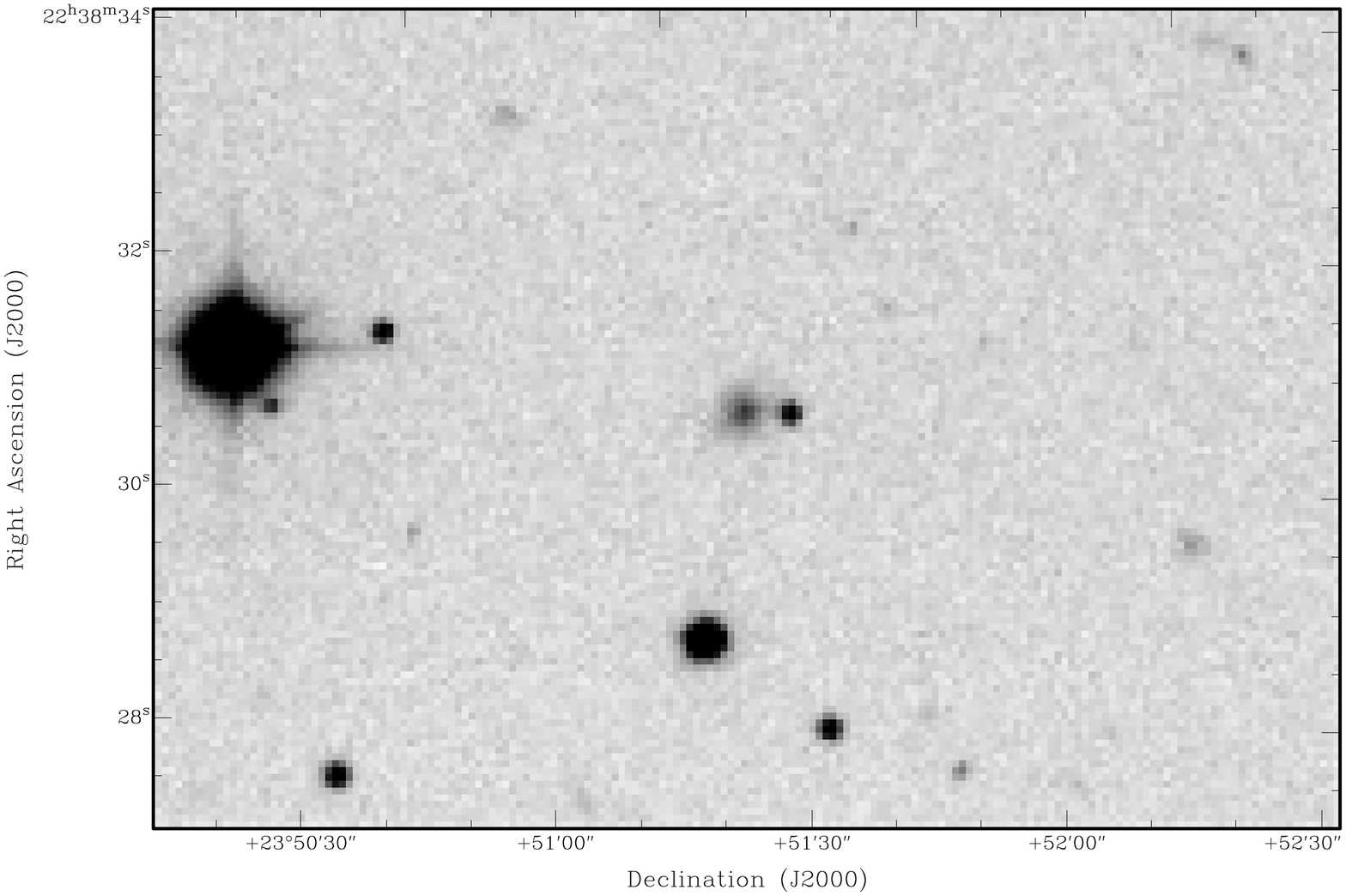}\\
\caption{H\,{\sc i} spectra (left) and Sloan Digital Sky Survey 
$g$-band images (right) of AGES J223627+234258 (top) and AGES J223829+235134 
(bottom).}
\label{newgals}
\end{figure}

The two new H\,{\sc i} sources reported here (AGES J223627+234258 and AGES 
J223829+235135) both lie in the same velocity range as 
NGC 7339, at around 1400 km\,s$^{-1}$ and outside the area mapped by Haynes
(1981).  The spectra and optical (SDSS) images of these galaxies are shown in 
Figure \ref{newgals}.
Both of the new galaxies appear to be narrow, single-peaked sources, which is
consistent with their being dwarf galaxies.  No tidal features can be 
definitively identified linking them to the NGC 7332/9 pair, but that these
could be present at even lower column-densities than we reach here cannot be
ruled out.

The data cube was examined at the positions of the candidate dSph galaxy
identified in our optical survey of the region (see \S \ref{optsurv})
and also at the positions of the candidate dSph galaxies KKR 73 and KKR 72 
identified by Karachentseva et al. (1999).  This search found no neutral 
hydrogen signals in the redshift range of the group.

\begin{figure}
\plottwo{fig12a.ps}{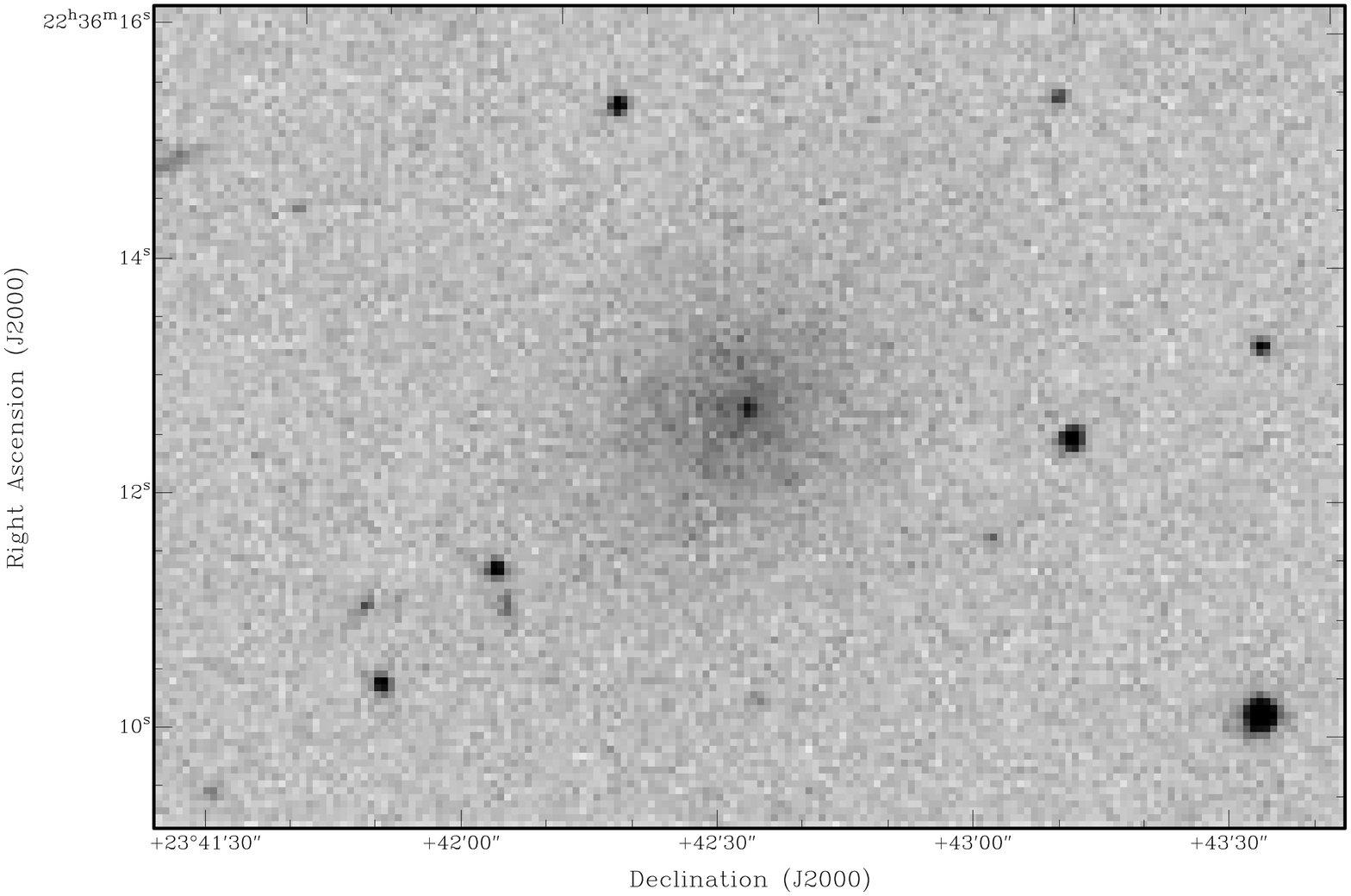}\\
\plottwo{fig12c.ps}{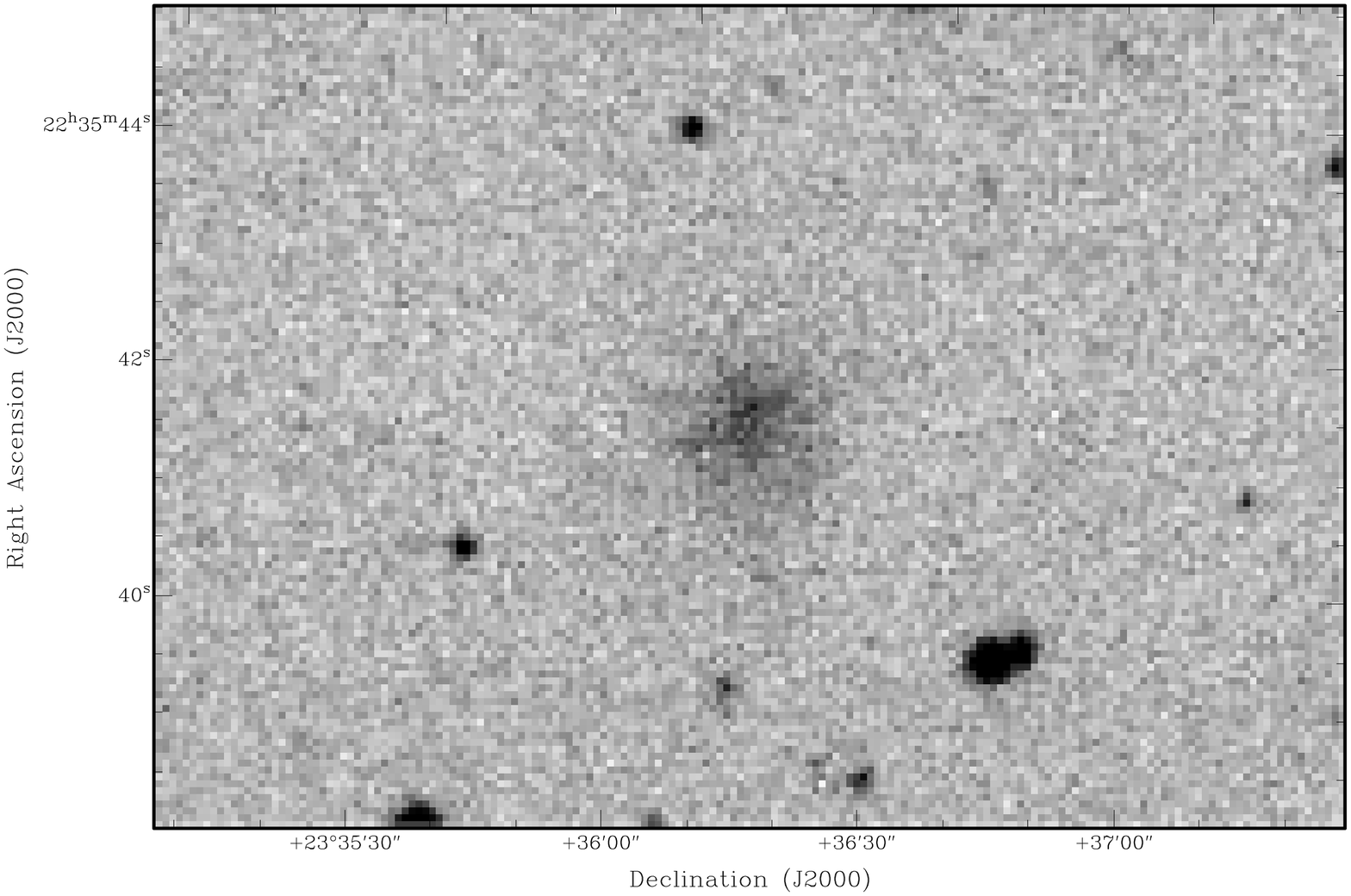}\\
\plottwo{fig12e.ps}{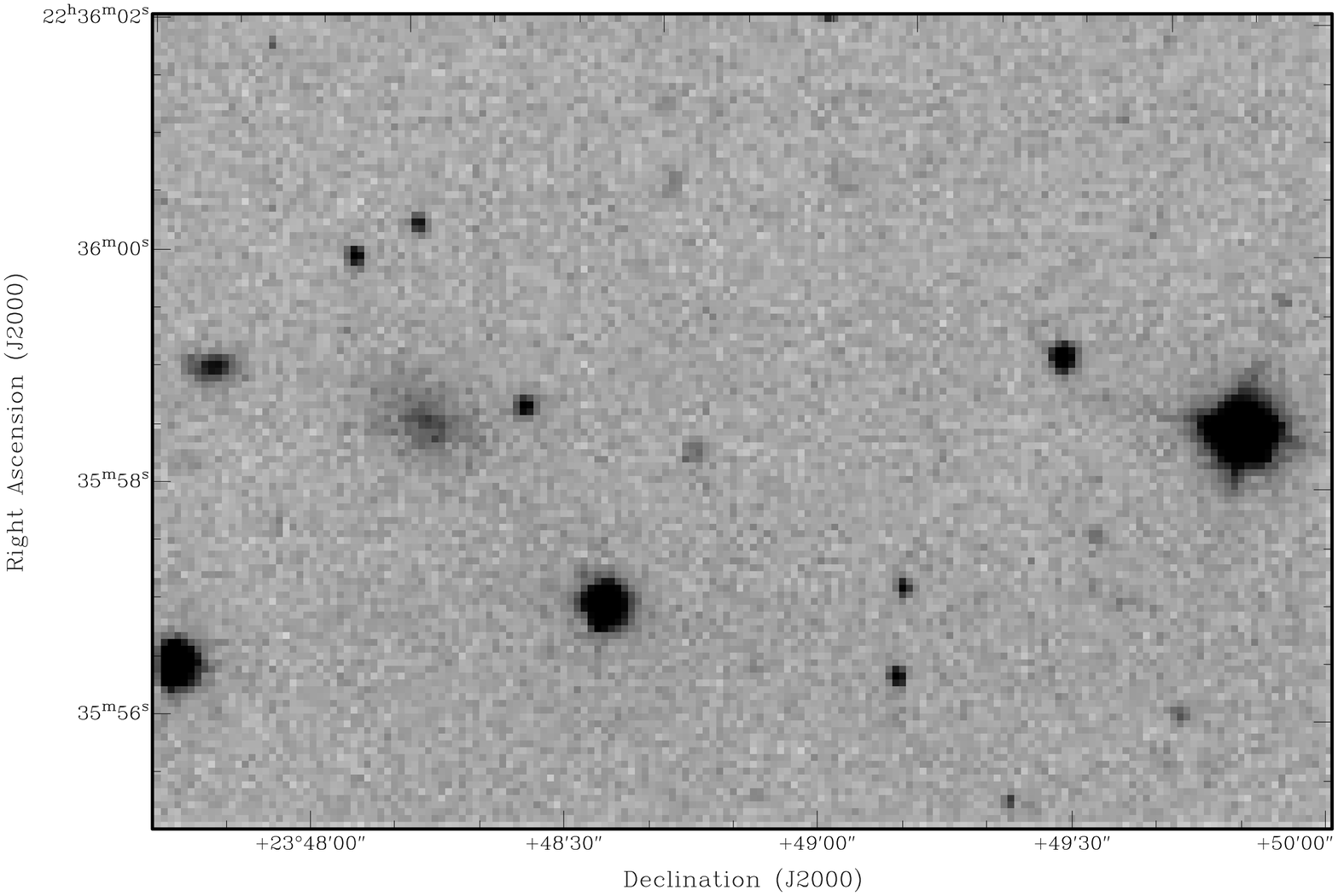}\\
\caption{Left: H\,{\sc i} spectra at the positions of the dwarf spheroidal 
candidates KKR 72 (top), KKR 73 (center) and J223558+234825 (bottom).  
These show no detection in the range 50 -- 3000 km\,s$^{-1}$.  Right: 
Sloan Digital Sky Survey $g$ band images of the galaxies.  Note that 
J223558+234825 is on the left of the image, due to its proximity to a scan
boundary.}
\label{dSphCandidates}
%\end{minipage}
\end{figure}

Our observations can place a limit on the H\,{\sc i} mass of these three
undetected galaxies (assuming them to be in the group) of
$1.4 \times 10^7 M_\odot$, almost an order of magnitude lower than the 
Huchtmeier et al. (2000) observations.  However, if these are dSph galaxies, 
as appears to be the case from their optical morphologies, then their 
non-detection at this limit is not surprising and we can neither confirm
nor rule out their being members of the group.  The spectra and DSS images
are shown in Fig. \ref{dSphCandidates}.

Our H\,{\sc i} mapping and optical search have brought the total number of
possible dwarf galaxies in the group to five, two of which are gas-rich and
are confirmed as group members.  This gives an overall ratio of small to large
galaxies of 1 -- 2.5 and of small to large H\,{\sc i}-rich galaxies of 2.  The
group appears, therefore, to be less dwarf-rich than would be expected from
the HIPASS H\,{\sc i} mass function (Zwaan et al. 2005), which (for our limit 
of $1.4\times
10^7 M_\odot$) predicts $5.0^{+1.5}_{-1.2}$ galaxies less than $10^8 M_\odot$
for each galaxy above $10^9 M_\odot$.  However, this under-density is less
than a 2$\sigma$ result and thus may simply be due to random fluctiations.

\subsubsection{Properties of the group galaxies}

All of the members and potential members of the NGC 7332 group have been
covered by the SDSS.  We have carried out ellipse fitting on the SDSS $g$-band
data using the Starlink
ESP package in order to find the surface-brightness profiles of the
dwarf galaxies (including those from Karachentseva et al. 1999) and have made
exponential fits to these profiles.  These fits were extrapolated to find the
total luminosity of the galaxy.  $g - r$ colours were found using apertures
fitted to cover as much of the galaxy as possible without including other
sources.  One galaxy (KKR 72) had a significant inner component on the
surface-brightness profile, this was fitted with a second exponential after
subtraction of the fit to the outer regions and was found to contribute 
negligibly to the overall luminosity.

The results of our fitting are shown in Table \ref{sdssprop}.  Column (1)
gives the ID of the galaxy.  Columns (2) and (3) give the optical R.A.
and declination, as found from the SDSS.  Columns (4) and (5) give the
central surface-brightness and the scale length for the exponential fit.
Column (6) gives the extrapolated total $g$-band magnitude.  Column (7) gives
the $g - r$ colour and Column (8) gives the $g$-band absolute luminosity, 
corrected for a $g$-band extinction of 0.15 mag. (from the IRSA extinction
calculator).

We find that AGES J223627+234258 has a fairly low H\,{\sc i} mass to light 
ratio for a dwarf irregular of $0.13 \pm 0.01 M_\odot/L_\odot$.  AGES 
J223829+235135 is more interesting, with a high H\,{\sc i} mass to light ratio 
of $2.0 \pm 0.2 M_\odot/L_\odot$, a very blue
colour ($g - r = 0.1\pm 0.1$), and a fairly low surface brightness (although
not as low as the dSph candidates).  From both its optical appearance and
its high gas fraction, this would appear to be an underevolved dIrr galaxy,
possibly similar to those studied in the Cen A group by Grossi et al. (2007).

The neutral hydrogen properties of the known or suspected members of the NGC 
7332 group are 
given in Table \ref{hiprops}.  Column (1) gives the ID of the galaxy.
Column (2) gives the velocity.  This is the H\,{\sc i} velocity measured from
our datacube where the source is detected and the optical velocity from Simien
\& Prugniel (1997) for NGC 7332.  Column (3) gives the velocity width at 
50\% of the peak flux.  For the detections, this is measured from our 
datacube, otherwise it is the  velocity width assumed in producing the 
flux limit (shown in italics).  Column (4) gives the
integrated flux, with an upper limit (based on a 3-$\sigma$ limit of 2.25 mJy
multiplied by the assumed velocity width) for non-detections and 
Column (5) gives this translated into a mass (or mass limit) at the 
adopted distance of 23 Mpc for the group.
Column (6) gives the H\,{\sc i} mass to light ratio derived from the measured
H\,{\sc i} mass and the SDSS $g$-band luminosity.  For NGC 7332 and NGC 7339,
the $g$-band luminosity has been calculated using the $B_T$ and $V_T$
magnitudes from the RC3 (de Vaucouleurs et al. 1991) and the transformation
$g = V + 0.60(B-V) - 0.12
\pm 0.02$ from the SDSS web pages, giving $g = 11.5\pm 0.3$ for NGC 7332 and
$g = 12.7\pm 0.3$ for NGC 7339.

\begin{deluxetable*}{llllllll}
%\begin{minipage}{150mm}
\tablecaption{SDSS properties of the dwarf galaxies in the NGC 7332 group.\label{sdssprop}}
\tabletypesize{\footnotesize}
%\renewcommand{\thefootnote}{\mbox{{${\alph{footnote}}$}}}
%\begin{tabular}{llllllll}
%\hline
\tablehead{
\colhead{ID}&
\colhead{RA}&
\colhead{Dec.}&
\colhead{$\mu_0 (g)$}&
\colhead{$h_g$}&
\colhead{$m_g$}&
\colhead{$g-r$}&
\colhead{$L_g$}\\
\colhead{}&
\colhead{(J2000)}&
\colhead{(J2000)}&
\colhead{(mag arcsec$^{-2}$)}&
\colhead{(arcsec)}&
\colhead{(mag)}&
\colhead{(mag)}&
\colhead{($10^7 L_\odot$)}\\
\colhead{(1)}&
\colhead{(2)}&
\colhead{(3)}&
\colhead{(4)}&
\colhead{(5)}&
\colhead{(6)}&
\colhead{(7)}&
\colhead{(8)}}
%\hline
\startdata
KKR 73&22:35:40.60&23:36:26.6&$23.51\pm 0.02$&$6.3\pm 0.2$&$17.5\pm 0.2$&$0.5\pm 0.1$&$6.7\pm 0.5$\\
J223558+234825&22:35:58.04&23:48:22.4&$23.51\pm 0.02$&$4.40\pm 0.15$&$18.3\pm 0.2$&$0.5\pm 0.1$&$3.2\pm 0.2$\\
KKR 72&22:36:11.75&23:42:43.46&$23.59\pm 0.02$&$12.4\pm 0.2$&$16.1\pm 0.1$&$0.4\pm 0.1$&$24.1\pm 1.6$\\ 
AGES J223627+234258&22:36:27.82&23:42:59.4&$21.84\pm 0.06$&$7.98\pm 0.16$&$15.3\pm 0.1$&$0.4\pm 0.1$&$49.8\pm 1.9$\\
AGES J223829+235135&22:38:29.69&23:51:31.1&$22.28\pm 0.02$&$2.26\pm 0.05$&$18.5\pm 0.1$&$0.1\pm 0.1$&$2.6\pm 0.1$\\
%\end{tabular}
%\end{minipage}
\enddata
\end{deluxetable*}

\begin{deluxetable*}{llllll}
%\begin{minipage}{150mm}
\tablecaption{H\,{\sc i} properties of the NGC 7332 group galaxies.
\label{hiprops}}
%\renewcommand{\thefootnote}{\mbox{{${\alph{footnote}}$}}}
%\begin{tabular}{llllll}
%\hline
\tablehead{
\colhead{ID}&
\colhead{Vel.}&
\colhead{$\Delta V_{50}$}&
\colhead{$F_{HI}$}&
\colhead{$M_{HI}$}&
\colhead{$M_{HI}/L_g$}\\
\colhead{}&
\colhead{(km\,s$^{-1}$)}&
\colhead{(km\,s$^{-1}$)}&
\colhead{(Jy km\,s$^{-1}$)}&
\colhead{($10^7
M_\odot$)}&
\colhead{$(M_\odot/L_\odot)$}\\
\colhead{(1)}&
\colhead{(2)}&
\colhead{(3)}&
\colhead{(4)}&
\colhead{(5)}&
\colhead{(6)}}
%\hline
\startdata
KKR 73&\nodata&{\it 50} \tablenotemark{a}&$<0.11$&$<1.4$&$<0.21$\\
J223558+234825&\nodata&{\it 50} \tablenotemark{a}&$<0.11$&$<1.4$&$<0.44$\\
KKR 72&\nodata&{\it 50} \tablenotemark{a}&$<0.11$&$<1.4$&$<0.06$\\
AGES J223627+234258&1409&58&$0.54\pm 0.02$&$6.7\pm 0.3$&$0.13\pm 0.01$\\
NGC 7332&1172&{\it 300} \tablenotemark{a}&$<0.68$&$<8.4$&$<0.005$\\
NGC 7339&1341&327&$9.1\pm 0.9$&$114\pm 11$&$0.19\pm 0.03$\\
AGES J223829+235135&1414&32&$0.41\pm 0.3$&$5.1\pm 0.4$&$2.0\pm 0.2$\\
%\hline
\enddata
\tablenotetext{a}{Value used in estimating upper limits}
%\end{tabular}
%\end{minipage}
\end{deluxetable*}

\subsection{NGC 1156 and the surrounding region.}

NGC 1156 was analysed in the same manner as NGC 7339 (above).  This gave
a beam-corrected integrated flux of $75.6 \pm 6.4$ Jy km\,s$^{-1}$, which
is consistent with the values of Swaters et al. (2002)
and Haynes et al. (1998).  This gives an H\,{\sc i} mass for the galaxy 
of $1.08 \pm 0.09 \times
10^9 M_\odot$.  The fitted gaussian has deconvolved major and minor axes of 
$272 \pm 16$ and $212 \pm 13$ arc seconds respectively, at a position angle
of $93 \pm 3$ degrees (north through east), which appears consistent with
the Swaters et al. WSRT map of the galaxy.  The fitted position of the galaxy
is 02:59:52.3, 25:14:30 (J2000), 0.3 arc minutes from the NED position (taken
from the 2MASS catalog, Jarrett et al. 2000).

\begin{figure}
\plotone{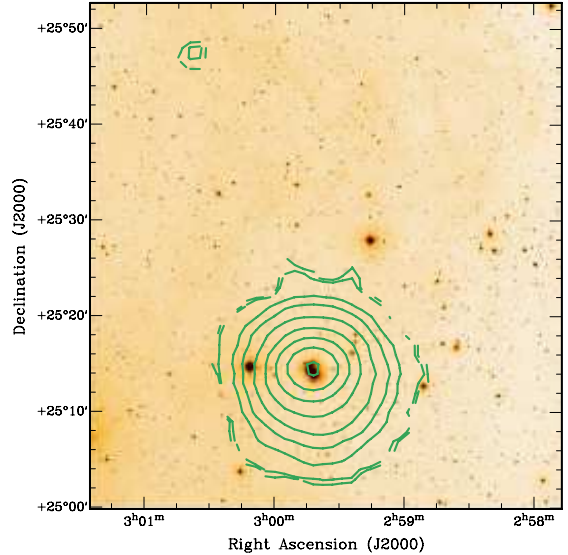}
\caption{DSS2 $B$-band image of the NGC 1156 region overlaid with contours
from a 3-sigma clipped integrated H\,{\sc i} (moment 0) map.  Contours are at 
0.05, 0.1, 0.2, 0.5, 1, 2, 4, 8, 16 and 32 Jy km\,s$^{-1}$.}
\label{n1156region}
\end{figure}

\begin{figure}
\plottwo{fig14a.ps}{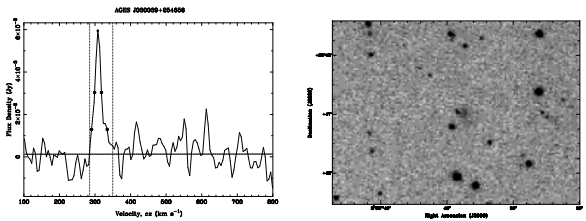}
\caption{H\,{\sc i} spectrum (left) and Super-COSMOS scanned POSS II $B_J$-band 
image (right) of AGES J030039+254656.}
\label{j0300+2456}
\end{figure}

AGES J030039+254656 is a small dwarf galaxy found 35 arc minutes 
north-north-east of NGC 1156 (80 kpc in projection), as can be seen in 
Figure \ref{n1156region} showing the integrated H\,{\sc i} (moment 0) map.  An 
optical counterpart is clearly visible on the digitized sky survey (see 
Figure \ref{j0300+2456}) at 03:00:38.6, 25:47:02 (J2000.  It has an H\,{\sc i}
flux of $0.114 \pm 0.032$ Jy km\,s$^{-1}$, giving it a neutral hydrogen
mass of $1.63 \pm 0.46\times 10^6 M_\odot$.  

As stated in \S \ref{newdwarfobs}, we measure a SFR of $8.7 \pm 0.9 \times 
10^{-4} M_\odot$ yr$^{-1}$ for AGES J030039+254656, compared to $0.71\pm 
0.07 M_\odot$ yr$^{-1}$ for NGC 1156.  The star-formation rate and 
H\,{\sc i} mass are thus both around three order of magnitudes lower than 
NGC 1156.  From the SuperCOSMOS
plate-scanning survey (see \S \ref{bggals}), we find $B_J = 20.14\pm 0.3$, 
$R = 19.69\pm 0.3$ and $B - R = 0.45 \pm 0.07$.  Using the extinction maps of 
Schlegel et al. (1998), we find a very blue corrected color of $B - R = -0.03$
and $L_B = 2.7 \pm 0.8 \times 10^6 L_\odot$, giving $M_{HI}/L_B = 0.61 \pm 0.26
M_\odot/L_\odot$.

The new galaxy is on the same side of the galaxy  as the possible 
tidal tail reported by Hunter et al. (2002) as having been seen in VLA 
observations.  This possible tail stretched 3$^\prime$ south-east
from an H\,{\sc i} complex to the north-east of the galaxy that has an
unsually large velocity spread.  The tail is not seen in the AGES
data, however as its length is less than the Arecibo beam size we would
not expect to be able to distinguish it from the main body of H\,{\sc i} in
NGC 1156.

Following Verley et al.
(2007), we can estimate the ratio of the tidal force being exerted currently 
by AGES J030039+254656 to the binding force of NGC 1156. If we assume that 
the H\,{\sc i} mass can be related to the total mass by
$M_{\rm tot} = \gamma M_{\rm HI}$ and use the H\,{\sc i} radius of
212 arcseconds measured for NGC 1156 by Swaters et al. (2002), then:

%\begin{eqnarray}
%Q_{i \rm p} & = \frac{F_{\rm tidal}}{F_{\rm bind}}\\
%& = \left(\frac{M_i}{M_{\rm p}}\right) \left(\frac{D_p}{S_{i \rm p}}\right)^3
%\end{eqnarray}

\begin{equation}
Q_{i \rm p} \; =\; \frac{F_{\mathrm tidal}}{F_{\mathrm bind}}\\
\; = \; \left(\frac{M_i}{M_{\rm p}}\right) \left(\frac{D_p}{S_{i \mathrm p}}\right)^3
\end{equation}

Where $M_i = \gamma_i M_{{\rm HI}, i}$ is the mass of the interactor (AGES 
J030039+254646), $M_{\rm p} = \gamma_{\rm p} M_{\rm HI, p}$
is the mass of the primary (NGC 1156), $D_{\rm p}$ is the diameter of the
primary, and $S_{i \rm p}$ is the separation between the two galaxies.  

This gives $Q_{i \mathrm p} = 1.2 \times 10^{-5} 
\frac{\gamma_i}{\gamma_{\mathrm p}} $, 
where a value of $\sim 10^{-2}$ or greater indicates significant tidal forces. 
The ratio $\frac{\gamma_i}{\gamma_{\rm p}}$ would therefore have to be of order
a thousand for this to be significant, that is AGES J030039+254656 would have 
to have a ratio of total mass to H\,{\sc i} mass a thousand times higher than 
that of NGC 1156.  It seems unlikely, therefore, that AGES J030039+254656, 
in its current position, could be exerting any significant tidal force on NGC
1156.  However, the possibility of a past interaction cannot be ruled out.

From the radio velocity difference ($77\pm 4$ km\,s$^{-1}$) and the
projected separation of 80 kpc between NGC 1156 and AGES J030039+254656,
the minimum dynamical mass of the pair is calculated to be $1.1\pm 0.1 
\times 10^{11} M_\odot$, giving $M_{dyn}/L_B \geq 65 M_\odot/L_\odot$.

Having only one (known) dwarf companion, NGC 1156 appears to be well below
the average.  Between our limit of $1.4\times 10^6 M_\odot$ and $10^8 M_\odot$,
the HIPASS H\,{\sc i} mass function (Zwaan et al. 2005) predicts 
$17.9^{+8.1}_{-5.5}$ dwarf galaxies for each galaxy of over $10^9 M_\odot$,
3$\sigma$ above what we observe in NGC 1156.

% v2.0 Revised by dvg on Jan 17 2010
% v3.0 dvg Mar 13 2010

\subsection{The AGES Volume}

\label{bggals}
%-----------------------------
The AGES cubes are bandpass-limited at a redshift of around 20,000 km\,s$^{-1}$
and have a combined area of 10 deg$^2$. This gives a comoving volume 
behind NGC 1156  and NGC 7332 of 24,000 Mpc$^3$. 
%%% DVG: \beta=v/c=0.0667 => z=0.06909=> V_comov=(10/41253)(4pi/3)D_c^3=24,000Mpc^3
In this volume we find a further 82 confirmed sources: 46 in the NGC 7332 cube
and 36 in the NGC 1156 cube. All the sources 
have been associated with optical counterparts 
(see Table~\ref{optidfinal}). Radio sources were correlated both with the 
APM\footnote{see \url{http://www.ast.cam.ac.uk/\~{}apmcat/}} 
on-line sky catalogue and with the SuperCOSMOS\footnote{see 
\url{http://www-wfau.roe.ac.uk/sss/pixel.html}} sky surveys, yielding
identifications within 85 arcseconds in all cases.  
Table~\ref{optidfinal} gives the positions of the optical counterparts, the
offsets from the H\,{\sc i} positions (H\,{\sc i}--optical), the 
$B_\mathrm{J}$ and $R$ magnitudes from the SuperCOSMOS survey (with errors of 
0.3 mag for m $>$ 15; Hambly, Irwin \& MacGillivray (2001), the H\,{\sc i} to 
$B_\mathrm{J}$ luminosity ratio, the literature name (if any), and the radial
velocity (where available) from the NED database\footnote{see 
\url{http://nedwww.ipac.caltech.edu}} (sources for quoted velocities are
given in Table \ref{nedredshifts}).  For some galaxies in the NGC 1156
region, photometry is available from CFHT observations in $g$ and $r$-bands
with the {\sc megacam} imager.  Aperture photometry for these sources is
given in footnotes to the table.  It should be noted that for low 
surface-brightness galaxies, the SuperCOSMOS magnitudes may dramatically
underestimate the luminosity (c.f. Disney \& Phillipps 1983); an example of 
this can be seen with AGES J030014+255335, which has $M_{HI}/L_B = 11.1\pm 3.8
M_\odot/L_\odot$ from the SuperCOSMOS photometry, while the deeper CFHT imaging
gives a much more feasible $M_{HI}/L_g = 2.0\pm 0.3 M_\odot/L_\odot$.  

\begin{figure}
\plotone{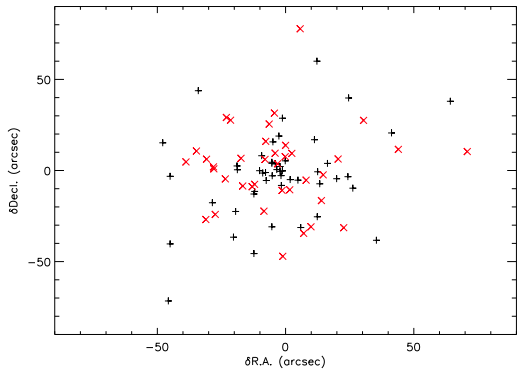}
\caption{Offsets (H\,{\sc i}--optical) for sources in the NGC 1156 field (red) 
and in the NGC 7332 field (black). 90\% of the sources have offsets 
smaller than 51 arcseconds.}
\label{offsetsnew}
\end{figure}

The distribution of separations can be seen in Fig. \ref{offsetsnew}.
The median offset between the H\,{\sc i} source and the identified optical 
counterpart is 20 arcsec, with 90 per cent of counterparts falling within 51 
arcsec of the H\,{\sc i} source.  It can be seen that the NGC 1156 field
appears to have a less centrally condensed distribution of offsets than the
NGC 7332 field, this is borne out by the statistics with the median offset
being 26 arcsec in NGC 1156 and only 19 arcsec in NGC 7332.  This may be 
simply natural variation, but could also be due
to the strong and variable optical absorption (A$_B$ = 0.89 to 1.45 mags)
in the NGC 1156 region which makes identification of optical counterparts
difficult.  Low surface-brightness counterparts could have been missed, or
areas of high absorption on spatial scales smaller than that mapped by
Schlegel et al. (1998) could have led to the correct counterpart being 
dismissed as being too faint.

Examining the corresondance between the errors on the H\,{\sc i} positions 
(Table \ref{catalogue} and the the offsets between the optical and H\,{\sc i} 
positions, it appears that 40/87 of the optical counterparts lie within the 
R.A. error and 47/87 within the declination error.  Treating the RA and dec
errors as defining an error ellipse rather than as indepdendent, only
22/87 sources fall within the error ellipse.  In order to enclose
69 per cent of the sources (e.g. 1 sigma), the error radius needs to be
increased by a factor of 2.95 -- it thus appears
that the error on the H\,{\sc i} position underestimates the true
positional uncertainty by around a factor of three.

% v2.3 - Revised - 22 mar 2010 - rfm
% v2.0 - Updated and corrected - 26 dec 2009 - dvg
% v2.1 - Corrected - 17 jan 2010 - dvg
% v2.2 - Final - 12 mar 2010 - dvg
\begin{deluxetable*}{lllrrrrrlr}
\tablecaption{Optical identifications of galaxies in the NGC 1156  and
NGC 7332 fields.\label{optidfinal}}
\tabletypesize{\footnotesize}
\tablewidth{0pt}
\tablehead{
          \colhead{AGES ID}&
          \colhead{R.A.}&
          \colhead{Dec}&
          \colhead{$\Delta \alpha$}&
          \colhead{$\Delta \delta$}&
          \colhead{$B_{\mathrm{J}}$}&
          \colhead{$R$}&
          \colhead{$M_{HI}/L_B$}&
          \colhead{Literature Name}&
          \colhead{Velocity}\\
          \colhead{}&
          \colhead{(J2000)}&
          \colhead{(J2000)}&
          \colhead{(sec)}&
          \colhead{(arcsec)}&
          \colhead{ }&
          \colhead{ }&
          \colhead{($M_\odot/L_\odot$)}&
          \colhead{}&
          \colhead{(km\,s$^{-1}$)}
          }
\startdata
J025512+243812&                  02:55:12.49 &24:38:05.9 &-0.59 &  6.1 &15.85 &15.59 &$0.14\pm 0.05$ &CGCG 484-024              &  6996 \\  
J025626+254614&                  02:56:27.73 &25:46:18.6 &-1.73 & -4.6 &20.29 &19.38 &$1.00\pm 0.56$ &                        	&       \\
J025737+244321&                  02:57:38.56 &24:43:20.1 &-2.06 &  0.9 &17.01 &16.27 &$0.48\pm 0.16$ &2MASX J02573860+2443194 	&       \\  
J025742+261755\tablenotemark{a}& 02:57:41.98 &26:16:37.2 & 0.42 & 77.8 &18.78 &18.04 &$3.19\pm 0.24$ &AGC 122807              	& 10370 \\
J025753+255737\tablenotemark{b}& 02:57:52.68 &25:58:11.5 & 0.52 &-34.5 &17.08 &16.69 &$0.65\pm 0.22$ &2MASX J02575260+2558114 	&       \\  
J025801+252556\tablenotemark{c}& 02:57:59.12 &25:25:26.9 &-1.70 & 29.1 &16.05 &15.50 &$0.44\pm 0.14$ &CGCG 485-002B           	&  6904 \\  
J025818+252711\tablenotemark{d}& 02:58:18.23 &25:26:54.9 &-0.57 & 16.0 &16.54 &15.61 &$0.63\pm 0.20$ &CGCG 485-004            	& 10510 \\   
J025826+241836& 		 02:58:22.35 &24:18:24.4 & 3.24 & 11.6 &18.01 &16.97 &$1.05\pm 0.36$ &2MASX J02582235+2418241 	&       \\
J025835+241844&                  02:58:37.36 &24:18:33.3 &-2.56 & 10.7 &17.21 &16.66 &$0.99\pm 0.23$ &2MASX J02583738+2418331 	&       \\  
J025836+251656\tablenotemark{e}& 02:58:35.51 &25:16:48.3 &-0.01 &  7.6 &16.05 &15.14 &$0.86\pm 0.28$ &UGC 02442               	& 10452 \\    
J025842+252348\tablenotemark{f}& 02:58:44.47 &25:23:41.8 &-2.27 &  6.2 &16.19 &15.24 &$0.39\pm 0.13$ &V Zw 298                	& 10375 \\   
J025843+254521& 		 02:58:43.83 &25:45:28.5 &-1.23 & -8.5 &18.74 &17.63 &$ 3.4\pm 1.1$  &UGC 02445  		&  7215 \\
J025902+253518& 		 02:59:00.29 &25:35:12.7 & 1.51 &  6.3 &18.31 &17.34 &$0.31\pm 0.12$ &    			&   	\\
J025917+244756\tablenotemark{g}& 02:59:17.48 &24:48:43.1 &-0.08 &-47.1 &19.80 &19.29 &$2.07\pm 0.74$ &  			&   	\\
J025930+255419& 		 02:59:31.58 &25:54:12.3 &-1.28 &  6.7 &17.02 &16.11 &$0.33\pm 0.11$ &2MASX J02593158+2554122 	&   	\\
J025936+253446\tablenotemark{h}& 02:59:34.97 &25:35:02.5 & 1.03 &-16.5 &19.22 &18.43 &$0.55\pm 0.22$ &      			&   	\\
J025942+251430\tablenotemark{i}& 02:59:42.30 &25:14:16.2 & 0.00 & 13.8 & --   & --   &$0.32\pm 0.06$ &NGC 1156  		&   375 \\   
J025953+254350\tablenotemark{j}& 02:59:51.33 &25:44:21.4 & 1.67 &-31.4 &20.43 &18.32 &$0.78\pm 0.37$ &    			&   	\\
J025954+241323& 		 02:59:54.98 &24:13:30.6 &-0.89 & -7.6 &16.25 &14.75 &$0.54\pm 0.18$ &UGC 02457  		& 10217 \\ 
J030008+241600& 		 03:00:08.12 &24:15:50.6 & 0.18 &  9.4 &17.43 &16.06 &$0.74\pm 0.28$ &2MASX J03000813+2415503   &   	\\
J030014+250315\tablenotemark{k}& 03:00:13.88 &25:03:25.6 & 0.12 &-10.6 &21.50 &--    &$11.1\pm 3.8$  &      			&   	\\
J030025+255335& 		 03:00:26.97 &25:53:32.9 &-2.07 &  2.0 &22.41 &20.58 &$ 7.3\pm 3.6$  &  			&   	\\
J030027+241301\tablenotemark{l}& 03:00:21.98 &24:12:50.6 & 5.22 & 10.4 &18.63 &17.88 &$1.52\pm 0.51$ &  			&   	\\
J030036+241156\tablenotemark{m}& 03:00:38.69 &24:12:22.9 &-2.29 &-26.9 &19.05 &18.45 &$1.13\pm 0.46$ &   			&	\\
J030039+254656& 		 03:00:38.71 &25:47:01.3 & 0.59 & -5.3 &20.14 &19.69 &$0.62\pm 0.26$ &  			&   	\\
J030112+242411& 		 03:01:12.46 &24:24:19.9 &-0.96 & -8.9 &16.92 &15.58 &$0.62\pm 0.20$ &2MFGC 02447  		&   	\\    
J030136+245602\tablenotemark{n}& 03:01:37.68 &24:55:34.4 &-1.58 & 27.6 &17.96 &17.12 &$0.24\pm 0.09$ &      			&   	\\
J030139+254442& 		 03:01:39.68 &25:44:16.5 &-0.47 & 25.5 &18.92 &18.37 &$2.27\pm 0.75$ &  			&   	\\
J030146+254314\tablenotemark{o}& 03:01:43.56 &25:42:46.5 & 2.24 & 27.5 &17.17 &16.02 &$0.66\pm 0.22$ &2MASX J03014354+2542474   &   	\\
J030200+250030& 		 03:02:00.32 &25:00:52.3 &-0.62 &-22.3 &17.68 &16.65 &$0.86\pm 0.28$ &2MASX J03020033+2500524   &   	\\    
J030204+254745& 		 03:02:04.30 &25:47:35.5 &-0.30 &  9.5 &17.16 &15.86 &$0.67\pm 0.22$ &FGC 0378   		&  6786 \\
J030234+244938& 		 03:02:34.20 &24:49:48.9 &-0.10 &-10.9 &20.03 &18.07 &$ 4.3\pm 1.4$  &    			&   	\\
J030254+260028& 		 03:02:54.43 &26:00:24.3 &-0.23 &  3.7 &18.33 &17.37 &$2.01\pm 0.66$ &LSBC F480-V04  		& 10618 \\
J030309+260407& 		 03:03:08.92 &26:03:35.5 &-0.32 & 31.5 &18.41 &17.90 &$0.63\pm 0.23$ &     			&   	\\
J030325+241510\tablenotemark{p}& 03:03:27.46 &24:15:05.3 &-2.86 &  4.7 &16.78 &15.81 &$0.33\pm 0.11$ &  			&   	\\
J030355+241922& 		 03:03:53.63 &24:19:24.4 & 1.08 & -2.4 &19.56 &18.94 &$2.23\pm 0.79$ &  			&   	\\
J030450+260045& 		 03:04:49.57 &26:01:15.8 & 0.73 &-30.9 &19.37 &18.39 &$0.88\pm 0.39$ &  			&   	\\ 
J030453+251532& 		 03:04:55.12 &25:15:56.1 &-2.02 &-24.1 &17.81 &17.07 &$0.60\pm 0.21$ &2MASX J03045528+2515558   & 	\\  
\enddata
\end{deluxetable*}

\addtocounter{table}{-1}
\begin{deluxetable*}{lllrrrrrlr}
\tablecaption{(continued)}
\tabletypesize{\footnotesize}
\tablewidth{0pt}
\tablehead{
          \colhead{AGES ID}&
          \colhead{R.A.}&
          \colhead{Dec}&
          \colhead{$\Delta \alpha$}&
          \colhead{$\Delta \delta$}&
          \colhead{$B_{\mathrm{J}}$}&
          \colhead{$R$}&
          \colhead{$M_{HI}/L_B$}&
          \colhead{Literature Name}&
          \colhead{Velocity}\\
          \colhead{}&
          \colhead{(J2000)}&
          \colhead{(J2000)}&
          \colhead{(sec)}&
          \colhead{(arcsec)}&
          \colhead{ }&
          \colhead{ }&
          \colhead{($M_\odot/L_\odot$)}&
          \colhead{}&
          \colhead{(km\,s$^{-1}$)}
          }
\startdata
J223111+234146& 		 22:31:11.09 &23:41:17.2 &-0.09 & 28.7 &18.15 &17.13 &$1.14\pm 0.42$ &      			&   	\\ 
J223122+230436& 		 22:31:21.98 &23:04:41.0 &  0.13& -5.0 &16.77 &15.97 &$0.45\pm 0.16$ &2MASX J22312197+2304404   &   	\\  
J223143+244513& 		 22:31:43.36 &24:44:58.0 & -0.36& 15.7 &16.02 &14.91 &$0.18\pm 0.06$ &MCG +04-53-003   	        &   	\\ 
J223213+232657& 		 22:32:13.17 &23:26:58.1 &-0.57 & -1.1 &16.76 &15.92 &$0.79\pm 0.26$ &2MASX J22321321+2326574   &   	\\    
J223218+235131\tablenotemark{q}& 22:32:19.88 &23:51:48.7 &-2.08 &-17.7 &15.98 &15.11 &$0.23\pm 0.08$ &CGCG 474-005  		& 11947 \\  
J223223+232613& 		 22:32:20.82 &23:26:51.3 & 2.58 &-38.3 &18.84 &17.87 &$ 3.2\pm 1.1$  &  	   		&   	\\
J223231+231601& 		 22:32:31.98 &23:16:37.6 &-1.48 &-36.6 &15.80 &11.03 &$0.24\pm 0.08$ &KUG 2230+230  		& 11782 \\  
J223236+235555& 		 22:32:36.55 &23:55:54.4 &-0.25 &  0.5 &16.41 &15.43 &$1.21\pm 0.39$ &UGC 12072 		&  7454 \\   
J223237+231209\tablenotemark{r}& 22:32:37.59 &23:12:39.9 &-0.39 &-30.9 &17.23 &16.63 &$0.44\pm 0.16$ &    			&   	\\  
J223245+243816& 		 22:32:44.30 &24:38:41.4 & 0.90 &-25.4 &17.98 &18.15 &$0.53\pm 0.21$ &  			&   	\\  
J223318+244545& 		 22:33:16.60 &24:45:41.1 & 1.19 &  3.9 &16.76 &15.75 &$0.62\pm 0.32$ &2MASX J22331661+2445408   &   	\\  
J223320+230400& 		 22:33:20.29 &23:04:00.2 &-0.09 & -0.2 &19.52 &18.72 &$ 4.0\pm 1.4$  &      			&   	\\  
J223329+231109& 		 22:33:27.31 &23:10:29.2 & 1.79 & 39.8 &17.51 &16.87 &$1.22\pm 0.41$ &   			&	\\   
J223342+242712& 		 22:33:37.42 &24:26:34.0 & 4.68 & 38.0 &17.49 &16.92 &$0.87\pm 0.30$ &  			&   	\\  
J223355+243114& 		 22:33:58.23 &24:32:25.6 &-3.33 &-71.6 &19.20 &18.29 &$1.63\pm 0.63$ &  			&   	\\  
J223415+233057& 		 22:34:18.28 &23:31:00.2 &-3.28 & -3.2 &17.07 &17.11 &$0.32\pm 0.12$ &  			&   	\\  
J223449+240744& 		 22:34:49.68 &24:07:55.5 &-0.88 &-11.5 &17.35 &16.71 &$0.35\pm 0.13$ &  			&   	\\   
J223502+235258\tablenotemark{s}& 22:35:00.25 &23:53:02.5 & 1.45 & -4.5 &17.61 &16.51 &$0.36\pm 0.15$ &   			&	\\
J223502+242131\tablenotemark{t}& 22:35:01.51 &24:20:31.0 & 0.89 & 60.0 &19.09 &18.04 &$1.88\pm 0.71$ &				&	\\
J223506+233707& 		 22:35:06.15 &23:37:12.5 &-0.55 & -5.5 &19.00 &18.51 &$ 3.9\pm 1.3$  &  			&   	\\
J223517+244317\tablenotemark{u}& 22:35:14.88 &24:43:26.7 & 1.91 & -9.7 &17.93 &17.17 &$0.47\pm 0.25$ &   			&	\\   
J223605+242407\tablenotemark{v}& 22:36:02.40 &24:24:12.0 & 2.30 & -5.0 &17.93 &16.67 &$0.26\pm 0.10$ &  			&   	\\    
& 		 		 22:36:03.47 &24:24:08.7 & 1.23 & -1.7 &18.03 &16.97 &$0.26\pm 0.10$ &  			&   	\\ 
J223613+243504& 		 22:36:13.39 &24:34:59.5 &-0.39 &  4.5 &16.75 &15.31 &$0.76\pm 0.25$ &2MASX J22361331+2434596 	&   	\\ 
J223627+234258& 		 22:36:27.85 &23:42:59.4 &-0.65 & -1.4 &17.04 &15.94 &$0.45\pm 0.15$ &      			&   	\\
J223628+245307& 		 22:36:30.28 &24:52:23.2 &-2.48 & 43.8 &19.29 &--    &$ 5.2\pm 2.1$  &  			&   	\\
J223631+240823&			 22:46:31.28 &24:08:14.9 &-0.68 &  8.1 &19.72 &18.27 &$ 3.6\pm 1.3$  &				&	\\
J223701+225532& 		 22:36:59.83 &22:55:39.3 & 0.97 & -7.3 &17.95 &17.88 &$2.25\pm 0.74$ &KUG 2234+226  		& 11511 \\ 
J223715+232957& 		 22:37:15.19 &23:29:38.1 &-0.19 & 18.9 &17.62 &16.76 &$0.48\pm 0.19$ &    			&  	\\   
J223739+244947& 		 22:37:38.28 &24:49:30.1 & 0.82 & 16.9 &17.73 &16.34 &$1.59\pm 0.54$ &  			&   	\\   
J223741+242520& 		 22:37:41.12 &24:25:28.2 &-0.12 & -8.2 &18.57 &18.13 &$3.66\pm 1.19$ &  			&   	\\ 
J223745+225309\tablenotemark{w}&  22:37:48.49 &22:52:53.8 &-3.49 & 15.2 &21.39 &19.37 &$11.3\pm 4.5$  &				&	\\
J223746+234712& 		 22:37:47.24 &23:47:12.1 &-0.74 & -0.1 &13.50 &11.60 &$0.30\pm 0.10$ &NGC 7339  		&  1313 \\   
J223823+245207\tablenotemark{x}&22:38:26.39 &24:51:14.4 & 3.01 & 20.6 &17.52 &16.22 &$0.71\pm 0.30$ &    			&  	\\ 
J223829+235135& 		 22:38:29.69 &23:51:31.4 &-0.29 &  3.6 &19.95 &--    &$ 4.3\pm 1.5$  &  			&   	\\ 
J223834+231114& 		 22:38:34.56 &23:11:11.7 &-0.16 &  2.2 &18.24 &16.93 &$1.82\pm 0.65$ &  			&   	\\ 
J223839+234247\tablenotemark{y}&22:38:39.38 &23:42:49.9 &-0.38 & -2.9 &17.89 &16.26 &$0.79\pm 0.29$ &   			&	\\ 
J223842+233156& 		 22:38:43.20 &23:32:08.9 &-0.90 &-12.9 &18.13 &16.91 &$1.60\pm 0.55$ &   			&	\\  
J223846+234923& 		 22:38:47.02 &23:49:45.5 &-1.42 &-22.5 &19.56 &18.76 &$ 3.8\pm 1.4$  &   			&	\\   
J223900+244752& 		 22:38:59.75 &24:47:57.3  &0.35 & -5.3 &16.75 &15.11 &$0.61\pm 0.21$ &2MASX J22385968+2447579 	&   	\\   
J223905+240651& 		 22:39:05.23 &24:06:53.8 &-0.13 & -2.8 &17.86 &16.48 &$1.85\pm 0.61$ &      			&   	\\   
J223946+242157\tablenotemark{z}&22:39:47.20 &24:22:42.6 &-0.90 & 45.6 &17.12 &15.45 &$0.32\pm 0.11$ &2MASX J22394715+2422428	&	\\
J224005+244154& 		 22:40:05.97 &24:41:53.6 &-1.37 &  0.4 &16.50 &14.94 &$1.99\pm 0.49$ &NSF J224006.18+244157.1 	& 11992 \\  
J224016+244658& 		 22:40:16.69 &24:46:54.1 &-0.39 &  3.9 &16.62 &15.22 &$0.14\pm 0.06$ &2MASX J22401669+2446537 	&   	\\   
J224024+243019& 		 22:40:25.38 &24:30:16.5 &-1.38 &  2.5 &19.08 &17.85 &$1.17\pm 0.46$ &      			&   	\\   
J224025+243925& 		 22:40:28.19 &24:40:05.3 &-3.28 &-40.3 &16.68 &15.07 &$0.32\pm 0.12$ &2MASX J22402816+2440055 	&   	\\
J224039+243229& 		 22:40:39.16 &24:32:29.5 &-0.16 & -0.5 &18.77 &18.06 &$1.74\pm 0.63$ &      			&   	\\    
J224052+234635&			 22:40:52.34 &23:46:29.3 &-0.01 &  5.3 &19.37 &19.01 &$2.36\pm 0.88$ &		        	&       \\  
J224110+243500& 		 22:41:09.47 &24:34:28.7 & 0.43 &-31.3 &18.22 &16.87 &$4.54\pm 1.49$ &  			&   	\\   
J224125+232228& 		 22:41:24.59 &23:22:28.7 & 0.91 & -0.7 &14.95 &14.09 &$0.22\pm 0.07$ &IC 5243  			&  7144 \\  
\enddata
\end{deluxetable*}

\addtocounter{table}{-1}
\begin{deluxetable*}{l}
\tablewidth{2\columnwidth}
\tablenotetext{a}{Fainter galaxy closer at 02:57:43.6, 26:17:54 with $B_\mathrm{J} = 21.3$.}
\tablenotetext{b}{Fainter galaxy closer at 02:47:53.0, 25:57:34 with $B_\mathrm{J} = 19.7$.}
\tablenotetext{c}{CGCG 484-026 at 6966 km\,s$^{-1}$ also lies within the beam at 02:58:03.2, 25:26:57 with $B_\mathrm{J} = 14.5$.  Spectrum appears to show two distinct components, consistent with two confused galaxies.}
\tablenotetext{d}{CFHT observations give $g = 14.52\pm 0.02$, $r = 14.87\pm 0.02$.}
\tablenotetext{e}{CFHT observations give $g = 15.04\pm 0.02$, $r = 14.76\pm 0.02$.}
\tablenotetext{f}{CFHT observations give $g = 15.09\pm 0.02$, $r = 14.59\pm 0.02$.}
\tablenotetext{g}{CFHT observations give $g = 18.31\pm 0.03$, $r = 18.03\pm 0.03$.  Fainter galaxy closer at 02:59:16.4, 24:48:30 with $B_\mathrm{J} = 20.4$, and other faint galaxies in the vicinity.}
\tablenotetext{h}{CFHT observations give $g = 17.99\pm 0.02$, $r = 17.76\pm 0.02$.}
\tablenotetext{i}{Too bright for SuperCOSMOS photometry, RC3 value of $12.08\pm 0.16$ used to calculate $M_{HI}/L_B$.}
\tablenotetext{j}{Fainter galaxy closer at 02:59:53.8, 25:43:58 with $B_\mathrm{J} = 22.5$.}
\tablenotetext{k}{CFHT observations give $g = 19.19\pm 0.03$, $r = 18.87\pm 0.03$.}
\tablenotetext{l}{Fainter galaxy closer at 03:00:25.6, +24:13:13 with $B_\mathrm{J} = 20.9$, and other even fainter galaxies closer than this.}
\tablenotetext{m}{Fainter galaxy closer at 03:00:37.3, +24:11:44 with $B_\mathrm{J} = 22.5$.}
\tablenotetext{n}{CFHT observations give $g = 17.24\pm 0.02$, $r = 16.60\pm 0.02$.  Fainter galaxy closer at 03:01:37.2, +24:56:13 with $B_\mathrm{J} = 20.8$.}
\tablenotetext{o}{CFHT observations give $g = 15.85\pm 0.02$, $r = 15.37\pm 0.02$.}
\tablenotetext{p}{May have star superposed.  Some much fainter galaxies lie closer.}
\tablenotetext{q}{A similar galaxy is found at 22:32:38.5 +23:11:41 with $B_\mathrm{J} = 18.3$.}
\tablenotetext{r}{Elongated image at 22:35:02.7 +23:52:55 with $B_\mathrm{J} = 22.6$.}
\tablenotetext{s}{May have star superposed.  Some much fainter galaxies lie closer.}
\tablenotetext{t}{Smaller and fainter galaxie s at 22:35:04.4, +24:21:57 with $B_\mathrm{J} - 20.9$ and at 22:35:17.5 +24:43:31 with $B_\mathrm{J} = 19.83$.}
\tablenotetext{u}{An elongated compact object lies at 22:38:29.65 +23:51:34.5, $B_J$=19.75, $R$=17.63, $I$=17.15}
\tablenotetext{v}{Pair of interacting galaxies.  $M_{HI}/L_B$ is average for the pair.}
\tablenotetext{w}{Multiple faint galaxies in the region.}
\tablenotetext{x}{Fainter galaxy closer at 22:38:23.1, +24:51:56 with $B_\mathrm{J} = 21.4$.}
\tablenotetext{y}{Fainter galaxy closer at 22:38:39.0, +23:42:56 with $R = 20.8$.}
\tablenotetext{z}{Fainter galaxy closer at 22:39:46.9, +24:21:33 with $B_\mathrm{J} = 21.63$.}
\end{deluxetable*}

\subsubsection{Large Scale Structure}

\begin{figure}
\plotone{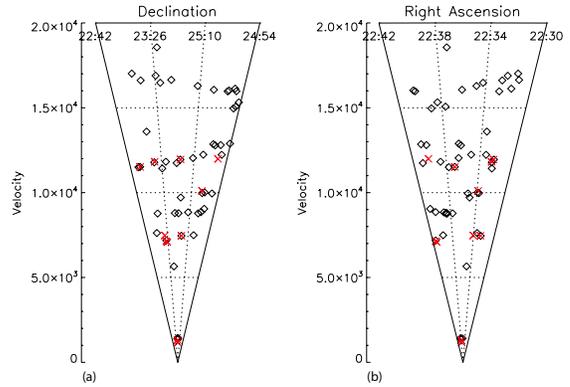}
\caption{Pie-slices of the large scale structure in the region behind NGC 7332: 
(a) Declination; (b) Right Ascension.  Literature galaxies (with redshifts) 
are shown as red crosses; AGES galaxies are shown as
black diamonds.  The horizontal axis has been expanded for clarity.}
\label{lss_all}
\end{figure}

\begin{figure}
\plotone{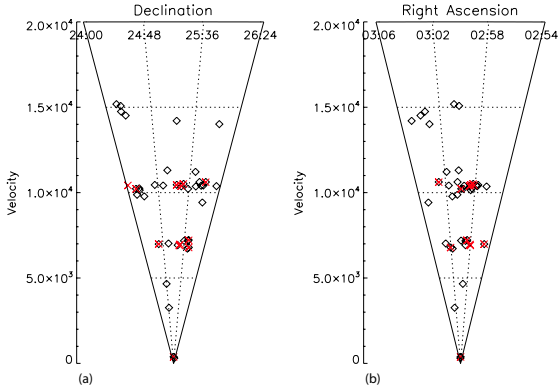}
\caption{Pie-slices of the large scale structure in the region behind NGC 1156: 
(a) Declination; (b) Right Ascension.  Literature galaxies (with redshifts) 
are shown as red crosses; AGES galaxies  are shown as black diamonds.  The 
horizontal axis has been expanded for clarity.}
\label{n1156lss}
\end{figure}

The large velocity range covered by AGES allows us to look at the large
scale structure behind the targetted regions.  Pie slices (Fig.
\ref{lss_all} for NGC 7332 and Fig. \ref{n1156lss} for NGC 1156) show the 
distribution of galaxies in declination and right ascension.  
These figures also show the previously known galaxies with redshifts, given in
Table \ref{nedredshifts}

\begin{deluxetable*}{lll}
\tablecaption{Literature redshifts for galaxies in the volume behind
NGC 1156 and NGC 7332\label{nedredshifts}}
\tablewidth{0pt}
\tablehead{
          \colhead{ID}&
	  \colhead{Redshift (km\,s$^{-1}$}&
	  \colhead{Reference}\\
	  }
\startdata
CGCG 484-024&            6996&  Springob et al. 2005\\
AGC 122807&		 10370& Saintonge et al. 2008\\
CGCG 485-002B&		 6904&  Springob et al. 2005\\
CGCG 484-026&		 7003&  Springob et al. 2005\\
CGCG 485-003&		 10429& Springob et al. 2005\\
CGCG 485-004&		 10510& Springob et al. 2005\\
UGC 02442&		 10452& Springob et al. 2005\\
UGC 02445&		 7215&  Springob et al. 2005\\
V Zw 298&		 10375& Springob et al. 2005\\
UGC 02457&		 10217& Springob et al. 2005\\
FGC 0378&		 6786&  Springob et al. 2005\\
LSBC F480-V04&		 10618& Saintonge et al. 2008\\
CGCG 474-005&            11947& Springob et al. 2005\\
KUG 2230+230&            11782& Springob et al. 2005\\
UGC 12072& 		 7454&  Springob et al. 2005\\
2MASX J22335064+2426393& 10103& Lawrence et al. 1999\\
IC 5231&		 7473&  Gregory et al. 2000\\
KUG 2234+226&		 11511& Springob et al. 2005\\
NSF J224006.18+244157.1& 11992& Aldering et al. 2007\\
IC 5242&		 7145&  Springob et al. 2005\\
IC 5243&		 7154&  Springob et al. 2005\\
\enddata
\end{deluxetable*}

From Fig. \ref{lss_all}, two structures appear to be present behind NGC
7332.  The first,
between 11000 and 12500 km\,s$^{-1}$ can be seen most easily in the 
declination slice.  The second, between 14000 and 18000 km\,s$^{-1}$, can
be seen in both slices.  Fitting surfaces to the distribution of galaxies
confirms that all 10 galaxies in the lower  redshift range and 12 out of 14
in the higher redshift range fall within 300 km\,s$^{-1}$ of the fitted 
surface.  The two in the higher redshift range that do not seem to be in the
structure both have R.A.$ > 22^h40^m$, while the other twelve are all at lower
R.A.s.

The redshift range 2000 -- 7000 km\,s$^{-1}$ behind NGC 7332 is identified
by Fairall (1990) as being through void regions.  This region does appear
to have a lower than normal galaxy density, with only one source (AGES 
J223506+233707) found.  AGES J223506+233707 is a nearly edge-on 
spiral at a recessional velocity of 5655 km\,s$^{-1}$, giving a distance of
80 Mpc for $H_0 = 71$ km\,s$^{-1}$ Mpc$^{-1}$.  It
does not seem disturbed either in its H\,{\sc i} profile
or in its optical (SDSS) appearance and has an H\,{\sc i} mass of $1.1 \pm 0.1
\times 10^9 M_\odot$, similar to NGC 1156 and the Large Magellanic Cloud.  
From aperture photometry performed on SDSS images, it has a $g = 18.7\pm 0.4$, 
giving $M_g = -15.8 \pm 0.2$ and $L_g = 2.3 \times 10^8 L_\odot$.  This
gives it a high $M_{HI}/L_g = 4.8 M_\odot/L_\odot$.  The color (again from 
aperture photometry on SDSS images) is $g-r = 0.2 \pm 0.6$, making this a
rather blue galaxy -- which is consistent with its high H\,{\sc i} mass to 
light ratio and makes this appear to be an under-evolved galaxy.  This may
be a good example of a galaxy that has evolved in a void region without
interacting significantly with other galaxies and thus has not evolved
as fast as other galaxies, but determining whether this is the case would
require much future study.

Figure \ref{n1156lss} shows large scale structure in the NGC 1156 region, which
is clearly present at $\sim 7000$ km\,s$^{-1}$ and just
about 10 000 km\,s$^{-1}$.  However, unlike NGC 7332, these are both visible
in literature redshift data, with galaxies around UGC 2445 making up the
closer structure and the further structure including the WBL 091 and PPS2
187 groups (White et al. 1999; Trasarti-Battistoni 1998).

\section{Conclusions}

One of the principle goals of the AGES is to investigate the population of
dwarf galaxies in different environments.  Our H\,{\sc i} mapping of the NGC 
7332 group has revealed two new dwarf 
members, doubling the number of confirmed group members.  An additional 
possible member has been found in an optical search, bringing the number of 
possible dSph companions without redshift information to three and the total 
number of possible galaxies in the group to seven.  In the NGC 1156 group,
we found a single dwarf companion that had not been previously catalogued. 
This does not appear to be large enough or currently close enough to NGC 1156
to have a significant effect on that galaxy.

In the background regions, we have detected 82 sources, 45 of which are entirely
new discoveries.  We find evidence of large scale structure in both cubes,
with newly-discovered structures at $\sim 12 000$ and $\sim 16 
000$ km\,s$^{-1}$ behind NGC 7332 and confirm structures visible in 
the literature redshifts at $\sim 7 000$
and $\sim 10 000$ km\,s$^{-1}$ behind NGC 7332.  We also find one galaxy in 
the void region between 2000 and 7000 km\,s$^{-1}$ behind NGC 7332.

\section*{Acknowledgements}

We thank Mike Irwin for keeping the archive of the APM surveys available,
Mark Calabretta and the ATNF for maintaining the Livedata and Gridzilla
multibeam processing packages and for modifying them to read Arecibo data and
Giuseppe Gavazzi and Luca Cortese for useful discussions and the optical 
spectroscopy observations presented in Section \ref{optspec}.  We also thank
the anonymous referee for his useful comments that helped improve the paper.

I.D.K. was partially supported by RFBR gants 10-02-00123 and
RUS-UKR 09-02-90414.

This research is partly based on observations obtained at the 
Canda-France-Hawaii Telescope (CFHT)
which is operated by the National Research Council of Canada, the Insitut
National des Sciences de l'Univers of the Centre National de la Recherche
Scientifique of France, and the University of Hawaii.

This research has made use of data obtained from the SuperCOSMOS Science
Archive, prepared and hosted by the Wide Field Astronomy Unit, Institute for
Astronomy, University of Edinburgh, which is funded by the UK Science and
Technology Facilities Council.

\clearpage

\begin{thebibliography}{}

\bibitem[a(1)]{Aldering07} Aldering G. et al., 2007, Central Bureau Electronic
Telegram, 991
\bibitem[a(1)]{Ajhar97} Ajhar E. A., Lauer T. R., Tonry J. L., Blakeslee J. P.,
 Dressler A., Holtzman J. A.; Postman M., 1997, AJ, 114, 626
\bibitem[a(1)]{Auld06} Auld R. et al., 2006, MNRAS, 371, 1617
\bibitem[a(1)]{Balkowski83} Balkowski C., Chamaraux P., 1983, A\&AS, 51, 331
\bibitem[a(1)]{Barazza01} Barazza F. D, Binggelli B., Prugniel P., 2001, A\&A,
373, 12
\bibitem[a(1)]{Barnes98} Barnes D. G. et al., 2001, MNRAS, 322, 486
\bibitem[a(1)]{Biermann79} Biermann P., Clarke J. N., Fricke K. J., 1979, A\&A,
75, 19
\bibitem[a(1)]{Bursteib87} Burstein D., Krumm N., Salpeter E. E., 1987, AJ, 94,
88
\bibitem[a(1)]{Condon98} Condon J., Cotton W. D., Greisen E. W., Yin Q. F., 
Perley R. A., Taylor G. B., Broderick J. J., 1998, AJ, 115, 1693
\bibitem[a(1)]{Cortese08} Cortese L. et al., 2008, MNRAS, 383, 1519
\bibitem[a(1)]{deVaucouleurs91} De Vaucouleurs G., de Vaucouleurs A., Corwin
H. G., But R. J., Paturel G., Fouque P., 1991, ``Third Reference Catalogue of 
Bright Galaxies'', Springer-Verlag, New York.
\bibitem[a(1)]{Disney83} Disney M. J., Phillpps S., 1983, MNRAS, 205, 1253
\bibitem[a(1)]{Fairall98} Fairall A. P., 1998, ``Large Scale Structures in the
Universe'', Wiley, New York.
\bibitem[a(1)]{Falcon04} Falc\'on-Barroso J. et al., 2004, MNRAS, 350, 35
\bibitem[a(1)]{Giovanelli05} Giovanelli R. et al., 2005, AJ, 130, 2613
\bibitem[a(1)]{Gregory00} Gregory S. A., Tifft W. G., Moody J. W., Newberry 
M. V., Hall S. M., 2000, AJ, 119, 545
\bibitem[a(1)]{Grossi07} Grossi M., Disney M. J., Pritzl B. J., Knezek P. M.,
Gallagher J. S., Minchin R. F., Freeman, K. C., 2007, MNRAS, 374, 107
\bibitem[a(1)]{Hambly01} Hambly N. C., Irwin M. J., MacGillivray H. T., 2001,
MNRAS, 326, 1295
\bibitem[a(1)]{Haynes81} Haynes M. P., 1981, AJ, 86, 1126
\bibitem[a(1)]{Haynes98} Haynes, M. P., van Zee, 
L., Hogg, D. E., Roberts, M. S., Maddalena, R. J., 1998, AJ, 115, 62 
\bibitem[a(1)]{Henning10} Henning P. A. et al. 2010, AJ, 139, 2130
\bibitem[a(1)]{Helou85} Helou G., Soifer B. T., Rowan-Robinson M., 1985, ApJ, 
298, L7
\bibitem[a(1)]{Huchtmeier00} Huchtmeier W. K., Karachentsev I. D., Karachentseva
V. E., 2000, A\&AS, 147, 187
\bibitem[a(1)]{Hunter02} Hunter D. A., Rubin V. C., Swaters R. A., Sparke L. S.,
Levine S. E., 2002, ApJ, 580, 194
\bibitem[a(1)]{James04} James P. A. et al., 2004, A\&A, 414, 23
\bibitem[a(1)]{Karachentseva73} Karachentseva V. E., Lebedev V. S., 
Shcherbanovskij A. L., 1973, Communications of the Special Astrophysical 
Observatory of the USSR AS, No. 8. 
\bibitem[a(1)]{Karachentseva99} Karachentseva V. E., Karachentsev I. D., 
Richter G. M., 1999, A\&AS, 135, 221
\bibitem[a(1)]{Karachentsev96} Karachentsev I,; Musella I., Grimaldi A., 1996, 
A\&A, 310, 722
\bibitem[a(1)]{Kent09} Kent B. R., Spekkens K., Giovanelli R., Haynes M. P., 
Momjian E., Cortes J. R., Hardy E., West A., 2009, Ap.J., 691, 1595
\bibitem[a(1)]{Knapp78} Knapp G. R., Kerr F. J., Williams B. A., 1978, ApJ, 
222, 800
\bibitem[a(1)]{Koribalski04} Koribalski B. S. et al., 2004, AJ, 128, 16
\bibitem[a(1)]{Lawrence99} Lawrence A. et al., 1999, MNRAS, 308, 897
\bibitem[a(1)]{Morganti06} Morganti R. et al., 2006, MNRAS, 371, 157
\bibitem[a(1)]{Moshir90} Moshir M. et al., 1990, ``IRAS Faint Souce Catalog,
 Version 2.0''
\bibitem[a(1)]{Plana96} Plana H., Boulesteix J., 1996, A\&A, 307, 391
\bibitem[a(1)]{Putman02} Putman M. E. et al., 2002, AJ, 123, 873
\bibitem[a(1)]{Saintonge08} Saintonge A., Giovanelli R., Haynes M., Hoffman, 
G. L., Kent B. R.,Martin A. M., Stierwalt S., Brosch N. , 2008, AJ, 135, 588
\bibitem[a(1)]{Schlegel98} Schlegel D. J., Finkbeiner D. P., Davis M., 1998, 
ApJ, 500, 525
\bibitem[a(1)]{Simien97} Simien F., Prugniel Ph., 1997, A\&AS, 126, 519
\bibitem[a(1)]{Springob05} Springob C. M., Haynes M. P., Giovanelli R., Kent 
B. R., 2005, ApJS, 160, 149
\bibitem[a(1)]{Tully88} Tully R. B., Fisher J. R., 1988, Catalog of Nearby 
Galaxies, CUP, Cambridge
\bibitem[a(1)]{Lister87} Staveley-Smith L., Davies R. D., 1987, MNRAS, 224, 953
\bibitem[a(1)]{Swaters02} Swaters R. A., van Albada T. S., van der Hulst J. M., 
Sancisi R.. 2002, A\&A, 390, 829
\bibitem[a(1)]{Tonry01} Tonry J. L. et al., 2001, ApJ, 546, 681
\bibitem[a(1)]{Trasarti-Battistoni98} Trasarti-Battistoni R., 1998, A\&AS, 130, 
341	
\bibitem[a(1)]{Verley07} Verley S., Leon S., Verdes-Montenegro L., Combes F.,
Sabater J., Sulentic J., Bergond G., Espada D., Garc\'ia E., Lisenfeld U., 
Odewahn S. C., 2007, A\&A, 472, 121
\bibitem[a(1)]{White99} White R. A., Bliton M., Bhavsar S. P., Bornmann P., 
Burns J. O., Ledlow M. J., Loken C., 1999, AJ, 118, 2014
\bibitem[a(1)]{Yun01} Yun M. S., Reddy N. A., Condon J. J., 2001, ApJ, 554, 803
\bibitem[a(1)]{Zwaan05} Zwaan M. A., Meyer M. J., Staveley-Smith L., Webster R.
L., 2005, MNRAS, 359, 30L

\end{thebibliography}
\end{document}